\documentclass[prb,showpacs,amssymb,amsmath,twocolumn,superscriptaddress]{revtex4}

\usepackage{graphicx}
\usepackage{bm}
\def\be{\begin{align}}
\def\ee{\end{align}}
\def\bea{\begin{eqnarray}}
\def\eea{\end{eqnarray}}

\renewcommand{\vec}{\mathbf}

\newcommand{\sgn}{\mathrm{sgn}}

\begin{document}
\title{Self-equilibration theorem in quantum-point contacts of interacting electrons: \\
time-dependent quantum fluctuations of tunnel transport \\ 
beyond the Levitov-Lesovik scattering approach}
\author{Gleb A. Skorobagatko}

\affiliation{%
\mbox{%
Institute for Condensed Matter Physics of National Academy of Sciences of Ukraine,
  Svientsitskii Str.1,79011 Lviv, Ukraine%
}}

\email{ <gleb.a.skor@gmail.com>}

\date{\today}

\begin{abstract}
Equilibration to  the steady state for a wide class of Luttinger liquid ballistic weakly linked tunnel contacts is extensively studied. Quantum fluctuations of tunnel current are considered in all orders in tunnel coupling and out of the equilibrium in the time domain. Especially, two important mathematical statements: Self-equilibration (SE-)theorem and Self-equilibration (SE-)lemma on the exact re-exponentiation of thermal average from the Keldysh-contour-ordered evolution operator for arbitrary weakly linked Luttinger liquid tunnel contact are proven. Demonstrated proof of SE-theorem and SE-lemma  represents first evidence of a novel emergent  phenomenon of \textit{"self-equilibration"} in the dynamics of quantum fluctuations of electron transport through ballistic tunnel junctions. This phenomenon and all related real-time full counting statistics are shown to be much more general as compared to the usual Levitov-Lesovik scattering approach for the non-interacting electrons, though SE-theorem also contains known results obtained within the Levitov-Lesovik scattering approach as corresponding  limiting case for lowest order cumulants at $ g=1 $. As the result, corresponding differential equation of "self-equilibration" for time-dependent Keldysh partition function of tunnel contact is derived and explained. It is shown that proven SE-theorem can be considered also as the dynamical version of well-known Jarzynski equality for the tunnel electron transport out of the equilibrium. As the result, for arbitrary weakly linked tunnel junction model of interest the exact time-dependent cumulant generating functional is derived being valid at arbitrary electron-electron repulsion in the Luttinger liquid leads of the junction, arbitrary temperature and arbitrary bias voltage. Respective general formula in its long-time asymptote turns into a non-perturbative Luttinger liquid generalisation of the Levitov-Lesovik type of detailed balance formula for cumulant generating function. As the consequence of obtained results, a universal character of  self-equilibration of tunnel current in one-dimensional weakly linked tunnel junctions is revealed and studied on the level of non-equilibrium Fano factor. As well, a new measure of disequilibrium in such systems - the "steady flow" rate - is also introduced and discussed.  
\end{abstract}

\maketitle

\section{Introduction}

Extensive development of contemporary experiments on qubit manipulations and quantum circuits engineering\cite{1,2,3,4,5,6} demands rigorous knowledge about the details of quantum electron transport in low-dimensional nanowires\cite{7,8,9,10}. Since the times of seminal Kane and Fisher papers\cite{11,12} on renormalzation group analysis of electron tunneling through different types of Luttinger liquid junctions, the features of non-equilibrium electron ballistic transport in one-dimensional electron systems has become more and more principal object of attention both in theory \cite{15,16,17,18,19,20,21}, \cite{26,27,28,29,30,31,32,33} and in the experiment\cite{2,3,4,5,6,7,8,9,10,14,37}. The common issue for such nanoelectronic devices is the one-dimensionality (1D) of chiral electron density excitations being responsible for respective ballistic electron transport in such structures at low enough temperatures\cite{11,12,13,15,16,17,18,19,20,21}. Here a well-known bosonization technique \cite{12,43} with basic relations between fermionic and bosonic field operators\cite{11,12,13,17,42,43} paves the way to effective theoretical description of e.g. quantum Hall edge states (QHE-states)\cite{18,19,20,21} and 1D quantum wires\cite{11,12,13,15,16,17,38,42,43} within Tomonaga-Luttinger liquid (TLL) model in terms of charge- and current densities of plasmonic excitations in such systems. 

From both fundamental and practical points of view generic bosonized model of quantum-point contact (QPC) is still far from its detailed and, at the same time, comprehensive understanding. However, due to their extreme practical importance, such QPC models stimulated a number of papers\cite{18,19,20,21,22,23,24,25,29,30,31,32,33,34,35,36,44,45,46,47} on each among two overlapped main approaches to their theoretical descriptions. Those approaches are: 1)various practical implementations\cite{18,19,20,30,31,32,33,36} -the latter consider various particular modifications of most general bosonized QPC models, but such treatment remains perturbative only up to the lowest orders in small tunnel coupling (for non-equilibrium effects\cite{19,20,36})or in small backscattering (e.g. for the problems of electron injection into QH edge states \cite{18,21,46,47})and 2) fundamental exact non-perturbative results\cite{15,16,22,23,24,25} on different aspects of general Luttinger liquid QPC models. Hovewer the latter class of related theoretical results being exact still has a limited number of implementations either due to extreme mathematical difficulties in the derivations of any simple analytical consequences from general too abstract solutions\cite{15,22,23}, or due to the uncontrollability (i.e. variance) of assumptions, one  needs to overcome\cite{16,25} in order to "distill" any  analytical non-perturbative answer from very general and very complicated expressions. 

First type of studies is more common than second one and includes a variety of different non-equilibrium results on fractional quantum Hall Eedge states (FQHES-) and single-electron transistor systems with 1D leads, which though remain valid only in the lowest orders in tunnel couplings\cite{26,27,28,29,30,31,32,33,34,35,36}. Somewhere in between the studies of the first (perturbative) and of the second (exact) type, the original method of functional bosonization\cite{21,45,46} is placed. Within latter framework one deals with scattering phases of relevant plasmonic excitations in the weakly constricted Luttinger liquid quantum wire or in fractional quantum Hall edge states \cite{21,44,45,46,47}. Such systems with weak electron backscattering one can treat exactly within the functional bosonization method as it was shown by Gutman, Gefen and Mirlin in the series of related papers \cite{21,44,45}. However, the latter is applicable, strictly speaking, only to those quasi-1D quantum wires having weak enough backscattering from the impurities rather than for quantum wires being interrupted by means of large enough tunnel barriers. This is because the Fredholm determinant factorization lying in the core of all the method implies a picture of multiple consecutive scatterings of quantum plasmonic excitations in the effective external time-dependent potential created by weak impurity (or weak constriction of the wire) being "dressed" into long-wavelength "classical" component of plasmonic charge density field. However, the latter picture seems to have no direct implementations for the electron transport problems involving weak electron tunneling. This is just because in the latter case the tunneling of a bare electron through a junction is a strongly non-perturbative process, which involves a destruction/creation of entire "cloud" of plasmonic excitations (the electron in bosonized picture is just a $ 2\pi $-kink of bosonic quantum field ). Therefore electron tunneling in the "weak link" (or alternatively, "weak tunneling" ) limit of a Luttinger liquid junction cannot be performed only just as forward scattering of  chiral plasmons delocalised throughout a junction as it takes place within the functional bosonization method of Refs.[21,44-47]. 
 In other words, functional bosonization is valid for 1D quantum wires which have quite "smooth" constrictions and/or inhomogeneties (as it takes places in FQH-edge states quantum dynamics) rather than for "sharp" boundary conditions which define common weakly linked tunnel junction. Hence, various functional bosonization results performed in Refs.[21,44-47] should be useful mostly in the cases of well-defined fractional-quantum-Hall effect (FQHE) edge states\cite{18,19,20,21,31,36}, where electrons with initially prepared distributions are injected into the edge states of FQHE systems and these wave packets interacts only weakly with each other.  At the same time, for Luttinger liquid tunnel junction of interest the functional bosonization method is problematic to use for the non-perturbative treatment because of superposition of two aspects\cite{12,13}: i) strong non-linearity of the tunnel Hamiltonian in its bosonic representation and ii)"sharp" boundary conditions in the vicinity of tunnel contact (i.e. strong backscattering from the tunnel barrier). 

In the studies of the second type (exact treatment), especially for the regime of "weak tunneling" in 1D junctions, people make use of the exact integrability methods, such as thermodynamic Bethe ansatz (TBA-) \cite{15,16,24,25} being applied to a cumbersome exact solution\cite{22,23} of a well-known boundary Sine-Gordon (BSG-)model, this case is reduced to. In more details, exactly solvable BSG-model\cite{22,23} allows one to obtain an infinite chain of connected exact integral equations, which might be decoupled from each other within TBA-method and, thus, solved explicitly\cite{16,24,25}. However, the latter is possible only under some specific assumptions about the eigenvalues and eigenstates of respective exactly solvable model\cite{24,25}. This research direction is represented by several basic papers by Ludwig, Fendley and Saleur\cite{15,23,24}, and ones by Komnik and Saleur\cite{16,25}, where, especially, the latter authors have pointed out on the importance of subtleties hidden in the implementations of TBA-method\cite{25}. In particular, they noticed that all the method is sensitive to the type of eigenstates (e.g. to different numbers of kink-antikink pairs, breathers, etc.) chosen for further analysis of TBA equations, because definite combinations of such eigenstates (e.g. kink-antikink pairs) allow for well-defined distribution functions only up to the calculation of real-time correlators of 3-d order in corresponding cumulant expansions, whereas already in 4-th order cumulant the kink-antikink pair becomes strongly fluctuating, which breaks the legitimacy of the decoupling of TBA chain of equations and, thus, all the TBA results for all higher order cumulants\cite{25}. 

On the other hand, the exact treatment of dual limiting case of weak scattering potential (i.e. weak backscattering case) (see e.g. \cite{12}) also faces certain difficulties since in this case (in the contrary to the opposite limit of weak tunneling we will consider here) there are no evident boundary condition for bosonic phase-field at the  position of "weak" impurity. Therefore, to decouple properly phase-field in the latter type of 1-dimensional interacting electron systems people turn to a bit different exactly solvable model of interacting resonant level (IRLM) in the one-dimensional tight-binding model \cite{49,51,52} which allows for exact solution in the bosonization picture by means of both analytical and numerical methods, especially, due to possibility of its re-fermionization at self-dual point. In particular, in the recent paper on the subject \cite{49} it was shown that such interacting resonant level model (IRLM) shares some common transport characteristics  with above-mentioned BSG model. 

Here it is important to emphasize that both mentioned here types of exact results on BSG- and IRLM-models of interacting 1D QPC still have very limited access to the exact desription of electron correlations in real time domain since exact analytical results via e.g. TBA method are for Green functions in $ k $-space and, thus, analytical results can be extracted only for steady state characteristics of electron-electron correlations (such as average current, zero-frequency shot-noise, and some higher-orders cumulants ) in the limit $ t \rightarrow \infty $   \cite{15,16,23,24,25} , whereas the real-time dynamics of respective correlators can be accessed only by means of numerical simulations involving finite number of diagrams and hence having approximate character\cite{49,51}. In this respect the study of exact real-time quantum evolution of weakly coupled tunnel junction of interacting electrons presented in below can be treated as very special first attempt to obtain an exact picture of real-time quantum dynamics in 1-D strongly correlated electron systems, thus, having only qualitative parallels with all mentioned above exact results obtained in previous literature for $ k $-space representation of correlators. At the same time, the results obtained below perfectly reproduce widely known \cite{12} steady state asymptotes for average current and shot-noise in the limit $ t \rightarrow \infty $.       

As one can see from the above, the complete description of real-time quantum dynamics out of the equilibrium for arbitrary Luttinger liquid QPC still remains a very challenging task. However, there are some theoretical hints\cite{28,30,31,32,33} on the fact, that even in the cases of interacting electrons being that complicated, respective charge-transfer statistics\cite{26,27,29} of a given tunnel junction still represents a much more simple and much more general picture. Especially, from the times of well-known Levitov and Lesovik papers\cite{26,27}, where authors derived in the lowest orders in tunnel coupling a remarkable generic (Levitov-Lesovik) formula for cumulant generating function of non-interacting electrons, it has been understood \cite{28} that above-mentioned extremely complicated (and strictly speaking, unknown in its details) picture of electron real-time correlations in any Luttinger liquid QPC should not be relevant for electron counting statistics at least in the long-time limit of their tunneling through the junction\cite{28}. Interestingly, this point of view is indirectly supported by a number of parallel investigations on another aspects of electron transport in quasi-one dimensional electron systems. These aspects are: 1)the validity of a linked cluster theorem for the orthogonality catastrophe calculation\cite{30}; 2)a common general relation between high order cumulants derived from the summation of selected linked-cluster diagrams in lowest order in electron tunneling\cite{31}; 3) simple expression for Fano factor from complicated exact solution of boundary-sin-Gordon-(BSG-)model  being reported in Ref.[24]; 4)Sukhorukov-Loss perturbative prediction\cite{32} about the universality of Fano-factor for interacting and non-interacting electrons; 5) the possibility to incorporate electron-electron interactions into functional bosonization description of weakly constricted 1D quantum wires reported in Refs.[21,44,45] together with the prediction \cite{35,36} of the electron charge  fractionalization in certain full counting statistics measurements; 6) derivation of universal noise formula for quantum Hall edge states from common arguments of quantum statistical mechanics \cite{33}. 7) Another recent important indirect evidences of the universality in non-equilibrium transport characterisitcs of tunnel junctions are the  universality established in Ref.[48] for finite-time dynamics of heat current in one-dimensional tunnel contact and, as well, the unexplained similarity between a picture of quantum fluctuations in resonant-level- and boundary-sin-Gordon models on a finite time-scales\cite{49}. As well, several analytical results out-of-the equilibrium  \cite{51,52}  on correlation effects in electron transport in the exactly solvable  resonant-level model, including recurrence relation (6) from Ref.[51], can also serve as indirect arguments for the possibility of exact treatment of time-dependent electron transport in interacting tunnel junctions.
 
However, all the above-mentioned separate hints on the universality of full counting statistics for interacting electrons in their "steady flow" (or, simpler, steady) state, on one hand, make use of general statistical arguments, such as e.g. the detailed balance or, more general, fluctuation-dissipation theorem, as hypothesises \cite{34,35}. For instance,  the limitations of the applicability of the scattering states involved into the Levitov-Lesovik approach were clearly demonstrated by Schönhammer \cite{54}.

In turn, it will be shown below that the detailed balance relation between "forward" and "backward" tunneling rates represents just a "steady state version" of more general fluctuation-dissipation theorem (or Jarzynski equality) for non-equilibrium tunneling processes\cite{39,40}. On the other hand,  all the results that general for interacting electrons in tunnel junction have remained, in fact, only perturbative. In other words, one is unable to claim, whether the detailed balance relations, or Jarzynski equality out of the equilibrium - remain valid also in arbitrary order of perturbation theory, or beyond the "steady flow" regime (i.e. on finite observation timescales). Moreover, in the derivation of classical Levitov-Lesovik formula\cite{26,27,28} for non-interacting electrons the detailed balance theorem for electron tunneling rates has been indirectly used in order to obtain correct renormalization of tunneling amplitudes by bias voltage\cite{28}. However, to be precise, such type of renormalization has been justified only for the non-interacting electrons, whereas for the case of interacting electrons in the leads (Luttinger liquid situation) all the above complicated quantum soliton-like dynamics\cite{24,25} should come into play with unknown details of respective finite-time evolution\cite{37}.

Therefore, this paper is intended to bring certainty into existing phenomenological hints on the universality of counting statistics and fluctuation-dissipation theorem for interacting and non-interacting electrons in tunnel junctions, thus, filling  the gap between existing fragmentary cases of exact treatment of the real-time non-equilibrium dynamics for arbitrary Luttinger liquid quantum-point contacts in the weak tunneling limit. Especially, here I calculate exactly the real-time propagator with all infinite Keldysh contour-ordered expansion of evolution operator for biased Luttinger liquid QPC in the weak tunneling limit on the finite timescales beyond the long-term asymptote, for arbitrary temperature, bias voltage and electron-electron interaction in junction electrodes. In order to obtain these results, I make use of two novel exact mathematical results derived in this paper: Self-equilibration (SE-)theorem and related Self-equilibration (SE-)lemma. These two mathematical statements represent a generalization of Summation (S-)theorem and related Summation (S-)lemma being formulated and proved by the author previously in the Ref.[38]. In addition, it will be shown below that performed rigorous proof of Self-equilibration (SE-) theorem, at the same time, is a rigorous mathematical proof of the detailed balance theorem for the steady state of tunnel current. As well, it is demonstrated in this paper that proven SE-theorem represents a dynamical version of well-known  Jarzynski equality - general re-exponentiation relation for generic quantum statistical system out-of the- equilibrium. This way, by means of SE-theorem, the relation of Jarzynski type becomes generalized on a new class of non-stationary  non-equilibrium quantum many-body systems, i.e. on the dynamics of interacting electrons in arbitrary Luttinger liquid tunnel junctions in the weak tunneling regime. All the obtained general results lead to a rigorous statement about the universal character of tunnel current equilibration process to the steady state in all weakly linked tunnel contacts. This, in turn, clearly points out on a novel phenomena of \textit{self-equilibration} in such systems, where the equilibration of \textit{all types} of quantum fluctuations in the system with time turns out to be governed only by one class of (minimal) four-fermion time-correlations (see main text below). 

This paper is organized as follows. After the Introduction in Section 1, the bosonized description of weakly linked tunnel junction is given in Section 2. Non-equilibrium Keldysh partition function and full-counting statistics formalism for ballistic tunnel contacts are described in Section 3. In Section 4  basic mathematical results: self-equilibration (SE-) theorem as well as related SE-lemma are given. In Section 5 the connection of SE-theorem with self-equilibration phenomenon and with non-equilibrium dynamical version of Jarzynski equality are established. The consequences of the obtained results for full counting statistics of tunneling electrons, detailed balance relations and equilibrium steady state characteristics of arbitrary bosonized weakly linked tunnel junctions are represented in Section 6 together with explicit calculations of the non-equilibrium Fano factor and non-equilibrium "steady flow" rate. In Section 7 common characteristic features of obtained self-equilibrated dynamics of quantum fluctuations in one-dimensional tunnel contacts are discussed. The conclusions are given in Section 8. All cumbersome calculations are given in three Appendices to the paper. In the Appendix A the regularization of a basic Luttinger liquid pair correlator is discussed. In the Appendix B the rigorous proof of SE-theorem and related SE-lemma are given. Calculation of basic time-integrals for the steady state transport characteristics in the long-term asymptote are shown in the Appendix C.

\section{Model}

A generic theoretical model of interest represents a tunnel contact in the regime of weak tunneling (i.e. with strong tunnel barriers)\cite{12,13} connecting two long enough quantum wires as electrodes, which in most general case of interest can be modelled as two (right (R) and left (L)) semi-infinite Luttinger liquids. The total Hamiltonian of our problem consists of two terms 
\begin{equation}
H_{\Sigma} = H_{LL} + H_{int},
\end{equation}

 where $H_{LL}$ represents the Hamiltonian of the left and right Luttinger liquids and $H_{int}$ stands for the tunnel interaction between two. If we consider the QPC to be located at $x=0$, then Hamiltonian $ H_{LL} $ will describe two one-dimensional quantum wires of length $ L $ each  (we suppose everywhere $ L  \rightarrow \infty$ ) with arbitrary local electron-electron repulsion constant $ U_{s} $. At the same time, the spectrum of relevant plasmonic excitations within the bosonized picture of free Luttinger liquid is linearized around the value $ E_{F} $ of common Fermi level of both electrodes and electron reservoirs at $ x \rightarrow \pm \infty $ (Fermi energy $ E_{F} $ of metallic quantum wires in common QPC devices - is typically of the order of several electron-volts\cite{38} ). Thus, Luttinger liquid Hamiltonian of the electrodes  (in the Schroedinger representation)  reads\cite{17,38} 

\begin{equation}
H_{LL} =\frac{1}{ 2 \pi} \sum_{j = L, R} v_g  \int^0_{- \infty} \left\{ g \left(\partial_x \varphi_j(x) \right)^2 + \frac{1}{g} \left( \partial_x \theta_j(x)  \right)^2 \right\} \rm{d}x
\end{equation}

, where $\theta_{L (R)}(x)=\pi\int_{-\infty}^{x}dx' \rho_{L ( R )}(x')$ and $\varphi_{L( R)}(x)=\pi\int_{-\infty}^{x}dx' j_{L ( R )}(x')$ are the usual charge- and phase- bosonic quantum fields ($ \langle \theta_{L (R)}(x,t) \rangle=0 $ and $ \langle \varphi_{L (R)}(x,t) \rangle=0 $) in the Luttinger liquid description of semi-infinite 1D quantum wire, those corresponding to fluctuating parts of charge- and current electron densities  
 \begin{equation}
 \rho_{L ( R )}(x)=\partial_x \theta_{L(R)}(x) = \sum _{c=+,-}: \Psi_{c, L(R)}^\dagger(x) \Psi_{c, L(R)}(x) :
\end{equation}
 and 
 \begin{equation}
 j_{L ( R )}(x)=\partial_x \varphi_{L(R)(x)}=\sum _{c=+,-}c: \Psi_{c, L(R)}^\dagger(x) \Psi_{c, L(R)}(x) : 
\end{equation} 

 in the QPC leads \cite{11,12,13,17,25,38,43}(here index $ c=+,- $ denotes chirality of right- or left-moving electron and symbol $:\ldots: $ denotes normal ordering of standard fermionic creation(annihilation) field operators  $ \Psi_{c, L(R)}^\dagger(x)(\Psi_{c, L(R)}(x)) $ - which create (annihilate) bare electron with fixed chirality $ c=+,- $ at spatial point $ x $ of the left (right) Luttinger liquid electrode of the junction). Here and below I put $ \hbar=1 $ and $ \vert e \vert=1 $. In the Hamiltonian $ H_{LL} $ of Eq.(2) for QPC electrodes $g$ is a dimensionless correlation parameter which is defined as $ g = \left( 1 + U_{s}/2E_{F} \right) ^{-1/2}$ (for repulsive interactions it fulfils $0 < g \leqslant 1$)  while $v_g$ is the group velocity of collective plasmonic excitations in the leads\cite{13,38,43}. The case $ g=1 $ corresponds to situation where $ U_{s}=0 $, i.e. it describes non-interacting electrons (Fermi liquid) in "bulk" electrodes of given QPC, while the case  $0 < g < 1 $ refers to situation of strong Coulomb repulsion between electrons ($ U_{s} \gtrsim E_{F} $) in the 1D Luttinger liquid electrodes of tunnel contact. 
 
 Evidently, Hamiltonian $ H_{LL} $ is quadratic, i.e. it is  effectively non-interacting in the bosonic representation of Eq.(2) even in the presence of very strong repulsion $ U_{s} \gtrsim E_{F} $ between real electrons in the leads. The latter fact represents important advantage of bosonized representation (2-4) which allows for the common theoretical description for the widest range of different quantum wires in the majority of realistic tunnel contacts. 
 
Further, within the "weak tunneling" approach\cite{12,13}  "charge"- bosonic fields: $ \theta_{L,R}(x=0,t) $ are pinned on the edges of respective QPC electrodes at point $ x=0 $ by means of the following "sharp" boundary condition\cite{12,13,17,38,43}

\begin{equation}
\theta_{L}(x=0,t)= \theta_{R}(x=0,t)=0.
 \end{equation} 

 Then the tunnel Hamiltonian $ H_{int} $ of our system in the approach of \textit{weak tunneling} reads

\begin{align}\label{eq:H_int_APP}
\begin{split}
  H_{int} &= t_t : \Psi_{+, R}^\dagger \Psi_{+, L}  (0,t) : +
  t^{\ast}_t : \Psi_{-, L}^\dagger \Psi_{-, R}  (0,t) : 
\end{split}
\end{align}
where $ t_{t} $ and $ t^{\ast}_{t} $ are bare tunneling amplitudes and fermionic field operators $\Psi_{c, j}^{\dagger} \left( x, t \right)$ [$\Psi_{c, j}\left( x, t \right)$] creates (anihilates) an electron at position $x$ and time $t$, with chirality $c = 1, 2$ and on side $j = L, R$ [note that $c=+$ ($c=-$) indicates moving towards (away from) the QPC] while symbol $:\dots:$ indicates normal ordering of these fermionic field operators. Such fermionic fields can be written in terms of the above-introduced bosonic quantum charge- and phase-fields (in the Schroedinger representation) according to well-known bosonization formula

\begin{align}\label{eq:Psis_APP}
  \Psi_{c, j}(x,t) = \frac{\eta_{c,j}}{\sqrt{2 \pi a_0}} e^{ic\left(k_F-\pi/L\right)x}e^{\mp i \rm{eV} t/2}e^{ i
  \left( c \theta_j(x)  + \varphi_j(x) \right)}
\end{align}

where these quantum fields are defined in 1+1 space (spatial coordinate + time) at each point $(x,t)$ of each (left or right) semi-infinite electrode and $\eta_{c,j}$ are Klein factors, $k_F$ is the Fermi momentum, while the signs $\mp$ in the exponential corresponds to $R$ (-) and $L$ (+). Here the quantity $\rm{eV}=\Delta \mu=\mu_{L}-\mu_{R}$ refers to the difference between chemical potentials of the Luttinger liquid leads being also a chemical potentials of two (left and right) remote reservoirs of electrons, which are coupled to two respective (left and right) Luttinger liquid electrodes of QPC at points $ x=\pm L \rightarrow \pm\infty $ as "source" and "drain" for tunneling electrons\cite{13}.(Notice, that for the difference between respective chemical potentials of QPC leads: $\Delta \mu=\mu_{L}-\mu_{R}  $ I have kept a common notation: $ \Delta \mu=eV $ with $ V $ being a bias voltage applied across given QPC.) Both chemical potentials $ \mu_{L(R)} $ are counted from the common Fermi energy of reservoirs $ E_{F} $ and of both electrodes). In realistic quantum point contacts one typically has $ eV \ll E_{F} $ which implies typical values of  $ eV $ varying from several $ \mu eV $ to several meV\cite{38} .

In fact, the boundary condition (5) and the form of weakly coupled tunnel Hamiltonian (6) are connected with each other because physically the "sharpness" of boundary conditions encoded in (5) indirectly requires also the smallness of related tunnel coupling constant $ \tilde{\lambda} $ as compared to all other energy scales of the model. However, it is widely known \cite{43} that taking into account both charge- and phase bosonic fields for infinite Luttinger liquid (i.e. when a whole system consists of just a single infinite Luttinger liquid lead without tunnel junction) results only in the following renormalization $ 1/g \rightarrow (1/g + g ) $ of the exponent in the power law of corresponded free Luttinger liquid pair correlator in time domain (see e.g. Giamarchi's book of Ref.[43] ). Thus, increasing the small parameter $ \tilde{\lambda}/\Lambda_{g} \ll 1 $ to some small extent one could expect the same type of renormalization for the power law time dependence in all involved pair correlators. But in such the situation for strongly interacting electrons with $ g \ll 1 $ one can neglect this sort of renormalization since  $ 1/g \gg g $ in this case. Therefore, formally, the requirement (5) can be relaxed to some extent for strongly repulsing electrons (i.e. for $ g \ll 1  $) as compared to condition (6) (or vice versa).   However, in what follows we will always presume the validity of both Eqs.(5,6) simultaneously, since they  both correspond to the case of weakly linked tunnel junction under consideration. Though, formally the validity of all basic results of the paper which will be derived below does not depend on the specific value of  $ \tilde{\lambda} $ parameter, the validity of Eqs.(5,6) for arbitrary $ g \leq 1 $ requires the smallness of tunneling coupling constant, i.e. inequality $ \tilde{\lambda}/\Lambda_{g} \ll 1 $ should always fulfil in our model.

Furhter, if one defines following non-local bosonic charge- and phase-fields\cite{17,38} 
\begin{eqnarray}
 \theta_{\pm}(q)=\left[\theta_L(q) \pm \theta_R(q)\right] \nonumber \\
 \varphi_{\pm}(q)=\left[ \varphi_L(q) \pm \varphi_R(q) \right], 
 \end{eqnarray} 
which fulfil standard commutation relations
\begin{align}
\begin{split}
\left[\theta_\alpha(q),\varphi_{\alpha'}(q')\right]&=-2 i\frac{\pi}{g}\sgn{(q-q')}\delta_{\alpha,\alpha'}\\
\left[\theta_\alpha(q),\partial_q \varphi_{\alpha'}(q')\right]&=-2 i\frac{\pi}{g}\delta(q-q')\delta_{\alpha,\alpha'}\
\end{split}
\end{align}
(here $ q=(x,t) $ is a generalized coordinate in the 1+1 - dimensional  quantum field description of 1d quantum wires) then one can rewrite the interaction Hamiltonian $ H_{int} $ of our model in the following bosonized form (in the Schroedinger representation, see also Refs.[17,38])

 \begin{align}\label{eq:H_int}
  H_{int} = \tilde{\lambda} \cos \left(
  \varphi_-(x)  + \rm{eV} t \right)]|_{x = 0} 
\end{align}
where  $\tilde{\lambda}=\vert t_{t}/(\pi a_0)\vert$, where $\vert t_{t} \vert$ is an absolute value of a bare tunneling amplitude for a given QPC. The parameter $a_0$ is the lattice constant of the model (which is practically of the order of $ 10^{-10} m $). This lattice constant also provides a natural high-energy cut-off $ \Lambda_{g}=v_{g}/a_{0} \simeq E_{F} $, and goes to zero in the continuum limit ($ a_{0}\rightarrow 0 $)\cite{43}. 

At this point one might ask whether it is possible to consider in a similar fashion to Eqs.(1-10) the dual case of weak impurity potential (i.e. weak backscattering case) \cite{12} simply by replacing in interaction Hamiltonian (10) the phase-field  $\varphi_{-}(x=0)$  by dynamically conjugated charge-field $  \theta_{+}(x=0) $ at point $ x=0 $. However, as it has been already mentioned in the Introduction, in this case one authomatically faces the problem of relevant boundary conditions for the phase-field  $\varphi_{-}(x=0)$  at the impurity position $ x=0 $ needed in order to decouple quantum dynamics of $\varphi_{-}$  and $  \theta_{+} $ quantum fields. Another obvious problem connected with the former is the impossibility to incorporate a difference $ eV $ between chemical potentials of the remote electron reservoirs simply as the time-dependent phase drop at $ x=0 $ as it can be safely done in the opposite weak link limit of Eqs.(5,10). Obviously, a source of both these technical problems is just a "smooth" boundary between left and right electrodes in the weak impurity potential case. As it was discussed in details in the Introduction section relevant approach would require different theoretical methods such as functional bosonization approach, which was successfully applied to such type of models \cite{21,44,45,46,47}. Due to this unobvious though principal difference between weak backscattering and weak link models of QPC it seems problematic to generalize starightforwardly all the model under consideration as well as results being  obtained in what follows on the dual weak backscattering limit of the QPC model. Therefore, in what follows I will not discuss the applicability of all the presented method as well as results obtained on the dual limiting case of weak imurity potential in the QPC except brief discussion in the next section.

In what follows it is convinient to work in the interaction representation with respect to $ H_{int} $ , where  for initial ground state $ \vert \phi_{LL}(0)\rangle $ of Luttinger liquid tunnel junction one has following temporal evolution (see also Ref.[17])
\begin{equation}
\vert \phi_{LL}(t)\rangle = \mathcal{T}_{t} e^{-i \int_{0}^{t}dt' H_{I}(t') }\vert \phi_{LL}(0)\rangle 
\end{equation}

where $ \mathcal{T}_{t} $ stands for chronological ordering of the operators in the exponent while  

\begin{equation}
H_{I}(t)= e^{i H_{LL}t}H_{int} e^{-i H_{LL}t}.
\end{equation}
is interaction (tunnel) Hamiltonian in the interaction representation. In this representation the dynamics of quantum fluctuating bosonic charge- and phase- fields $ \theta_{\pm} (x, t) $ and $ \varphi_{\pm} (x, t) $ is governed by free Heisenberg quantum equations of motion (QEOMS) \cite{17} 

\begin{align}
 \dot{\varphi}_{\pm} (x, t)& =-\frac{v_g}{g} \partial_x \theta_{\pm} (x, t) \\
 \dot{\theta}_{\pm} (x, t) &= -g v_g   \partial_x \varphi_{\pm} (x, t). 
\end{align}

Obviously, Eqs.(13,14) together with "sharp" boundary condition (5) and canonical commutation relations (9) yield a solution which can be decomposed in Fourrier series over secondary quantized creation and annihilation operators of free propagating plasmonic modes

\begin{align}\label{eq:phi_Fourier}
\begin{split}
  \varphi_\pm  \left(x, t \right) = i\sqrt{\frac{\pi}{2Lg}} \sum_{k \neq 0} \frac{e^{ - \left| k \right| \alpha_0 / 2}}{\sqrt{\left| k \right|}} \times \\
   \left[e^{-ipx} b^\dagger_{k ,\pm} \left(
  t \right) - e^{ipx}b_{k ,\pm} \left( t \right)
  \right]\,
\end{split}
\end{align}
where we have introduced standard
high-momentum cut-off $\exp \left( - \left| k \right| \alpha_0 / 2 \right)$ of given Luttinger liquid model on the energy scale: $ \Lambda_{g}=\frac{v_{g}}{\alpha_{0}}\approx \varepsilon_{F}/g $ . The bosonic creation (annihilation) operators $b^\dagger_{k} \left( t
\right)$ [$b_{k} \left( t
\right)$] are of the form \cite{17} 
\begin{eqnarray}
  & b^\dagger_{k ,\pm} \left( t \right)= b^\dagger_{k,\pm} \left( 0 \right) e^{i v_g t \left| k \right|} & 
  \nonumber\\
  &  b_{ k, \pm} \left( t
  \right) = b_{ k, \pm} \left( 0 \right)
  e^{- i v_g t \left| k \right|} & 
\end{eqnarray}
where $b^\dagger_{k,\pm}\left( 0
\right)$ and $b_{k,\pm}\left( 0
\right)$ fulfill standard bosonic commutation relations ($\alpha,\alpha'=\pm$)
\begin{align}
  \left[ b_{k, \alpha} ( 0 ), b^\dagger_{k', \alpha'} ( 0 ) \right] = \delta_{k, k'} \delta_{\alpha, \alpha'} \,.
\end{align}
Free Hamiltonian of the Luttinger liquid in this representation is simply \cite{17} 
\begin{align}
  H_{L L} = \sum_{\alpha = \pm ; k = 0}^{\infty} v_g \left| k \right| b^{+
  }_{k, \alpha} \left( 0 \right) b_{ k, \alpha} \left( 0 \right)
\end{align}
and the vacuum expectation value of definite spatially delocalized plasmonic mode 
\begin{align} \label{eq:nb}
\langle b^{+
}_{k, \alpha} \left( 0 \right) b_{
k' , \alpha'} \left( 0 \right) \rangle = n_b \left(k  \right)  \delta_{k, k'} \delta_{\alpha, \alpha'} 
\end{align}
where $n_b \left( k  \right)= \left[ \exp \left( v_g \left| k \right| / T \right)
- 1 \right]^{- 1}$ is the usual Bose-Einstein distribution at temperature $T$, with $k_B=1$. Since all quantum mesoscopic devices such as QPCs of interest make practical sense only in the low-temperature regime where they can provide a good enough $ I/V $ -characteristics, in order to minimize thermal fluctuations one should provide $ T \ll \Lambda_{g}, E_{F} $, in typical mesoscopic experiments \cite{5,6} the ratio  $ T / \Lambda_{g} $ is of the order of $ 10^{-5} $ provided that typical temperatures should be of the order of $ 10^{1} mK$ (see also Appendix A in Ref.[38] for discussion of related experimental conditions). Though in the model under the consideration both high- and low-temperature regimes with respect to applied bias voltage $ V $ can be realized. That means all formulas in what follows are valid in both limits $ T \ll eV \ll  \Lambda_{g}, E_{F} $ and $ eV \leq T \ll  \Lambda_{g}, E_{F}$.

\section{Non-equilibrium Keldysh partition function and full counting statistics}

In order to detect the  state of charge transport through our Luttinger liquid tunnel junction (or QPC) we should examine the full counting statistics (FCS) of our system\cite{27,28,29}. First, the possibility to extract all moments of tunnel current distributions - i.e. the  full counting statistics of charge transfer through QPC - from the evolution of coupled spin-1/2 galvanometer was demonstrated by Levitov and co-authors in Ref.[27]. Especially, in Ref.[27] it has been shown how one should modify overall evolution (11) of QPC  by interaction with counting field $ \xi $ in order to calculate from the first principles  so-called Keldysh partition function (KPF) $ \tilde{\chi}(\xi,t) $ which, in turn, gives access to all irreducible moments (cumulants) of charge transferred (-or tunnel current) distributions. The FCS method for ballistic tunnel junctions is a well-developed technique, hihglighted in many original papers and reviews (see e.g. \cite{28,29,51} ) and detailed discussion of this method is beyond the scope of this paper. However, I will emphasize below several subtle key-points of the FCS ideology which appear to be relevant in what follows and at the same time, to my knowledge, were not paid much attention to in related literature. As well, for further details of common FCS method one can see e.g. review \cite{29}.

It is obvious that each physical electron while tunelling through the tunnel barrier of QPC represents an elementary pulse of tunnel current $ I_{t} $. Suppose there is an isolated 1/2-spin ( i.e. spin-1/2 galvanometer \cite{27,29,51}) it creates corresponding vector potential $ \vec{a}(\vec{r}) $ in the close vicinity of tunnel contact\cite{29}. Let us suppose at the moment $ t=0 $ spin starts to "feel"  single-electron tunnel current through QPC via corresponding coupling Hamiltonian \cite{29,51} $ H_{\sigma}=-\frac{\chi(t)}{2e}\hat{\sigma}_{z}\hat{I}_{t} $, where $ \hat{I}_{t} $ is an operator of tunnel current through the junction, whereas $ \chi(t)=\chi f(t) $ represents time-dependent coupling constant (in the simplest measurement protocol function $ f(t)$ is just a square "pulse" of duration $ t_{m} \leq t $). Then this additional Hamiltonian $  H_{\sigma}$ should be added to $ H_{int} $ in Eq.(12) to be substituted into overall quantum evolution (11) of tunnel junction. 

And here we come to the subtle point: due to the presence of complicated current operator $ \hat{I}_{t} $ and, as well, due to the initial (at $ t=0 $) and final (at $ t=t_{m} \leq t $) points in evolution of coupled spin the additional Hamiltonian $ H_{\sigma} $ becomes explicitly time-dependent (on the concrete protocol of measurement determined by the shape of $ f(t) $ function ) and operator-valued (in general not commuting with $ H_{int} $ and $ H_{LL} $ at different moments of time in the T-exponent of Eq.(11)). And all this, generally speaking, should make impossible exact calculations of respective Keldysh partition function $  \tilde{\chi}(\xi,t) $ beyond the model of "classical" spin \cite{29,51} which is common within FCS framework.  However, we will see that such difficulty can be circumvented by using very general arguments. Due to quantum evolution of Eqs.(11,12) the quantum (operator-valued) phase  associated with additional Hamiltonian $ H_{\sigma} $  of QPC interaction with measurement device reads $ \hat{\theta}(t) =\int_{0}^{t}dt'H_{\sigma}(t')=\chi \int_{0}^{t}dt'f(t')\hat{I}_{t'}/e $ which is equivalent to $ \hat{\theta}(t) =\xi \hat{N}_{t} $ where $ \hat{N}(t)= \hat{Q}(t)/e $ is an operator of electrical charge (in the units of $ e $ elementary charge of electron) transferred through QPC by the moment $ t $ . It is quite clear that in the most general case the observables associated with quantum operators  $ \hat{\theta}(t) $ and $ \hat{N}(t) $ should strongly fluctuate. 

On one hand, as soon as it is possible (by definition of any macroscopic measurement device) to measure different eigenvalues $ \theta_{i}$ of quantum operator $ \hat{\theta} $ one \textit{should} obtain from these well-defined measurements also a quantization of classical electrical charge being transferred through given QPC by the moment $ t $: $ Q_{i}=eN_{i} $, where $ N_{i} $ are different eigenvalues of $ \hat{N} $ operator. This establishes one-to-one correspondence between two spectra of eigenvalues: $ {\theta_{i}} $ which contains \textit{all possible values} of spin 1/2-galvanometer precession angles being measured macrospopically \cite{29} and $ {N_{i}} $ - of all possible numbers of electrons being transferred through given QPC by given moment $ t>0 $ of observation. Thus, proper arrangements of eigenvalues in these two spectra should give $ \xi =const $ for the ratio $ \theta_{i}/N_{i} $ of any two $ i $-th eigenvalues of $  \hat{\theta} $ and $ \hat{N} $ operators. Obviously, constant quantity $ \xi $ can be interpreted as \textit{ angle of precession for spin-1/2 galvanometer per single tunneling event in given QPC provided that $ \xi $ is well-defined counting field}. 

On the other hand, the appearance of spin-interacting Hamiltonian "dressing" $ \bar{\mathcal{T}}_{t}\exp(i\int_{0}^{t}dt'H_{\sigma}(t')) H_{I}(t) \mathcal{T}_{t}\exp(-i\int_{0}^{t}dt'H_{\sigma}(t'))  $ for interaction Hamiltonian (10) in quantum evolution T-exponent of Eq.(11) results in operator reonrmalization of bare tunneling amplitudes $ t_{t} \Rightarrow t_{t} \exp(-i\hat{\theta}(t)) $ and $ t_{t}^{*} \Rightarrow t_{t}^{*} \exp(i\hat{\theta}(t)) $  with operator-valued phase $ \hat{\theta}(t) =\int_{0}^{t}dt'H_{\sigma}(t') $ in the tunnel Hamiltonian (10). Obviously, since our tunnel Hamiltonian (6,10) describes only single-electron tunneling the eigenvalues of $ \pm\hat{\theta}(t) $ operators  should correspond to a fixed magnetic phase acquired by a single electron during its tunneling through tunnel junction from left to right electrode (or vice versa) in the effective magnetic field of spin-1/2 galvanometer. Since all single-electron tunneling events are identical due to sharp boundary conditions (5) in weak link limit, corresponded  single-electron magnetic phases should be equal for all single-electron tunneling events as well. The only realization of such possibility is when $ t_{t} \exp(-i\hat{\theta}(t)) =t_{t} \exp(-i\xi)  $ and $ t_{t}^{*} \exp(i\hat{\theta}(t)) = t_{t}^{*} \exp(i\xi)  $ with $ \xi $ being the angle of precession for spin-1/2 galvanometer per single tunneling event in given QPC or simply a stationary counting field.

This way, in weak tunneling limit of QPC one can get rid  of quantum fluctuations both in magnetic phase for tunneling electron (due to sharp boundary conditions at the junction) and in the spin-1/2 galvanometer (due to charge quantization). The equivalence of magnetic phase of tunneling electron to the counting field (precession angle of spin galvanometer per one tunneling electron) provides the required equivalence of spin-current interaction in the spin-1/2 galvanometer- and tunneling electron's frames of reference.  

Thus, having in hands enough outcomes of spin 1/2 galvanometer precession angle measurements, one can construct Fourier series with probability distributions (see e.g. Ref.[29]): $\mathcal{P}(N,t)$ (as Fourier coefficients)  for all possible amounts of  total electrical charge $q=eN$ ($ N=1,2,..\infty $) being transmitted through QPC during fixed interval of time $t$. This infinite series is identical to one for all possible outcomes $ \theta_{i} $ ($ i=1,2,..\infty $) of spin-1/2 galvanometer precession angles. Such series is nothing but Fourier decomposition of so-called Keldysh partition function (KPF) $ \tilde{\chi}(\xi,t) $. For stationary counting field the latter reads \cite{29}

\begin{align}
\begin{split}
\tilde{\chi}(\xi,t)=\sum_{n=-\infty}^{\infty}\mathcal{P}(N,t)e^{i\xi n}=e^{-\mathcal{F}(\xi,t)}
\end{split}
 \end{align}
where $ N=\vert n \vert $ and $ \mathcal{F}(\xi,t) $ is called cumulant generating function (CGF)\cite{28,29}.

 Notice, that the fact that counting field $ \xi $ in Eq.(20) is time-independent (or stationary ) is completely due to accepted weak tunneling approach as it was explained in the above. It is important to mention here that all the above considerations on the fixed magnetic phase of tunneling electron are seemed to be not valid in the opposite limit of QPC with weak impurity potential (weak backscattering limit) in such the case all the described above FCS scheme should become ill-defined beyond the special super Fermi-liquid case \cite{52} due to strong quantum fluctuation of magnetic phase of electron scattering by weak impurity potential at point $ x=0 $.

Therefore, if one manages somehow to find exact expression for CGF $ \mathcal{F}(\xi,t) $ in the weakly linked model of QPC with interacting electrons in the leads, then the latter functional will give access to exact calculation of all "irreducible" moments (cumulants) of the time-dependent distribution: $\mathcal{P}(N,t)$ as well as to exact time-dependent distribution function $ \mathcal{P}(N,t) $ of tunneling electrons by itself. Hence, in according with common definition \cite{27,29,51} one has for the cumulant of arbitrary order $ k $ 
\begin{align}
\begin{split}
\langle\langle N^{k}(t)\rangle\rangle=-(-i)^{k}\frac{\partial^{k}\mathcal{F}(\xi,t)}{\partial \xi^{k}}\mid_{\xi=0, k=1,2,..}.
\end{split}
 \end{align}

From  the context of common FCS theory \cite{27,28,29,51}, from definitions (10,11,12,20,21) and above discussion it follows that Keldysh partition function (KPF) $ \tilde{\chi}(\xi,t) $ in the weak link limit (5) of QPC model can be calculated in the interaction representation as the averaged Keldysh contour-ordered T-exponent (or evolution operator), where a finite-time integration is performed over the complex Keldysh contour (from $ 0 $ to $ t $ on real axis) with respective (Keldysh contour-dependent via stationary counting field $ \xi $) tunnel Hamiltonian \cite{28,29}. In our case (see also Refs.[17,38]) corresponding formula reads     
\begin{align}\label{eq:GenFunc}
\tilde{\chi}(\xi,t)=\left\langle \mathcal{T}_K\exp(-i\tilde{\lambda}\int_{\mathcal{C}_K(t)}A_{\xi(\tau)}(\tau){\rm d}\tau)\right\rangle,
\end{align}
here the symbol $ \langle \ldots \rangle $ stands for thermal averaging performed over the ground state of QPC leads $\langle \phi_{LL}(0) \vert \ldots \vert \phi_{LL}(0) \rangle $ which is completely characterized by Eqs.(9,13-19) together with standard Gibbs factors for Bose gas originating from Eqs.(17-19). In Eq.(22) symbol $\mathcal{T}_K$ marks the sum over all possible time-orderings on the complex-time Keldysh contour  $\mathcal{C}_K(t) \in (0\pm i0;t)$. Notice, that chosen Keldysh contour differs a bit from a more conventional one with its lower branch being confined in the limits: $ (0 - i\beta;t) $, where $ \beta=1/T $ - is the inverse temperature (here and below the temperature is in the units of energy). The latter difference is compensated here by the presence of conventional bosonic Gibbs factors in the procedure of averaging in Eq.(22) \cite{53} . 

In general, within the "bosonic language", one can introduce time-dependent "quantum potential" $ A_{\xi(\tau)}(\tau)$ into Eq.(22) as follows
\begin{align}\label{eq:GenFunc}
 A_{\xi(\tau)}(\tau)=\tilde{\lambda}^{-1}H_{I}(\tau, \xi(\tau))=\cos{\left[\varphi_{-}(\tau)+f_{\xi(\tau)}(\tau)\right]}
\end{align}
with the time- and counting field-dependent function $ f_{\xi(\tau)}(\tau) $, where $\xi(\tau)=\pm\xi  $ is a counting field defined on the upper (lower) branch of the Keldysh contour $\mathcal{C}_K\in (0-i\beta;t)$. Such a counting field $  \xi(\tau)\in (-\pi;\pi) $ - physically represents a fixed phase difference being acquired by the tunneling physical electron due to its interaction with an external detector (spin 1/2 galvanometer) of charge transfer through QPC . The case $ \xi(\tau)>0 $ describes electron tunneling "forward in time" (or, say, from left to the right lead), while $ \xi(\tau)<0 $ case corresponds to electron tunneling "backward in time" (or, say, from right lead to the left one). Evidently, both  $ A_{\xi(\tau)}(\tau) $ and function $ f_{\xi(\tau)}(\tau) $ are defined in terms of two different quantum potentials $ A_{\pm\xi}(\tau) $ and two different single-valued functions $ f_{\pm\xi}(\tau) $ defined solely on the upper (lower) branch of Keldysh contour $\mathcal{C}_K\in (0;t)$
\begin{align}
\begin{split}
A_{\xi(\tau)}(\tau)=\left\{ 
 \begin{matrix}
A_{+\xi}(\tau)=\cos{\left[\varphi_{-}(\tau)+f_{+\xi}(\tau)\right]} , \textit{Im}\lbrace \tau \rbrace > 0 \\
A_{-\xi}(\tau)=\cos{\left[\varphi_{-}(\tau)+f_{-\xi}(\tau)\right]} , \textit{Im}\lbrace \tau \rbrace < 0
        \end{matrix} \right.
 \end{split}
\nonumber 
 \end{align}
\begin{align}
\end{align} 
where
\begin{align}
f_{\pm\xi}(\tau)=[eV\tau\pm\xi]
\end{align}
are single-valued time- and counting field-dependent functions with $ \xi \in (0;\pi) $ being an absolute value of stationary counting field $\xi $ on each branch of the Keldysh contour . As it  follows from all the above formulas (20-25) give rise to the real-time description of full counting statistics of electrons tunneling through weakly-linked QPC in terms of stationary counting field $ \xi $ irrespective of the details of a quench protocol for interaction of QPC with its measuring device (spin-1/2 galvanometer).

Within the framework of Eqs.(12,22,23) quantum field operators which enter the Keldysh contour-ordered T-exponent (such as quantum fields $ \varphi_{\pm}(x,t) $ in our case) should be considered as the exact solutions of respective Heisenberg quantum equations of motions (QEOMS) (13,14) together with boundary condition (5) and commutation relations (9) for the non-interacting part $ H_{LL} $ of total Hamiltonian, see also for the details Ref.[17]. Hence, the solution of Eqs.(13-14) with boundary condition (5) are expressions (15-17).  Notice, that such sort of interaction representation (5,12-17,22-24) for decoherence- and quench problems in 1D quantum contacts can be effective (even if to keep T-exponent only up to the lowest orders in respective interaction constant) only if one is able to solve exactly respective QEOMS for all quantum fields entering evolution operator of interest, while in all situations, where it is impossible to find the exact solutions of "non-interacting" QEOMS for $ H_{LL} $, one should use another methods to handle the problem. The latter is exactly the case for many QPC models, which either have "smooth" boundary conditions (where Luttinger liquid correlation parameter $ g $ becomes a certain function of spatial coordinate: $ g(x) $) \cite{21,44,45,46,47}, or e.g. for FQH edge states systems where one deals with a propagation of electron wave packet along certain 1D edge "channel" \cite{18,19,20}. Therefore, in the cases of the first type \cite{21,44,45,46,47}, in order to solve the entire problem (even exactly, meaning numerically) one needs to assume "weak backscattering" of propagating electron (this is the opposite limit to a "weak link" approach I use here) and then one needs to treat all the problem within the functional bosonization method\cite{21,44,45,46,47}. Whereas for the problems of the second type \cite{18,19,20} the presumptions about simple configurations of incident electron wave packets and, as well, about the weakness of wave packets backscattering (the latter assumption allows for perturbative treatment) are necessary. Anyway, one can conclude from the above, that the "price" for complication of the mathematical description of QPC electrodes is either a breakdown of exact solvability of such QPC models (like it takes place for FQH edge states and respective Mach-Zender interferometry, which were typically solved only in the lowest orders of perturbation theory in tunnel couplings), or sufficient limitations on allowed types of boundary conditions for given 1D system (e.g. only "smooth" boundaries with weak backscattering are allowed for applicability of the functional bosonization method \cite{21,44,45,46,47}) and, as well, on the forms of incident electron wave packets.  All mentioned limitations are the restrictions prescribed by the functional bosonization method applicability \cite{21,44,45,46,47}. 

However, all the above-mentioned complications do not take place in the present QPC model with "sharp" boundary conditions at $ x=0 $ corresponding to "weak link" limit of the junction: respective boundary condition $ \theta_{\pm}(0,t)=0 $ which "pins" bosonic charge-fields at the high enough tunnel barrier simplifies  the description of 1D leads of QPC sufficiently. As the result, one can treat the leads as homogeneous semi-infinite Luttinger liquids with definite constant correlation parameter $ g $, where relevant phase- and charge- bosonic fields ($ \varphi_{\pm}(x,t) $ and $ \theta_{\pm}(x,t)$) are described by  "free" solutions (15-17,19) of respective Heisenberg QEOMS in the interaction representation (12). 
 In what follows we will widely use the pair- (or, equally, two-time)  correlator between "free" phase-field $ \varphi_{\pm}(0,t) $ of Eq.(15) taken at two different moments of time, say, $ t $ and $ t' $. Such the exact pair correlator of "free" phase field $ \varphi_{\pm}(0,t) $ in time domain is given in the Appendix A to this paper (see integral of Eq.A1), and was discussed previously in Refs.[17,38] as well.

\section{Self-equilibration (SE-) theorem and related SE-lemma for arbitrary bosonized tunnel junction}

Here I formulate and prove (see related proofs in the Appendix (B) below) the validity of following  important mathematical statements: Self-equilibration (SE-) theorem and related Self-equilibration (SE-) lemma which are the generalizations of S-theorem and S-lemma being proven by the author earlier in Ref.\textbf{[38]} on the case of interaction with counting field $ \xi(\tau) $ in the bosonized tunnel Hamiltonian (10).

$ \lozenge $ \textit{SE-Theorem}: \textit{"The case of exact re-exponentiation of Keldysh contour time-integration of the T-exponent with counting field-dependent non-linear bosonic quantum potential"}

$\blacktriangle $ \textit{For any non-linear bosonic operator of the form:}
\begin{align} \nonumber
 A_{\xi(\tau)}(\tau)=\cos{\left[\varphi_{-}(\tau)+f_{\xi(\tau)}(\tau)\right]}.
\end{align}
\textit{where}
\begin{align}
\begin{split}  \nonumber
f_{\xi(\tau)}(\tau)=\left\{ 
 \begin{matrix}
f_{+\xi}(\tau), \textit{Im}\lbrace \tau \rbrace > 0 \\  \nonumber
f_{-\xi}(\tau), \textit{Im}\lbrace \tau \rbrace < 0
        \end{matrix} \right.
 \end{split}
\end{align}
\textit{with property}
 \begin{align}
\begin{split}  \nonumber
\vert f_{i}(\tau_{1})-f_{j}(\tau_{2})\vert=\vert f_{i}(\tau_{2})-f_{j}(\tau_{1})\vert
\end{split}
 \end{align} 
\textit{ $ i,j=\pm\xi $ - fulfilled for $f_{\pm\xi}(\tau)$, which are certain functions of time being defined as $ f_{+\xi}(\tau) $ ($ f_{-\xi}(\tau) $) on the upper (lower) branch of Keldysh contour in the complex plane and $ \varphi_{-}(\tau) $ is time-dependent bosonic field with zero mean $ \langle \varphi_{-}(\tau)\rangle = 0 $ , which fulfils commutation relation of the form:}
\begin{align}
\begin{split}  \nonumber
\left[\varphi_{-}(\tau_{n}),\varphi_{-}(\tau_{n'})\right]=-2i\vartheta_{g}\sgn{\left[\tau_{n}-\tau_{n'}\right]}
\end{split}
 \end{align}
 \textit{where $ \vartheta_{g}=const $ (notice, that in our particular case: $ \vartheta_{g}=\pi/g $.)}

$ \blacktriangledown $ \textit{it follows for time-dependent average of Keldysh contour-ordered exponential from the non-linear quantum potential with Keldysh contour-dependent counting field $ \xi(\tau) $}
\begin{align}
\begin{split}  \nonumber
\tilde{\chi}(t)=\left\langle \mathcal{T}_K \exp(-i\tilde{\lambda}\int_{\mathcal{C}_K(t)}A_{\xi(\tau)}(\tau){\rm d}\tau)\right\rangle \\  \nonumber
=\exp \left\lbrace - \mathcal{F}(\xi,t) \right\rbrace
\end{split}
\end{align} 
\textit{where for the functional $\mathcal{F}(\xi,t)$ one has}
\begin{align}
\begin{split}  \nonumber
\mathcal{F}(\xi,t)=\left(\tilde{\lambda}\right)^{2}\times \\  \nonumber
\left\lbrace\frac{1}{2} \int\int_{\mathcal{C}_K(t)} d \tau_{1}d \tau_{2}\langle \mathcal{T}_K A_{\xi(\tau)}(\tau_{1}) A_{\xi(\tau)}(\tau_{2})\rangle\right\rbrace 
\end{split}
\end{align}
\textit{or, alternatively} 
\begin{align}
\begin{split}  \nonumber
\mathcal{F}(\xi,t)=\left(\tilde{\lambda}\right)^{2}\frac{1}{2}\int^{t}_{0}d \tau_{1} \int^{t}_{0} d \tau_{2}u(\tau_{1}-\tau_{2})\times \\ \left\lbrace {e^{i \vartheta_{g}}\Delta\kappa_{+}(\tau_{1}-\tau_{2})+e^{-i \vartheta_{g}}\Delta\kappa_{-}(\tau_{1}-\tau_{2})}\right\rbrace
\end{split}
\end{align}
\textit{with}
\begin{align}
 \begin{split}   \nonumber
\Delta\kappa_{\pm}(\tau_{1}-\tau_{2})=[\tilde{\kappa}_{\pm}(\tau_{1}-\tau_{2})-\kappa_{\pm}(\tau_{1}-\tau_{2})]
\end{split}
 \end{align} 
\begin{align} 
 \begin{split}  \nonumber
\tilde{\kappa}_{\pm}(\tau_{1}-\tau_{2})=\cos{[f_{\pm\xi}(\tau_{1})-f_{\mp\xi}(\tau_{2})]}\\  \nonumber
\kappa_{\pm}(\tau_{1}-\tau_{2})=\cos{[f_{\pm\xi}(\tau_{1})-f_{\pm\xi}(\tau_{2})]}
\end{split}
 \end{align}
\textit{with function $ u(\tau_{1}-\tau_{2}) $ being a pair correlator: }
\begin{align}
 \begin{split}   \nonumber
u_{g}(\tau_{1}-\tau_{2})=e^{\frac{-\left\langle \left[\varphi_{-}(\tau_{1})-\varphi_{-}(\tau_{2}) \right]^2\right\rangle}{2}}
\end{split}
 \end{align}
 \textit{properly regularized at the point $\tau_{1}=\tau_{2} $.}
 $ \blacksquare $

\medskip

The above SE-theorem, in turn,  makes use of the following SE-lemma (being the generalization of S-lemma from Ref.[17] on the case of counting field) which is also proven in the Appendix B below:

 $ \lozenge $ \textit{SE-Lemma}: \textit{"The case of the exact factorization of connected time-integrals from the average of counting field-dependent non-linear quantum potential. (Wick theorem for time-dependent local non-linear interaction between non-local free bosons)"}

$\blacktriangle $ \textit{For any non-linear bosonic operator of the form:}
\begin{align}
 A_{\xi(\tau)}(\tau)=\cos{\left[\varphi_{-}(\tau)+f_{\xi(\tau)}(\tau)\right]}.
 \nonumber
\end{align}
\textit{where}
\begin{align}
\begin{split} 
f_{\xi(\tau)}(\tau)=\left\{ 
 \begin{matrix}
f_{+\xi}(\tau), \textit{Im}\lbrace \tau \rbrace > 0 \\
f_{-\xi}(\tau), \textit{Im}\lbrace \tau \rbrace < 0
        \end{matrix} \right.
 \end{split}
\nonumber 
\end{align}
\textit{where the property}
 \begin{align}
\begin{split}   \nonumber
\vert f_{i}(\tau_{1})-f_{j}(\tau_{2})\vert=\vert f_{i}(\tau_{2})-f_{j}(\tau_{1})\vert
\end{split}
 \end{align} 
\textit{with $ i,j=\pm\xi $- is fulfilled for  $f_{\pm\xi}(\tau)$ being certain functions of time and defined as $ f_{+\xi}(\tau) $ ($ f_{-\xi}(\tau) $) on the upper (lower) branch of Keldysh contour in the complex plane and $ \varphi_{-}(\tau) $ is time-dependent bosonic field with zero mean $ \langle \varphi_{-}(\tau)\rangle = 0 $ , which fulfils commutation relation of the form:}
\begin{align}
\begin{split}
\left[\varphi_{-}(\tau),\varphi_{-}(\tau')\right]=-2i\vartheta_{g}\sgn{\left[\tau-\tau'\right]}
 \nonumber
\end{split}
 \end{align}
 \textit{where $ \vartheta_{g}=const $ (notice, that in our particular case: $ \vartheta_{g}=\pi/g $)}
 
$ \blacktriangledown $ \textit{it follows for the sequence of connected time integrals from the average of Keldysh contour-ordered product of non-linear quantum potentials coupled to Keldysh contour-dependent counting field $ \xi(\tau) $}
\begin{align}
\begin{split}
\int^{t}_{0} d \tau_{1}\int^{\tau_{1}}_{0} d \tau_{2}\ldots \int^{\tau_{k-2}}_{0} d \tau_{k-1}\int^{\tau_{k-1}}_{0} d \tau_{k}\\
 \int^{t}_{0} d \tau'_{1}\int^{\tau'_{1}}_{0} d \tau'_{2}\ldots \int^{\tau'_{(n-k)-2}}_{0} d \tau'_{n-k-1}\int^{\tau'_{(n-k)-1}}_{0} d \tau'_{n-k}\\
\times\langle A_{+\xi}(\tau_{k})\ldots A_{+\xi}(\tau_{1})A_{-\xi}(\tau'_{n-k})\ldots A_{-\xi}(\tau'_{1})\rangle_{S}\\
=\frac{(-1/2)^{n/2}}{(n/2)!}\left\lbrace\int^{t}_{0} d \tau_{1} \int^{t}_{0} d \tau_{2}u(\tau_{1}-\tau_{2})\Delta\kappa_{+}(\tau_{1}-\tau_{2})\right\rbrace^{k}\\
\times \left\lbrace\int^{t}_{0} d \tau_{1} \int^{t}_{0}d \tau_{2}u(\tau_{1}-\tau_{2})\Delta\kappa_{-}(\tau_{1}-\tau_{2})\right\rbrace^{(n/2-k)}
\nonumber
\end{split}
\end{align}
\textit{where}
\begin{align}
 \begin{split} 
\Delta\kappa_{\pm}(\tau_{1}-\tau_{2})=[\tilde{\kappa}_{\pm}(\tau_{1}-\tau_{2})-\kappa_{\pm}(\tau_{1}-\tau_{2})]
\nonumber
\end{split}
 \end{align} 
\begin{align}
 \begin{split} 
\tilde{\kappa}_{\pm}(\tau_{1}-\tau_{2})=\cos{[f_{\pm\xi}(\tau_{1})-f_{\mp\xi}(\tau_{2})]}\\
\kappa_{\pm}(\tau_{1}-\tau_{2})=\cos{[f_{\pm\xi}(\tau_{1})-f_{\pm\xi}(\tau_{2})]}
\nonumber
\end{split}
 \end{align}
\textit{with function $ u(\tau_{1}-\tau_{2}) $ being a pair correlator:}
\begin{align}
 \begin{split} 
u_{g}(\tau_{1}-\tau_{2})=e^{\frac{-\left\langle \left[\varphi_{-}(\tau_{1})-\varphi_{-}(\tau_{2}) \right]^2\right\rangle}{2}}
\nonumber
\end{split}
 \end{align}
 \textit{properly regularized at the point $\tau_{1}=\tau_{2} $.}
 $ \blacksquare $

\section{Self-equilibration (SE-) equation and dynamical Jarzynski relation for Luttinger liquid tunnel junction}

Now from the above Eqs.(22-25) together with proven in the Appendix B SE-theorem it is evident that following relation   
\begin{align}
\begin{split}
\tilde{\chi}(t,\xi)=\left\langle \mathcal{T}_K \exp \left\lbrace -i\int_{\mathcal{C}_K(t)} H_{I}(\tau,\xi(\tau)){\rm d}\tau \right\rbrace  \right\rangle = \\
\exp \left\lbrace - \mathcal{F}(\xi,t) \right\rbrace=\\
\exp \left\lbrace - \int\int_{\mathcal{C}_K(t)} \frac{d \tau_{1}d \tau_{2}}{2}\left\langle  \mathcal{T}_K H_{I}(\tau_{1},\xi(\tau_{1})) H_{I}(\tau_{2},\xi(\tau_{2}))\right\rangle  \right\rbrace\\
\end{split}
\end{align}
with $ H_{I}(\tau,\xi(\tau)) $ from Eqs.(23-25) and $ \varphi_{-}(t) $ bosonic field from Eqs.(15-17) represents \textit{exact} mathematical statement. 

Here I would like to stress out once more that the underlying model of Eqs.(1-25) of arbitrary weakly linked tunnel junction under consideration represents a reasonable approach to the description of real tunnel junctions constructed out of quantum wires of finite length and width. The limits of the applicability of such the approach have been extensively discussed in the above, in Sections (I-III) as well as in the author's previous paper \cite{38} on the subject. However, as soon as one accepts the theoretical model of Eqs.(1-25) for the description of a certain quasi-one dimensional tunnel junction, one should treat the re-exponentiation result of Eq.(26) proven by means of SE-theorem as the exact mathematical relation between two different ways of construction and summation of all possible kinds of diagrams which can appear in the l.h.s. and r.h.s. of Eq.(26), respectively.  In order to convince in the validity of Eq.(26) for all orders in tunnel Hamiltonian $ H_{I}(\tau,\xi(\tau)) $ (or, in coupling constant $ \tilde{\lambda} $ with respect to Eqs.(22-25)) one needs just to expand exponential functions in both sides of Eq.(26) into exact infinite power series and convince that all contributions from the odd powers of the coupling constant $ \tilde{\lambda} $ in the l.h.s. of Eq.(26) will cancel out whereas the rest of diagrams in the l.h.s. of Eq.(26) one can "wrap" into the exponential function in the r.h.s. of Eq.(26). That is, basically, the subject of SE-theorem proven in the Appendix B of this paper. In this sense, the mathematical relation of Eq.(26) is exact irrespectively to a specific value of $ \tilde{\lambda} $ tunnel coupling constant. Therefore, the relation (26), being valid at arbitrary time scale larger than the inverse high-enery cutoff $ \Lambda_{g} $ of Luttinger liquid model, reveals new remarkable features in the behaviour of tunnel current quantum fluctuations, which are beyond the scope of standard Levitov-Lesovik type of scattering approach\cite{26,27,28}. Especially, SE-theorem of Eq.(26) is a fingerprint of a very specific quantum evolution in one-dimensional interacting electron systems which I call in what follows as "self-equilibration regime" in which some sorts of quantum fluctuations compensate each other at arbitrary time scales. Below the novel effect of self-equilibration in one-dimensional interacting tunnel junctions will be explored in details.

Remarkably, real-time re-exponentiation formula (26) can be rewritten in the form of \textit{classical differential equation of self-equilibration} for the real-time Keldysh partition function which is established here to be valid for ballistic electron transport in arbitrary one-dimensional tunnel contacts.

\begin{align}
\begin{split}
\partial_{t} \tilde{\chi}(t,\xi) = - \left\lbrace \partial_{t} \mathcal{F}(\xi,t)\right\rbrace  \tilde{\chi}(t,\xi) \\
= - \left\lbrace\frac{1}{2} \int_{0}^{t} d \tau_{1} \left\langle [H_{I}(t,\xi), H_{I}(\tau_{1},\xi)]_{+}\right\rangle  \right\rbrace  \tilde{\chi}(t,\xi)\\
\end{split}
\end{align}
where $ [H_{I}(t,\xi), H_{I}(\tau_{1},\xi)]_{+} $ - represents anti-commutator of tunnel Hamiltonians (23)in the interaction representation (12) with counting field $ \xi \in [0;\pi] $ for two different moments of time. 

Especially, now from formulas (6,7,10,23) and Eqs.(26-27), it becomes qualitatively clear why the dynamics of quantum fluctuations proceeds in the regime, which can be called as "\textit{self-equilibrated}" one, preserving a universal character of equilibration to the steady state of tunnel current for a broad range of interacting one-dimensional ballistic tunnel junctions. Indeed, according to proven SE-theorem the temporal evolution of Keldysh partition function is completely defined by the time-dependent coefficient $ \partial_{t} \mathcal{F}(\xi,t) $ in Eq.(27) which, in turn, is proportional to thermal averages of the kind $  \langle:\Psi_{c, j}^\dagger(0,t) \Psi_{c, j'} (0,t)\Psi_{c', j'}^\dagger(0,t) \Psi_{c',j } (0,t):\rangle $ (where $ c,c'=+,- $ and $ j,j'=L,R $). The latter averages involves only four fermionic field operators at the point $ x=0 $ (i.e. at the tunell-coupled edges of each 1d electrode). Moreover, these averages contains only one pair of creation- and one pair of annihilation fermionic field operators on the edges of quantum wires. 

Therefore, from proven validity of SE-theorem of Eqs.(26,27) it follows that \textit{all quantum fluctuations of tunnel current in arbitrary  ballistic one-dimensional weakly-linked tunnel junction can be described as the superposition (i.e. exponential of Eq.(26) ) of all possible combinations of two-fermion tunneling processes. Such two-fermion tunneling correlations appear to be quasi-independent from each other because of their effective self-equilibration with their strongly interacting surroundings at each instant of time.} The latter observation can serve as the \textit{qualitative definition of any self-equilibrated regime of quantum fluctuations saturation}. At the same time, the qualitative reason behind the relevance of only two-electron correlations between tunneling events in the weak tunneling approach of the interest is rooted in the Pauli exclusion principle for tunneling bare electrons in combination with presumed "weak link" boundary condition (5) at the tunnel-coupled edges of the junction. These two facts evidently restrict the number of bare electrons which affect each other most strongly during the tunneling by the amount of two, because \textit{due to the Pauli exclusion principle, only two bare electrons can be strongly localized simultaneously and most closely to each other on the two tunnel-coupled edges of the junction with point-like boundary condition}. This way, self-equilibrated regime (26,27) of quantum fluctuations saturation to the steady state could explain several recently revealed common features in quantum charge-fluctuations dynamics which appear in a wide range of one-dimensional tunnel junctions\cite{48,49} and resonant level models\cite{49}.   

Therefore, \textit{one can consider the validity of differential equation (27) as very common definition of a novel self-equilibrated (SE-) regime of quantum fluctuations saturation to the steady state} for very broad range of many-body quantum field models even beyond one-dimensional Luttinger liquid tunnel junction model which is considered here. In what follows all the remarkable consequences of SE-regime of quantum fluctuations will be revealed. Hence, the proof of the validity of equations (26-27) and analysis of their consequences both represent main findings of this paper.

Here I would like to stress out once more, that in order to obtain central exact relations of Eq.(26-27) no assumptions additional to requirements of SE-theorems were made (see Appendix B below). These requirements are minimal: one needs to have only the interaction Hamiltonian (being standard bosonized tunnel Hamiltonian with arbitrary tunnel coupling in our case) in the form a fixed nonlinear ($ \cos  $ ) function of quantum (i.e. with zero mean) non-local bosonic (pase-)field ( $ \varphi_{-}(0,t) $ in our case ), freely propagating along  both semi-infinite 1D leads of the junction (the latter picture describes interaction representation implemented here), while conjugated (charge-) bosonic field  (i.e. $\theta_{\pm}(0,t)  $ in our case) should be pinned at point $ x=0 $ by means of "weak link" boundary condition: $\theta_{\pm}(0,t)=0  $ . In addition, fixed explicitly time-dependent phase shifts ( $ eVt $ for S-theorem of Ref.[38] and $ eVt + \xi(\tau) $ - for SE-theorem from Appendix B) are allowed (here $ \xi(\tau)=\pm\xi $ - is branch-dependent counting field defined on complex Keldysh contour: $ \xi(\tau)=\xi $ for $ Im{\tau}>0 $ and $\xi(\tau)=-\xi $ for $ Im{\tau}<0 $ - see Appendix B). 

But, on the other hand, starting right from the tunnel Hamiltonian definitions of Eqs.(10,23) in order to prove SE-theorem one strictly speaking does not have to require the smallness of a related coupling constant $ \tilde{\lambda} $, meaning that this coupling constant can be \textit{arbitrary} (i.e. it might not be small,in contrary to the majority of QPC models with  tunnel coupling). This observation could cufficiently enlarge the number of potential applications of obtained results. As well, as it has already been mentioned in the previous section, the resulting "gaussianity" of thermal fluctuations is an emergent phenomenon of accepted model, due to the existence of well-defined quasi-particles (two sorts of non-local free plasmons) in the system. Mathematically, the latter fact is the consequence of SE-lemma, derived in the Appendix B from above-mentioned general requirements of SE-theorem (the same is true also for S-lemma and S-theorem proven earlier by the author in Ref.[38]).

\textit{Remarkably, the exactness of Eq.(26) has been proven here by means of SE-theorem, can also serve as the manifestation and proof of well-known Jarzynski equality\cite{39,40} for a given quantum system with fluctuating quantum bosonic field $ \varphi_{-}(t) $  in the real-time regime.} 
Indeed, conventional Jarzynski equality\cite{40} (i.e. the one for imaginary time-domain, where $ \tau \in (0;0-i\beta) $ and $ \beta=1/T $ -is inverse temperature) for abstract statistical system out of the equilibrium reads
\begin{align}
\left\langle e^{-\beta W}\right\rangle_{tr} = e^{-\beta\Delta F},
\end{align}  
here $ \beta=1/T $ is inverse temperature, $ W $ is certain fluctuating quantity (say, external "work" on the system along its stochastic trajectory); the averaging in the l.h.s. of Eq.(28) is performed over the total number of different stochastic trajectories of the system (e.g. over different "external measurement" realizations) and $ \Delta F $ in the r.h.s. of Eq.(28) represents trajectory-independent change in system's free energy. 

Therefore, from the comparison of Eq.(26) and SE-theorem with Eq.(28) one can conclude that in our case fluctuating quantity $ W $ corresponds to a fluctuating action of interacting quantum field $ \varphi_{-}(t) $ in the Keldysh contour-ordered T-exponent, while the averaging $ \left\langle \ldots \right\rangle_{tr} $ over all possible stochastic trajectories in the l.h.s. of Eq.(28) in "conventional" Jarzynsky case\cite{39,40} corresponds in our case to summations in the l.h.s. of Eq.(26) over all possible (i.e. stochastic) "histories" of time-evolution of our system on the complex time Keldysh contour, for single spatial point  $ x=0 $, accompanied by system interaction with external (counting) field. At the same time, the r.h.s. of Eq.(26) represents just a particular realization of r.h.s. of Eq.(28), because the power in the exponent in the r.h.s. of Eq.(26) signals which particular type among all possible "histories" of the system in the complex time domain (at the point $ x=0 $) can be "wrapped" into the exponent function of respective average, in order to represent overall system's stochastic evolution (at point $ x=0 $) in the complex time domain. Therefore, such fixed type of system's histories (or, equally, a fixed type of skeleton diagrams) governs system's temporal evolution (at $ x=0 $) whereas other types of system's (stochastic) histories sum up exactly to zero contribution to r.h.s. of Eq.(26).  

Thus, from the comparison of Eqs.(26,27) with Eq.(28) it follows that SE-theorem represents a particular (dynamical) case of non-equilibrium Jarzynski equality (28) where electron's "imaginary work" along its  certain ($ j $-th) stochastic trajectory  in the time domain is
\begin{align}
W_{j}( \beta, \xi, t)=\left[ W(\beta, \xi, t) \right]^{(j)} =  \left[ \frac{i}{\beta}\int_{\mathcal{C}_K(t)} H_{I}(\tau,\xi(\tau)){\rm d}\tau \right]^{(j)} 
\end{align} 
(here $ j=1,2,\ldots, \infty $) which represents eigenvalue of electron's tunneling imaginary action associated with certain ($ j $-th) type of electron evolution during its tunneling through QPC. The latter is because one can always decompose respective quantum state at arbitrary moment of time as follows

\begin{align}
 \frac{i}{\beta} \int_{\mathcal{C}_K(t)} H_{I}(\tau,\xi(\tau)){\rm d}\tau \vert \phi_{LL}(0)\rangle_{T} =\sum_{j=1}^{\infty}W_{j}(\beta, \xi, t) \vert  j \rangle_{\beta}
\end{align}
in the basis of certain orthonormal and Gibbs-weighted eigenvectors $ \vert j \rangle_{\beta} $ with  corresponded eigenvalues $ W_{j}(\beta, \xi, t)  $. Then for our case the analog of trajectory-independent change in system's free energy $ \Delta F $ from the r.h.s. of Eq.(28) will be following

\begin{align}
\Delta E_{qf}(\beta,\xi,t)=\frac{1}{2\beta} \times \nonumber \\
   \int\int_{\mathcal{C}_K(t)}d \tau_{1}d \tau_{2} \left\langle  \mathcal{T}_K H_{I}(\tau_{1},\xi(\tau_{1})) H_{I}(\tau_{2},\xi(\tau_{2}))\right\rangle
\end{align}  
Thus,  in our case the latter quantity (31) describes time-dependent (i.e. dynamical) thermally averaged change in the energy of quantum fluctuations of the electron in 1-dimensional quantum wire due to electron tunneling through the tunnel barrier. This way Eqs.(29,31) -translate SE-theorem of Eqs. (26) to the quantum statistical "language" of non-equilibrium Jarzynski equality (28). Therefore, for arbitrary 1-dimensional tunnel junction at arbitrary temperature, bias voltage and time (if $ t > \Lambda_{g}^{-1} $ where $ \Lambda_{g} $ is high-energy cutoff of given continuous Luttinger liquid model) one can claim the validity of following \textit{dynamical version} of conventional Jarzynski equality 

\begin{eqnarray} 
\langle \exp\left\lbrace -\beta W(\beta, \xi, t)\right\rbrace \rangle_{\beta} = \sum_{j=1}^{\infty} \langle j \vert \exp\left\lbrace -\beta W_{j}( \beta , \xi, t)\right\rbrace  \vert  j \rangle_{\beta}  \nonumber \\
 = \exp\left\lbrace -\beta \Delta E_{qf}(\beta,\xi,t)\right\rbrace \nonumber 
\end{eqnarray}
\begin{equation}
\end{equation}  
with quantities $ W_{j}( \beta , \xi, t) $ and $ \Delta E_{qf}(\beta ,\xi,t) $ from Eqs.(29,31).

Evidently, revealed correspondence between Eq.(28) and Eq.(32) can be a novel (time-dependent and quantum) example of selective relevance of different stochastic trajectories similarly to the general non-equilibrium statistical effect being established in Ref.[50] for a wide enough class of statistical ensembles out-of-the equilibrium. 

\section{Self-equilibration of quantum fluctuations and FCS cumulant generating functional}

One can easily convince from the SE-theorem of Eqs.(22-26) that one can always represent the time-dependent cumulant generating functional $\mathcal{F}(\xi,t)$ in the following explicit and \textit{exact} form 
\begin{align}
\begin{split}
\mathcal{F}(\xi,t)= -\ln \left\lbrace \tilde{\chi}(t,\xi) \right\rbrace =\left(\tilde{\lambda}\right)^{2}\times \\
\frac{1}{2}\int^{t}_{0}d \tau_{1} \int^{t}_{0} d \tau_{2}u_{g}(\tau_{1}-\tau_{2})\times \\
\left\lbrace {\cos\left[\pi/g + eV(\tau_{1}-\tau_{2})\right]}\right\rbrace\left(e^{2i\xi\sgn{(\tau_{1}-\tau_{2})}}-1\right)\\ +\left(\tilde{\lambda}\right)^{2}\frac{1}{2}\int^{t}_{0}d \tau_{1} \int^{t}_{0} d \tau_{2}u_{g}(\tau_{1}-\tau_{2})\times\\
\left\lbrace {\cos\left[\pi/g - eV(\tau_{1}-\tau_{2})\right]}\right\rbrace\left(e^{-2i\xi\sgn{(\tau_{1}-\tau_{2})}}-1\right) 
\end{split}
\end{align}
with symmetric pair correlator
\begin{align}\label{eq:Corr_2}
\begin{split}
u_{g}(\tau-\tau')=\langle e^{i\varphi_-(\tau)-i\varphi_-(\tau')}\rangle \\ 
  =\left[ \frac{\pi T/\Lambda_{g}}{\sinh\left[\pi T \vert \tau - \tau' \vert \right]}\right]^{2/g}\,
\end{split}
\end{align}
which should be regularized properly at $ \vert\tau - \tau'\vert \rightarrow 0 $ with respect to corresponded time integrations of Eq.(33)(see Appendix A below). Formulas (33,34) together with Eqs.(20-22) and (26) -represent \textit{exact} general expression for the time-dependent Luttinger liquid QPC cumulant generating functional. 

One can see from Eqs.(33,34) as well as from Eqs.(22-26) and from SE-theorem of Sec.(IV) that all-cumulants generating functional $ \mathcal{F}(\xi,t)= -\ln \left\lbrace \tilde{\chi}(t,\xi) \right\rbrace  $ in the form (33,34) contains much more information about the real-time dynamics of quantum fluctuations in one-dimensional weakly linked QPC than the standard scattering approach of Levitov-Lesovik type \cite{26,27,28} can provide. Indeed, Keldysh partition function of the form (22-26) or, alternatively, corresponding  generating functional of the form (33,34) is the result of exact calculation of Keldysh contour ordered T-exponent (22) at arbitrary time-scale (larger than $ \hbar/\Lambda_{g} $ -the inverse high-energy cutoff of underlying Luttinger liquid model) in all orders of tunnel coupling constant $ \tilde{\lambda} $, in the whole range of bias voltages and temperatures (which though should be lesser than high-energy cutoff for Luttinger liquid model of electrodes) for all accesible values of Luttinger liquid correlation parameter $ 0 < g \leq 1 $.

Indeed, it was already stated by Levitov and Lesovik from the very beginning in their generic paper \cite{26} that their famous Levitov-Lesovik formula (see e.g. f-la (9) in Ref.[26]) had been obtained for the case of non-interacting electrons in the large-timescale asymptote (i.e. for time-independent steady state at $ t \rightarrow \infty $) within the scattering approximation in the lowest (i.e. second) order in scattering matrix elements for each energy channel of elastic electron scattering provided that all inelastic channels of electron scattering were neglected. In addition,  the limitations of the Levitov-Lesovik scattering approach in the real-time dynamics of fluctuations were studied later numerically by  Schönhammer in Ref.[54] for finite-sized quantum wires of non-interacting electrons. In particular, he had shown that the most relevant contribution to the entire logarithm of the Levitov-Lesovik formula at near-perfect electron tranasmission through QPC goes from the term being proportional to the square of tunneling amplitude (i.e. to $ \tilde{\lambda}^{2} $ in our notations) as if Levitov-Lesovik result would be valid only  to the lowest, i.e. second order of tunneling coupling constant or as if all logarithmic corrections of the form (5) from Ref.[26] were, in fact, artifacts of energy-independent (i.e. wide band-) scattering approximation and hence these logarithmic contributions "die off" in the large timescale steady state limit (see e.g. Eqs.(51,52) in Ref.[54] and related discussion in the text before and after these equations). 

On one hand, it is quite remarkable that the Levitov-Lesovik result of Eq.(5) from Ref.[26] for non-interacting electrons including corresponded finite-timescale logarithmic corrections  for the second order cumulant (i.e. for the shot-noise) can be easily obtained from Eqs.(33,34) when $ g=1 $  (non-interacting electrons in the leads) by expanding (33,34) with respect to a small quantity $ Tt / \hbar \ll 1 $ ( in this case the ratio $ (\Lambda_{1}/\hbar) $ with $ \Lambda_{1} $ being a high-energy cutoff of Luttinger liquid model of non-interacting electrons (Fermi liquid case)) would play a role of typical conductance dispersion frequency $ \omega_{0} $ from Eq.(5) of Ref.[26]). Moreover, the validity of generating functional (33,34) for weakly linked tunnel junctions  was indirectly supported (at least for the case of non-interacting electrons) in the later paper \cite{28} by Levitov and Reznikov where they had written the generating function (see Eq.(1) in Ref.[28]) as well as the general form of all cumulants (see Eq.(3) in Ref.[28] ) for the tunnel junction of non-interacting electrons in the steady state. Especially, the  relations (1,3) from Ref.[28] appear to be just a particular case of more general steady state relations of Eqs.(36-40) derived in below from Eqs.(33,34) for the steady state regime of arbitrary weakly linked Luttinger liquid tunnel junction.

On the other hand, SE-theorem and Eqs.(22-26) and (33,34) appear to be much more general than the Levitov-Lesovik scattering approach \cite{26,27,28,54}. Indeed, from Eqs.(22-26) as well as from the detailed proof of SE-theorem in the Appendix B, it is evident, that left- and right hand sides of Eq.(26) contain all possible types of diagrams which contribute to the tunneling including those describing all possible inelastic virtual processes which are beyond the scope of Levitov-Lesovik elastic scattering approach for non-interacting electrons. The latter is because within the bosonized description of one-dimensional interacting electron systems the relevant elementary excitations with linear dispersion are not bare electrons, but plasmons which represent non-linear collective excitations of interacting electrons in one-dimensional systems due to bosonization relations of Eqs.(7-9). Therefore, all possible inelastic virtual processes of scattering between bare interacting electrons are already taken into account in Eqs.(26,33) by means of all powers in $ \tilde{\lambda} $ in the exact power series expansion of Keldysh contour ordered T-exponent from Eqs.(22-25) with bosonized tunnel Hamiltonian of Eqs.(12-19). 

From the above it is clear that the SE-theorem proven and Eqs.(1-34) represent  real-time generalisation of a full-counting statistics scheme on the case of interacting electrons beyond the Levitov-Lesovik scattering approach, though containing all respective steady state results and finite timescale logarithmic corrections for the lowest cumulants as corresponded limiting case. To see this, at first, it is reasonable to consider the steady-state limit of tunnel  current at $  t \rightarrow \infty $. In this limit one will obtain 

\begin{equation}
\lim_{t\rightarrow \infty} \mathcal{F}(\xi,t)= W_{B}(\xi)\cdot t
\end{equation}
where, calculating analytically two time-integrals $ J_{C,B}=\lim_{t \rightarrow \infty}J_{C}(t) $ and $ J_{S,B}=\lim_{t \rightarrow \infty}J_{S}(t) $ (see Appendix C below) one can obtain for $  W_{B}(\xi) $ following \textit{exact} Luttinger liquid generalization of well-known "Levitov-Lesovik" formula 
\begin{align}
\begin{split}
W_{B}(\xi)=\left\lbrace {\Gamma_{g,+}\left[e^{2i\xi}-1\right]+\Gamma_{g,-}\left[e^{-2i\xi}-1\right]}\right\rbrace, 
\end{split}
\end{align}
with
\begin{align}
\begin{split}
 \Gamma_{g,\pm}=\Gamma_{g,\pm}(eV,T)=\left(\tilde{\lambda}^{2}/4\Lambda_{g}\right)F_{g}(eV,T)e^{\pm eV/2T}, 
\end{split}
\end{align}
where
\begin{eqnarray}
F_{g}(eV,T)=\frac{\vert\Gamma\left(1/g+i\left[eV/2\pi T\right]\right)\vert^{2}}{\Gamma\left(2/g\right)}\left[\frac{2\pi T}{\Lambda_{g}}\right]^{(2/g-1)}
  \end{eqnarray}
is our "Luttinger liquid"- (or "interaction") factor. Equations (35-38) give rise to the \textit{exact} long-time description of electron transport in the Luttinger liquid tunnel junction within the framework of its full counting statistics (FCS). Obviously, from the validity of Eqs.(35-38) it follows the fundamental property for the steady state limit of tunnel transport of both interacting and non-interacting electrons, being demonstrated previously only in the lowest order of perturbation theory in tunnel coupling and only for the case of noninteracting electrons (Fermi liquid) in the electrodes of tunnel junction \cite{28}  
 \begin{align}
\begin{split}
 \frac{\Gamma_{g,\pm}(eV,T)}{\Gamma_{g,\mp}(eV,T)}=e^{\pm eV/T}.
\end{split}
\end{align}
One may conclude that Eq.(39) represents just a "steady state version" of more general dynamical Jarzynski equality of Eq.(32) derived in the above by means of SE-theorem. Here it is important to highlight the fact, that due to non-perturbative nature of Eqs.(26,32), quantities $ \Gamma_{g,\pm}(eV,T) $ derived by means of Eqs.(37,38) represent exact tunneling rates for a given Luttinger liquid tunnel junction in its steady state regime (either for left-to-the-right electron tunneling for $ \Gamma_{g,+}(eV,T) $, or for electron tunneling in the opposite direction for $ \Gamma_{g,-}(eV,T) $ ). The latter means, that at $  t \rightarrow \infty $ , having large enough ensemble of identical QPC's (or, alternatively, repeating the same experiment enough times) one should have a perfect matching between experimentally measured values of $ \Gamma_{g,\pm}(eV,T) $ and theoretical predictions of Eqs.(37-39) with arbitrary accuracy. In principle, one is able to measure tunneling rates $ \Gamma_{g,\pm}(eV,T) $ directly, for example, by combining charge transfer detection (i.e. average current measurements) with the detection of electron tunneling events by means of counting field $ \xi $, i.e. by means of a precession of a spin being weakly coupled to QPC, see e.g. Ref.[29] for further details ).

Thus, important \textit{exact} result, which is proven here states that: \textit{In the "long-term" perspective, i.e. in the limit $ t \rightarrow \infty $ arbitrary repulsive electron-electron interactions in the leads of Luttinger liquid tunnel contact are unable to break the detailed statistical balance between physical "bare" electrons being transmitted through and reflected from such a tunnel contact.} Obviously, the latter statement represents direct consequence of self-equilibration regime of quantum fluctuations revealed in the SE-theorem from the above. 

Now substituting formulas (35,36) into Eq.(21) one easily obtains following exact relations (notice, that "minus" sign in the r.h.s. of equation reflects the sign of charge carriers - electrons. Here and everythere below index $ B $ is a shortcut of "balanced" meaning either detailed balance in the steady state or self-equilibration on the finite time-scale)
\begin{align}
\nonumber
 \lim_{t\rightarrow\infty}\langle\langle N^{k}(t)\rangle\rangle=\\
 \langle\langle N^{k}\rangle\rangle_{B}=\left\{ 
 \begin{matrix}
   -\left( \Gamma_{g,+}+\Gamma_{g,-}\right)t &, k=even\\
    -\left( \Gamma_{g,+}-\Gamma_{g,-}\right)t &, k=odd \\
        \end{matrix} \right.
\end{align}
The latter represent a Luttinger liquid generalization of similar expressions have been derived earlier only in the perturbative regime (i.e. only in the lowest non-vanishing order $ \tilde{\lambda}^{2} $ in tunnel coupling constant $ \tilde{\lambda} $) by Levitov and Reznikov for the case noninteracting electrons\cite{28}. In particular, for the average current from obtained general formulas (36-38,40) it follows that $ \bar{I}_{B}(eV,T)=-\left(\tilde{\lambda}^{2}/ \Lambda_{g}\right)F_{g}(eV,T)\sinh(eV/2T) $ as well, for respective noise power $ \bar{S}_{B}(eV,T)=-\left(\tilde{\lambda}^{2}/ \Lambda_{g}\right)F_{g}(eV,T)\cosh(eV/2T) $ with $ F_{g}(eV,T) $ from Eq.(38). Obviously, these expressions give correct "high-" and "low-temperature" asymptotes, which coincide with well-known Kane-Fisher scaling\cite{12} for arbitrary values of Luttinger liquid correlation parameter: $0 < g \leq 1$. Therefore, for the ratio of two subsequent irreducible correlators for arbitrary Luttinger liquid tunnel junction in the "detailed balance"- (or, equally, "steady flow"-) regime of tunneling of strongly correlated electrons one has
\begin{align}
\begin{split}
\lim_{t\rightarrow\infty} F_{B}(t) = \lim_{t\rightarrow\infty} \dfrac{\langle\langle N^{2k}(t)\rangle\rangle_{B}}{\langle\langle N^{2k-1}(t)\rangle\rangle_{B}}\\
= F(eV,T)=\frac{\bar{S}_{B}}{\bar{I}_{B}}=\coth(eV/2T). 
\end{split}
\end{align} 
(for $ k=1,2,... $), where $ F(eV,T) $ represents a \textit{universal} Fano-factor (i.e. noise power to current ratio), which remarkably coincides with the Fano-factor $F_{0}(eV,T)$ derived by Sukhorukov and Loss for noninteracting electrons in the leads\cite{32}. 

Now we have everything to study the most general non-equilibrium case of the problem at arbitrary $ t $. First, from Eqs.(21,33,34) a finite-time generalization of Eq.(40) becomes straightforward \cite{55} 

\begin{align}
 \langle\langle N^{k}(t)\rangle\rangle_{B}=\left\{ 
 \begin{matrix}
   -\left( \Gamma_{g,+}(t)+\Gamma_{g,-}(t)\right) t &, k=even\\
    -\left( \Gamma_{g,+}(t)-\Gamma_{g,-}(t)\right) t &, k=odd \\
        \end{matrix} \right.
\end{align}

At the same time from Eqs.(21,42) one can extract also the \textit{non-equilibrium Fano factor} $ F_{B}(t) $ which remains \textit{universal} for any pair of $ 2k $-th and $ 2k-1 $-th cumulants ($ k=1,2,\ldots $) as well as its steady-state analog from Eq.(41)
\begin{align}
\begin{split}
 F_{B}(t) =  \dfrac{\langle\langle N^{2k}(t)\rangle\rangle_{B}}{\langle\langle N^{2k-1}(t)\rangle\rangle_{B}}= -i \partial_{\xi} \ln \lbrace \partial_{\xi} \mathcal{F}(\xi,t) \rbrace \mid_{\xi=0} , 
\end{split}
\end{align} 
This formula, obviously, can be also rewritten explicitly as follows
\begin{align}
\begin{split}
 F_{B}(t) = \dfrac{\left( \Gamma_{g,+}(t)+\Gamma_{g,-}(t)\right)}{\left( \Gamma_{g,+}(t)-\Gamma_{g,-}(t)\right)}=\frac{\tilde{J}_{C}(t)}{\tilde{J}_{S}(t)}. 
\end{split}
\end{align} 

In Eqs.(42,44) one has $ \Gamma_{g,+}(t)\pm \Gamma_{g,-}(t)=\tilde{\lambda}^{2}J_{C(S)}(t) $, where $ J_{C}(t)=\dfrac{\tilde{J}_{C}(t)}{2} \left(\dfrac{\pi T }{\Lambda_{g}}\right)^{2/g}$ and $ J_{S}(t)=\dfrac{\tilde{J}_{S}(t)}{2} \left(\dfrac{\pi T }{\Lambda_{g}}\right)^{2/g}$ with
\begin{align}
\begin{split}
\tilde{J}_{C}(t)=\int^{t}_{0} ds \dfrac{\cos\left(eV s \right)\cos\left[ \pi/g - 2\eta(s)/g \right] }{\left[\sinh^2\left(\pi T s \right)+\left(\pi T /\Lambda_{g}\right)^2\cosh^2\left(\pi T s \right)\right]^{1/g}} \\
\end{split}
\end{align}
and
\begin{align}
\begin{split}
\tilde{J}_{S}(t)=\int^{t}_{0} ds \dfrac{\sin\left(eV s \right)\sin\left[ \pi/g - 2\eta(s)/g \right] }{\left[\sinh^2\left(\pi T s \right)+\left(\pi T /\Lambda_{g}\right)^2\cosh^2\left(\pi T s \right)\right]^{1/g}} \\
\end{split}
\end{align}
with $ \eta (t)=\arctan[ (\pi T/\Lambda_{g})\coth(\pi T t)] $. The presence of a phase factor $  \eta (t) $ in Eqs.(45,46) is important since it allows e.g. for finite time-scale logarithmic correction of the type of Eq.(5) from Ref.[26] for the second cumulant (shot-noise). To see that, one can expand  Eq. (45) (including the pahse factor $  \eta (t) $) with respect to the small parameter $ \pi T t \ll 1 $ and substitute the result into Eq.(42) for $ k=2 $ and $ g=1 $.

\begin{figure}
\includegraphics[height=14 cm,width=9 cm]{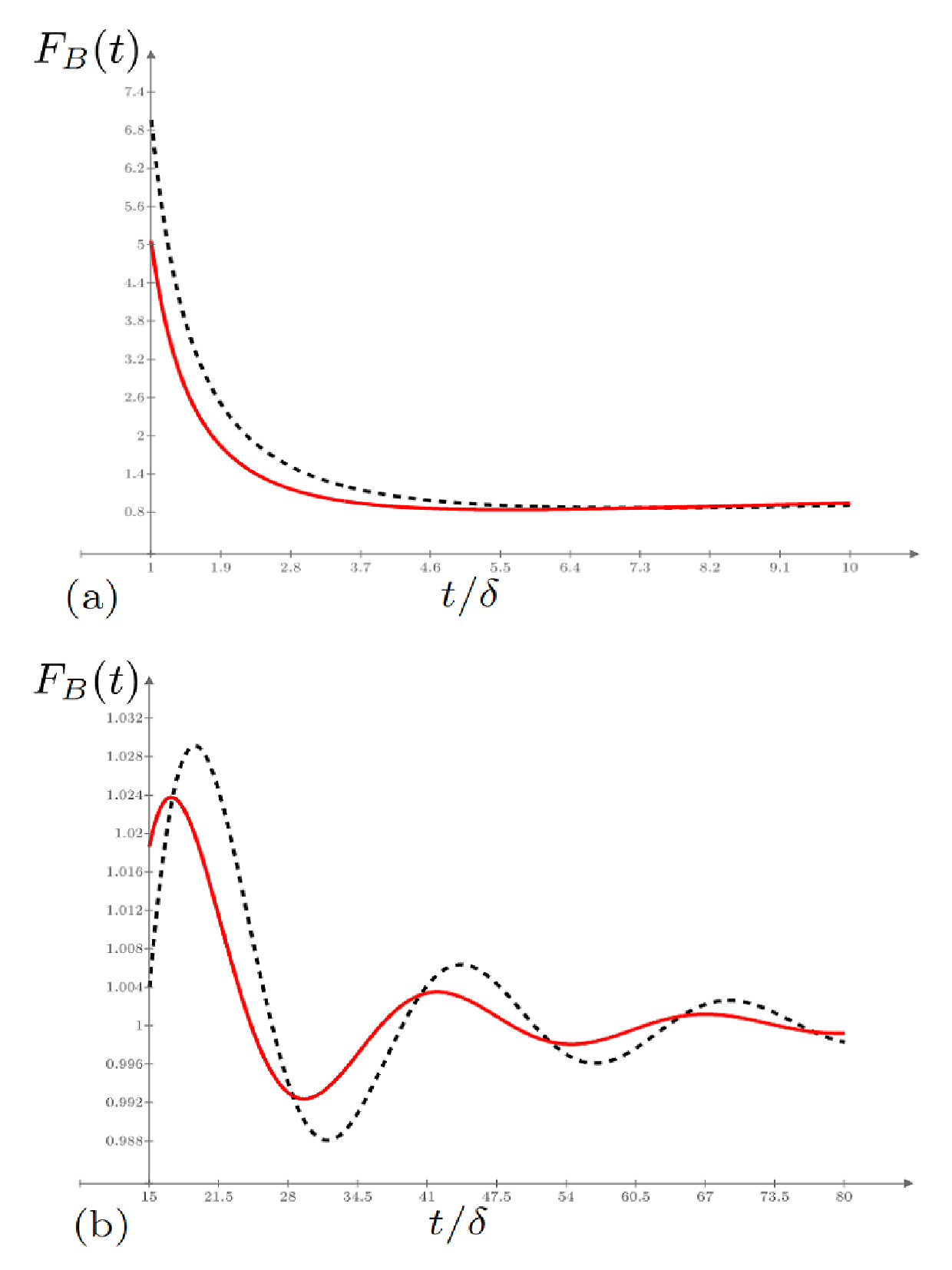}
\caption{Non-equilibrium Fano-factor $ F_{B}(t) $ as the function of time $ t / \delta $, measured in the units of $ \delta \approx \Lambda^{-1} $ for weakly interacting case (when $ g \lesssim 1$). On Figs.1(a,b) the  noninteracting (Fermi liquid) case: $ g=1 $ is described by the black dashed curves while red solid curves on both figures 1(a,b) are plotted for the case: $g=0.85 $. On Fig.1(a) one can see early stages of the (self-)equilibration to the steady state of tunnel current, where non-equilibrium Fano factor is superpoissonian, while on Fig.1(b) the small decreasing oscillations around the steady state poissonian value of Fano factor (being equal to $ 1 $) for the late stages of equilibration process are depicted.  For both Figs.1(a,b) I put $ T=0.001 \Lambda $ and $ eV=0.25 \Lambda $, while $ \tilde{\lambda}^{2}=0.025 \Lambda $.}
\end{figure} 

\begin{figure}
\includegraphics[height=14 cm,width=9 cm]{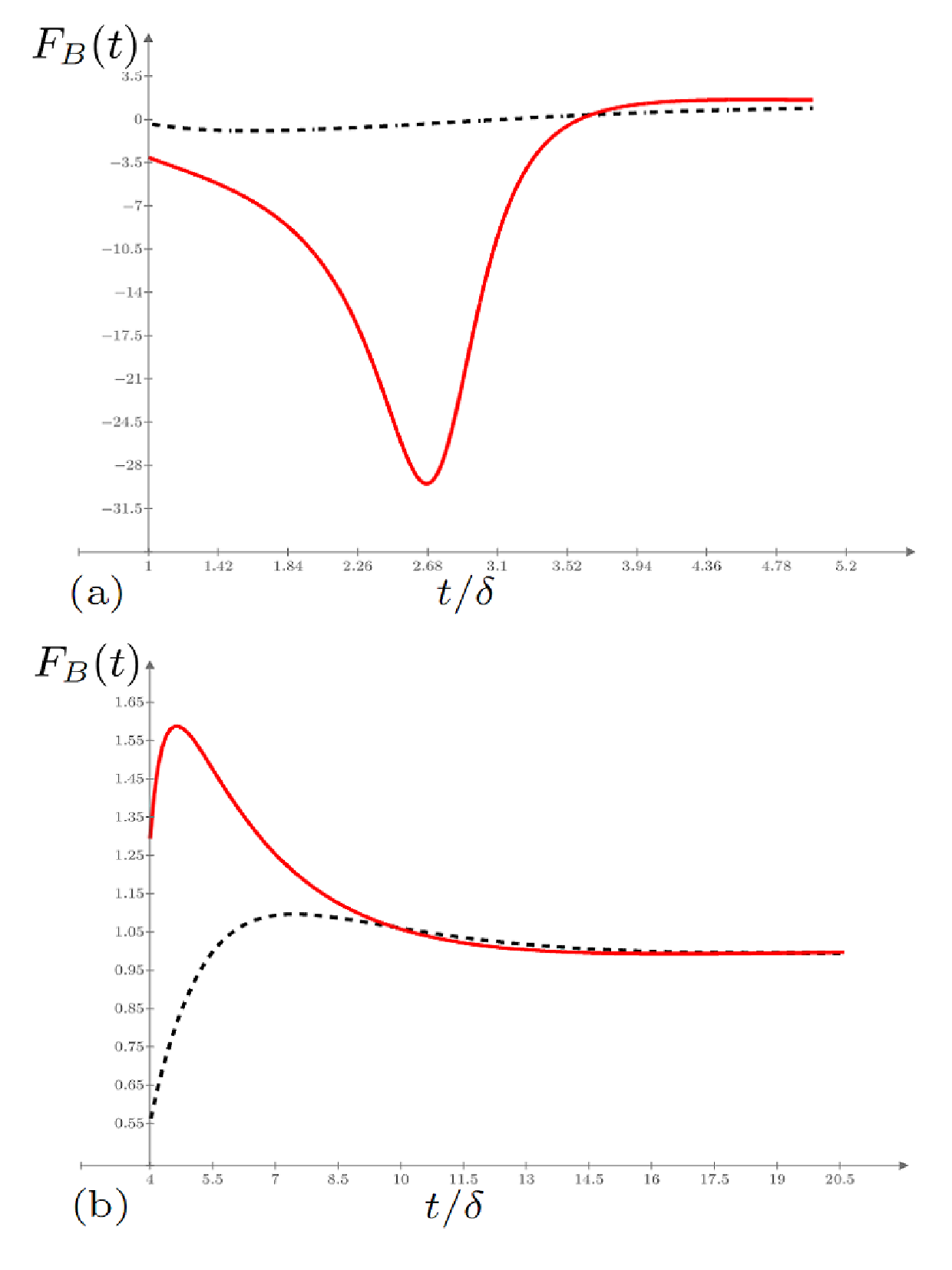}
\caption{Non-equilibrium Fano-factor $ F_{B}(t) $ as the function of time $ t / \delta $, measured in the units of $ \delta \approx \Lambda^{-1} $ for strongly interacting case (when $ g < 1$). On Figs.2(a,b)  black dashed curves correspond to the case $ g=0.5 $,  while red solid curves on both figures 2(a,b) are plotted for the case: $g=0.45 $. On Fig.2(a) one can see early stages of the (self-)equilibration to the steady state of tunnel current, where non-equilibrium Fano factor is superpoissonian (and even negative due to non-equilibrium fluctuations), while on Fig.2(b) small decreasing oscillations around the steady state poissonian value of Fano factor (being equal to $ 1 $) for the late stages of equilibration process are depicted.  All other parameters on Figs.2(a,b) are the same as ones on Figs.1(a,b). One can notice much faster (self-)equilibration to the steady state of the tunnel current with much weaker amplitude of charge fluctuations for strongly interacting electrons as compared to the case of weakly interacting ones. Though, one can notice a qualitative similarity between (self-)equilibration processes for weakly and strongly interacting electrons.}
\end{figure}

Obviously, Eqs.(41-46) enable one to introduce one more time-dependent \textit{measure of disequilibrium} for any moment of time $ t $ after the beginning of the tunnel transport measurements (though the lower bound on the relevant timescales is not zero but the inverse high-energy cutoff $ 1/\Lambda \simeq 1/E_{F} $) which though would have an explicit probabilistic meaning. Below I introduce such new quantity and call it a \textit{"steady flow" rate}. Since this quantity should not be strongly fluctuating at all times, should have well-defined limit in the steady state regime at $ t \rightarrow \infty $ and should vary between $ 0 $ and $ 1 $ in order to have a clear probabilistic meaning (see Discussion section  below), I have chosen for it following expression
\begin{align}
\begin{split}
R_{SF}(t)=\frac{1}{2}\left( \dfrac{t-\langle\langle N^{2k}(t)\rangle\rangle}{t-\langle\langle N^{2k-1}(t)\rangle\rangle}\right) \\
=\frac{1}{2}\left( \dfrac{1-\left( \Gamma_{g,+}(t)+\Gamma_{g,-}(t)\right)}{1-\left( \Gamma_{g,+}(t)-\Gamma_{g,-}(t)\right)}\right)=\frac{1}{2}\left( \dfrac{1-\tilde{\lambda}^{2}J_{C}(t)}{1-\tilde{\lambda}^{2}J_{S}(t)}\right), 
\end{split}
\end{align}
where $ J_{C}(t)=\dfrac{\tilde{J}_{C}(t)}{2} \left(\dfrac{\pi T }{\Lambda_{g}}\right)^{2/g}$ and $ J_{S}(t)=\dfrac{\tilde{J}_{S}(t)}{2} \left(\dfrac{\pi T }{\Lambda_{g}}\right)^{2/g}$ with $ J_{C(S)} $ from Eqs.(45,46).

One can see from Eqs.(45-47) that introduced measure of disequilibrium (or "steady flow" rate) $ R_{SF}(t) $ of Eq.(47) has  following remarkable property
\begin{align}
\begin{split}
\lim_{t\rightarrow \infty}R_{SF}(t)\approx \frac{F_{B}}{2}=\frac{1}{2}\coth(eV/2T) \approx \frac{1}{2}.
\end{split}
\end{align}

Notice, that quantities $ \Gamma_{g,\pm}(t) $  in Eqs.(42,44) have the same statistical meaning of exact (but non-equilibrium) tunneling rates as their steady state analogs: $ \Gamma_{g,\pm} $. Due to the exactness of underlying Eqs.(26,33,34) at arbitrary timescales, starting from the inverse high-energy cutoff of a given Luttinger liquid model in the QPC electrodes, these non-equilibrium rates can be measured directly with arbitrary accuracy in the electron transport experiments, analogously to $ \Gamma_{g,\pm} $ \cite{29}, if one manages to repeat the same electron transport experiment with given QPC for enough number of times, each time performing all needed tunneling measurements during the fixed period time interval $ (0;t) $ of chosen length $ t $ (the possible alternative could be the same transport measurements during the same $ (0;t) $ period of time on the large enough ensemble of identical QPC's). 

Naturally, the "steady flow" rate $ R_{SF}(t) $ defined by means of Eqs.(45-48) represents a clear \textit{measure of disequilibrium} for electron transport in \textit{any} 1D tunnel junction, if to understand the \textit{equilibrium} in such system as the steady state with certain constant average transport characteristics of Eqs.(35-41).

\begin{figure}
\includegraphics[height=14 cm,width=9 cm]{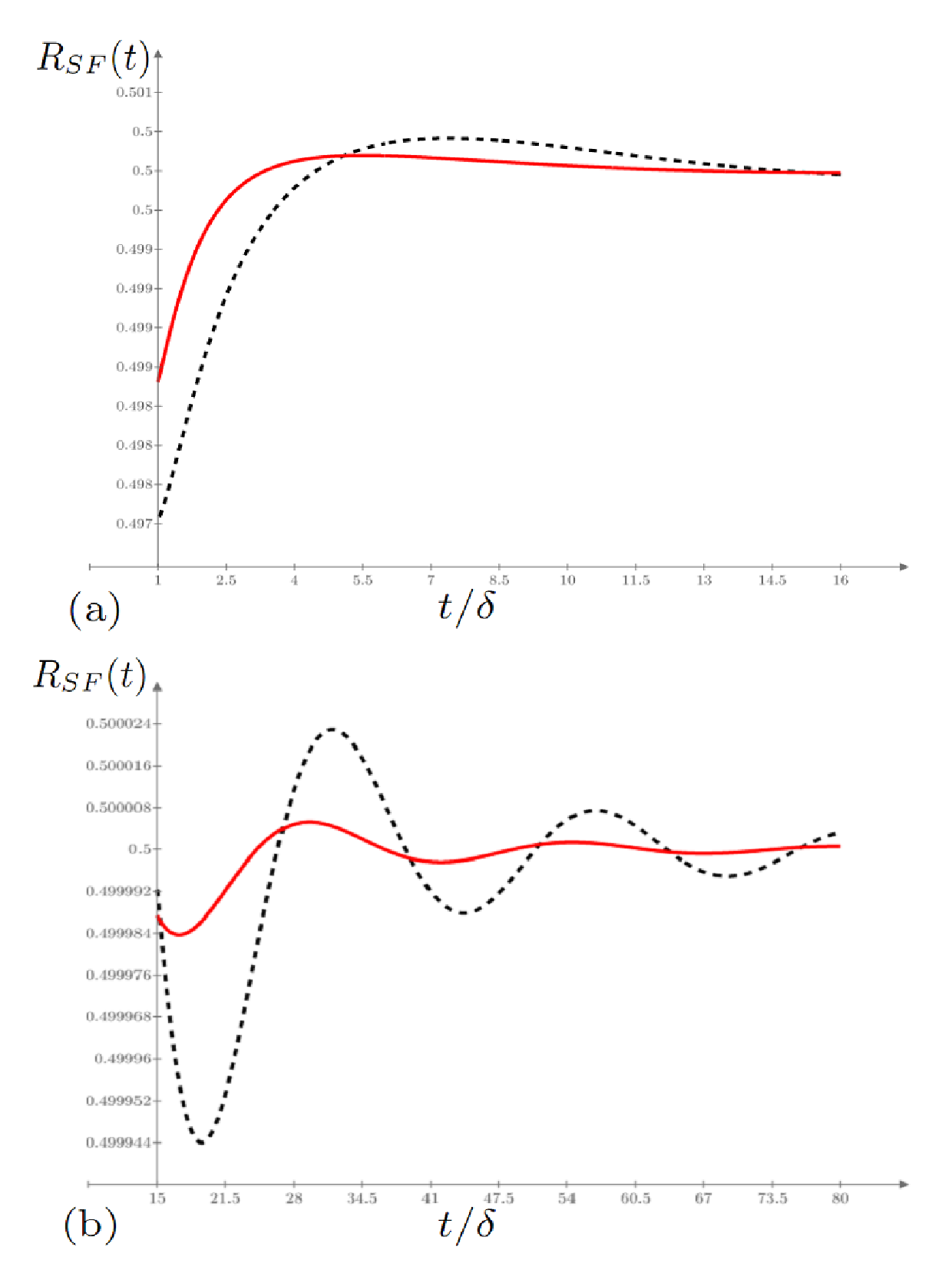}
\caption{Low-temperature "steady-flow rate" $ R_{SF}(t) $ as the function of time $ t / \delta $, measured in the units of $ \delta \approx \Lambda^{-1} $. On Figs.3(a,b) the noninteracting (Fermi liquid) case: $ g=1 $ is described by the black dashed curves  while red solid curves on both figures 3(a),(b) are plotted for the case: $g=0.85 $. All other parameters on Figs.3(a,b) are the same as ones on Figs.1(a,b). From Fig.3(a) one can see early stages of (self-)equilibration to the steady state of tunnel current, while on Fig.3(b) one can see small decreasing oscillations around the equilibrium steady state value of $ R_{SF}(t) $ being equal to $ 1/2 $ at the late stages of (self-)equilibration process  (see also the discussion section in the main text). Notice also much smaller amplitudes of quantum fluctuations for "steady-flow rate" $ R_{SF}(t) $ as compared to the low-temperature non-equilibrium Fano factor $ F_{B}(t) $ from Figs.(1,2) on all timescales. }
\end{figure} 

\begin{figure}
\includegraphics[height=14 cm,width=9 cm]{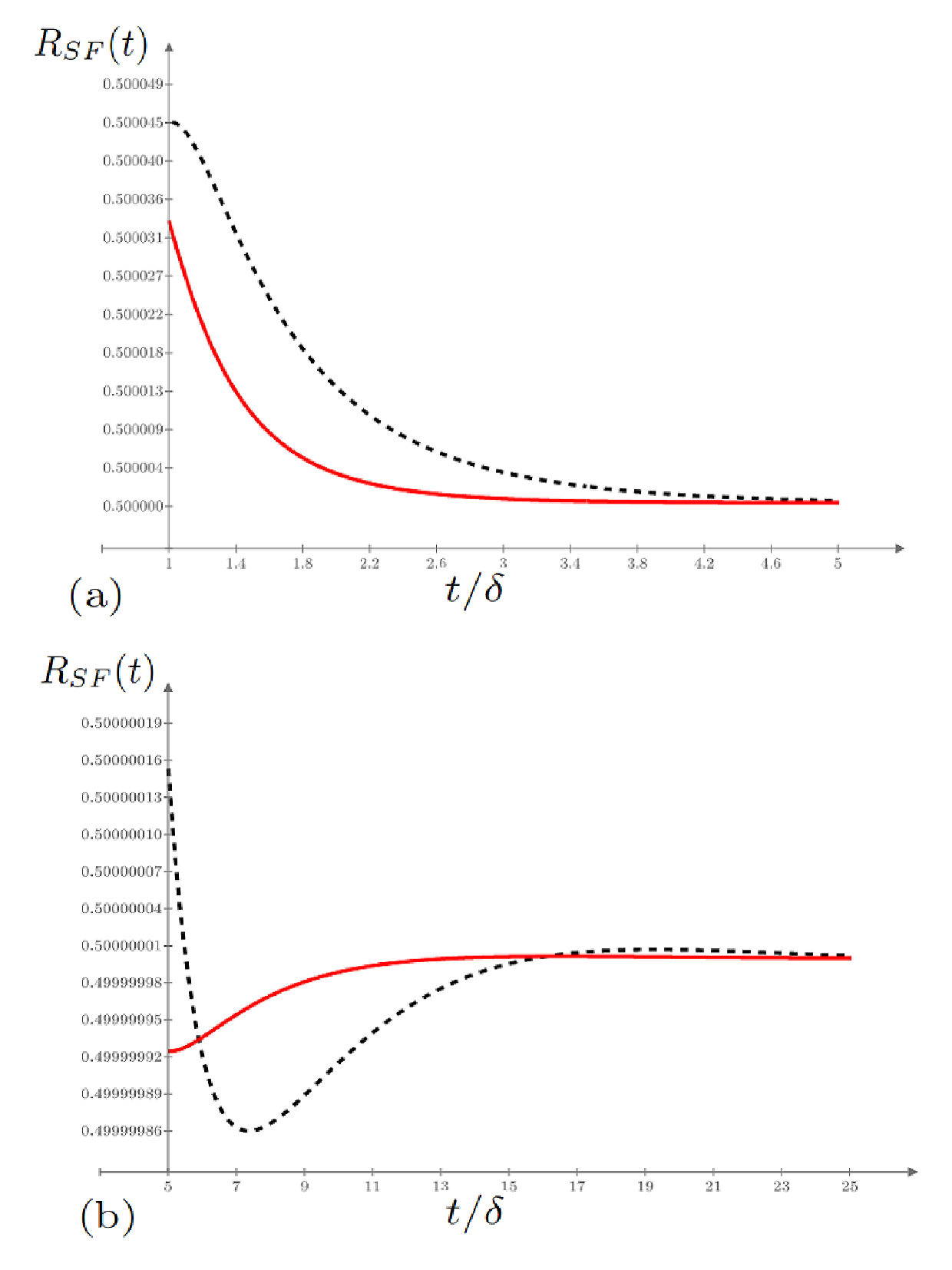}
\caption{Low-temperature "steady-flow rate" $ R_{SF}(t) $ as the function of time $ t / \delta $, measured in the units of $ \delta \approx \Lambda^{-1} $. On Figs.4(a,b) the case: $ g=0.5 $ is described by the black dashed curves, while red solid curves on both figures 4(a,b) are plotted for the case: $g=0.45 $. All other parameters on Figs.4(a),(b) are the same as ones on Figs.1(a,b). From Fig.4(a) one can see early stages of (self-)equilibration to the steady state of tunnel current, while on Fig.4(b) one can see small decreasing oscillations around the equilibrium steady state value of $ R_{SF}(t) $ being equal to $ 1/2 $ at the late stages of (self-)equilibration process (see also the discussion section in the main text). Here one also can notice the qualitative similarity in (self-)equilibration to the steady state of tunnel contact between weakly and strongly interacting electrons, although in the latter case such (self-)equilibration dynamics takes much smaller time.}
\end{figure} 

One can see that universality of relations (40-44) and (47-48) between two consquent irreducible correlators of $ 2k $-th and $ 2k-1 $-th order at arbitrary moment of system's temporal evolution (as well as in its steady state with respect to tunnel current) represent distinctive features of the \textit{universal self-equilibration behaviour} of quantum fluctuations in any (both interacting and non-interacting) one-dimensional tunnel contact. \textit{This established universal self-equilibration behavior is a direct consequence of SE-theorem and has been marked in the above as "self-equilibration of quantum fluctuations" phenomenon}. Several disconnected hints on the revealed universality of transport characteristics in low-dimensional tunnel contacts have been established earlier\cite{32,35,36,48,49} by different authors (see also the Introduction in the above). The overall qualitative picture behind discovered self-equilibration phenomenon will be illustrated in details in the next section.  

\section{Discussion}

On Figs.1,2(a,b) one can see plots of the time dependence of non-equilibrium Fano-factor $ F_{B}(t) $  in the low-temperature limit. Especially, on Figs.1(a,b) such graphs are plotted for non-interacting ($ g=1 $ black dashed curve) and weakly interacting ($ g=0.85 $ - red solid curve) electrons in the leads of Luttinger liquid QPC. Whereas, on Figs.2(a,b) there are the same time-dependences but for the case of strongly interacting electrons (for $ g=0.5 $ - black dashed curve and for $ g=0.45 $ - red solid curve, correspondingly). By comparing Figs.1(a),(b) with Figs.2(a,b) one can conclude about the \textit{universal character of (self-)equilibration to the steady state of tunnel current for both weakly - and strongly interacting electrons in the leads of arbitrary weakly linked tunnel contact}. Especially, in both cases of weakly and strongly interacting electrons in QPC leads, on the early stages of equilibration process the non-equilibrium Fano-factor $ F_{B}(t) $ quickly decreases from the large super-poissonian absolute values ($\vert F_{B}(t)\vert \gg 1 $) to vanishing oscillations aroud its low-temperature equilibrium poissonian value (being equal to $ 1 $) on the late stages of the (self-)equilibration process. The large absolute values of non-equilibrium Fano factor on the early stages of equilibration process in both "strongly interacting" and "non-interacting" cases should be refered to \textit{strong quantum fluctuations of all even correlators of tunnel current} at the beginning of equilibration process. Such a super-poissonian behavoir of Fano factor out-of-the equilibrium has been recently demonstrated for similar QPC models\cite{49}. Surprisingly, for non-interacting and weakly interacting electrons in tunnel junction the overall (self-)equilibration takes much longer time($ \sim 15-20\delta $ for early stage and $ \sim 10^{2} \delta $ - for the late stage of equilibration) than in the case of strongly repulsing electrons in the leads ($ \sim 3-5 \delta $ for early stage and $ \sim 10-50 \delta $ - for the late stage of equilibration). The reasons behind such the distinction as well as the issues of revealed universality of equilibration processes in any ballistic "weak-linked" tunnel contacts are discussed below.

Especially, one can follow the same type of (self-)equilibration behavior as for non-equilibrium Fano factor in the above on the new quantity $ R_{SF}(t) $ -"steady flow rate" being introduced by means of Eqs.(47,48). Here I demonstrate that such quantity has clear meaning in terms of different kinds of conditional probabilities for tunneling electrons. Therefore, the overall picture of the (self-)equilibration to the steady state of the tunnel current while being analyzed in terms of  $ R_{SF}(t) $ appears to be more intuitively clear than the same picture in terms of non-equilibrium Fano-factor dynamics.

Thus, on Figs.3,4(a,b) the quantity $ R_{SF}(t) $ is plotted as the function of time (in the units of $ \delta \simeq \Lambda^{-1}$), according to obtained exact Eqs.(45-48) for four distinct values of Luttinger liquid correlation parameter $ g $ in the low-temperature regime of electron tunneling. Figs.3(a,b) depict non-interacting and weakly interacting cases ($ g=1 $ and $ g=0.85 $), while Figs.4(a,b) refer to strongly interacting case ($ g=0.5 $ and $ g=0.45 $, correspondingly). On both Figs.3,4 figure (a) corresponds to early stage of (self-)equilibration process, while figure (b) depicts saturation of the non-equilibrium quantity $ R_{SF}(t) $ to its steady-state value (being equal to $ 1/2 $). One can see from Figs.3,4(a) that for any strength of electron-electron repulsive interactions in the electrodes (i.e. for any possible value of parameter $ 0< g \leq 1 $ ) the respective (self-)equilibration represents a fast enough and, at the same time, a tiny process, which evolves to the equilibrium on the timescales which are not bigger than $ \simeq 10^{2}\delta_{g} $: its first and most "disequilibrated" stage takes relatively short time $\sim 15-20\delta_{g}  $ - for weakly interacting electrons and even shorter (of the order of $\sim 3-5\delta_{g} $) - for strongly interacting electrons in the QPC leads. Besides that, as one can see from Figs.3,4(b)  on the later stages of system's quantum dynamics (of the order of $\sim 10-50\delta_{g}  $) the quantity $ R_{SF}(t) $ - slightly oscillates with a very small amplitude around its low-temperature steady-state- (or "steady-flow" - ) value being approximately equal to $ 1/2 $ , while the amplitude of such oscillations quickly decays to the values beyond the accuracy of its steady-state ($ 1/2 $) value detection. - These tiny oscillations represent a reminiscence of two different effects: First one is governed by bias voltage (i.e. by means of $ cos(eVt) $ and $ sin(eVt) $ factors in Eqs.(45,46)) - and it has the same nature as one for a non-stationary Josephson effect in the superconducting squids \cite{41} (since in both cases one has fluctuating phase-field $ \varphi(t) $ which enters Hamiltonian via $ cos(\varphi(t)+eVt) $ term). The second contribution to these tiny oscillations goes from the oscillating factors $ cos(\pi/g - 2\eta(t)/g) $ and  $ sin(\pi/g - 2\eta(t)/g) $ in Eqs.(45,46) and it is similar to the "Friedel oscillations" phenomena\cite{42} though realized here in the time domain and the latter is compatible with similar effects in the weak backscattering limit of constricted 1D quanutm wires. 

The most interesting effect of electron-electron interaction being evident from Figs.1-4 is that the steady state regime (i.e. the "steady flow" of strongly interacting electrons in the leads) establishes in the QPC system much faster than it takes the system of weakly interacting (and/or non-interacting) electrons in the QPC leads to evolve to the same regime of a "steady flow". This feature is remarkable just because in the steady state of ballistic electron transport "interacting" Luttinger liquid tunnel junctions provide much slower electron tunneling than "non-interacting" ones due to well-known Kane-Fisher effect of suppression of electron density of states on the edges of two tunnel-coupled 1D quantum wires\cite{11,12,17,38}. Here, as it can be seen from Figs.1-4, Kane-Fisher suppression of electron tunneling is manifested in much lower amplitudes of non-equilibrium  oscillations of "steady flow" rate $ R_{SF}(t) $ for interacting electrons, however the latter does not contradict much faster decay of these non-equilibrium oscillations to the low-temperature steady flow value of $ R_{SF}(t)\approx 1/2 $ being revealed here for arbitrary interacting  electrons (when $ g<1 $) in the tunnel junction. Moreover, it seems to be quite natural when strong electron-electron correlations in the system "enforce" its evolution to the  steady state more effectively than "weak" correlations in the non-interacting and weakly interacting cases can provide. 

Here one might ask, why revealed non-equilibrium effects on the early stages of ballistic electron transport in tunnel junctions are so small by magnitude if to consider respective "steady flow" rate quantity? - The answer is in the statistical meaning of introduced $ R_{SF}(t) $ rate with respect to weak tunneling regime under consideration. The point is, one may treat $ R_{SF}(t) $ as the \textit{probability} to find the electron in its "steady current-carrying" (or chiral, or  "steady-flow") quantum state, but "living" in one (left or right) electrode of QPC.  One can see from all graphs on Figs.3,4 that for large enough timescales this probability(i.e. the function $ R_{SF}(t) $) tends to its value $\approx 1/2 $ in the steady flow regime, which corresponds to the situation where mentioned "steady-flow" state of the electron has been already formed (with probability equal to one) and, thus, probability to find an electron on fixed (left or right) electrode of the junction becomes to be equal simply to $ 1/2 $ (if to exclude all thermal fluctuations, i.e. for low enough temperatures). 

Equally, one may think about introduced "steady flow" rate $ R_{SF}(t) $ as about the time-dependent \textit{conditional probability} for bare electron to be in its "steady flow" (or chiral) quantum state in any among two QPC electrodes but, at the same time, not to be involved into the (both virtual and real) tunnelings through the tunnel contact of interest. Obviously, such conditional probability also can be a definite "marker" of system (self-)equilibration to its steady flow state. Then from Figs.3,4 one can see that such the probability fluctuates only weakly due to assumed weak tunneling regime and (because of the same reason) - it saturates quickly to its equilibrium low-temperature value $\approx 1/2 $. The latter means, that \textit{in the "steady flow" regime of electron transport all electrons of the junction tend to form a collective chiral (or current-carrying) quantum state delocalized throughout the junction. Therefore, the probability to find any bare electron in one among two available QPC leads tends to be equal to $ 1/2 $ in the described "steady flow" regime of electron tunneling}. 

Obviously, the revealed special regime of real-time quantum fluctuations in tunnel junctions with the formation of collective "steady flow" state of tunneling electrons is beyond the scope of common Levitov-Leovik steady state scattering approach. 

Here it is worth to mention that similar physics with a formation of collective chiral "steady flow" quantum state might take place also for the electron transport in the (close to BSG \cite{49} )  self-dual resonant level model  \cite{49,51} where the  effective charge  of charge carriers can vary from $ 2e $ to $ e/2 $ depending on model parameters.

Evidently, all the graphs on the Figs.(1-4) as well as underlying analysis confirm the universality of self-equilibration behaviour in a wide range of ballistic tunnel contacts with both interacting and non-interacting electrons. This way plotted graphs properly illustrate self-equilibrated regime of quantum fluctuations being revealed here by means of Self-equilibration (SE-)theorem. 

\section{Conclusions}

To conclude, due to rigorous mathematical statements: Self-equilibration (SE-)theorem and Self-equilibration (SE-)lemma have been proven in the above, it becomes possible to calculate exactly the time-dependent cumulant generating functional in all orders of tunnel coupling for arbitrary Luttinger liquid tunnel junctions in the regime of a "weak link", for arbitrary temperatures and bias voltages. It has been also demonstrated that SE-theorem and all related calculations are beyond the common scattering approach which leads to Levitov-Lesovik formula for non-interacting electrons. However, one can extract all respective Levitov-Lesovik steady-state results for the lowest order cumulants as well as the finite-timescale logarithmic corrections to the former for the particular case of non-interacting electrons from all general formulas being derived in the above. The resulting exact formulas describe explicitly all real-time correlations in the system at any instant of time and in the limit of long observation timescales latter cumulants take generalized Levitov-Lesovik structure, similarly to the case of non-interacting electrons in the leads. The emergent universal behaviour of real-time quantum fluctuations of tunnel current and underlying SE-theorem allows to introduce and descibe, for the first time, a novel phenomenon of \textit{self-equilibration of quantum fluctuations} to the steady state, which appear to be a common issue for any one-dimensional tunnel junction due to derived \textit{self-equilibration differenttial equation}. The distinctive features and explanations of this novel phenomenon of quantum dynamics of many-body electron systems were studied in details. The exactness of obtained results being proved by means of SE-theorem at the same time proves the validity of two important theorems from statistical physics for arbitrary Luttinger liquid tunnel junction. These are: 1)  the time-dependent version of non-equilibrium Jarzynski equality and 2) the detailed balance theorem for tunnel current of interacting electrons in the long-term limit (i.e. in the steady state of tunnel current). On the other hand, exact time-dependent correlators of all orders (obtained in the above from the exact cumulant generating functional) allow to introduce and to study a new measure of disequilibrium in the system: a "steady flow" rate, which describes the character of the Luttinger liquid QPC \textit{self-equilibration} to its current-carrying steady state (here I call latter  \textit{"steady flow" state}). As the result, the universality of self-equilibration behaviour for ballistic electron transport in arbitrary  Luttinger liquid tunnel junction has been studied explicitly. Especially, it was found that strong electron-electron interactions in the QPC electrodes brings QPC to its steady state much faster than the same occurs in the case of non-interacting and weakly interacting electrons in the tunnel junction. All the obtained results are believed to be important for a wide range of problems in modern low-dimensional quantum mesoscopics and quantum statistical physics. 

Author would like to thank L.S.Levitov, who drew author's attention to C.Jarzynski result. This research was partially supported by the research grant No.09/01-2020 from the National Academy of Sciences of Ukraine.

\appendix
 
\section{ Basic real time pair correlator}\label{sec:Corr_Intt}

Let us pay more attention to pair correlator of Eq.(34). The basic ingredient in its calculation is   well-known integral over $ k $  with nonzero high-energy cutoff $e^{-\alpha_0 k}$, which  can be calculated exactly (see e.g. Refs.[17,38,43]) and the answer is
\begin{align}\label{eq:Corr_time}
\begin{split}
\left\langle\left[\varphi_{\pm}(t)-\varphi_{\pm}(0)\right]^2\right\rangle=I(t)/g\\  
=\frac{4}{g}\mathcal{P}\int_{0}^{\infty}\frac{dk}{k} e^{-\alpha_0 k}\left[1-\cos\left(v_{g}k\,t\right)\right] n_{B}(v_{g}k) \\ 
=\frac{2}{g} \log\left[\frac{[\Gamma(1+\alpha_0 T/v_{g})]^{2}}{\Gamma(1+\alpha_0 T/v_{g}-iT t)\Gamma(1+\alpha_{0} T/ v_{g}+iT t)}\right]
\end{split}
\end{align}
where $\Gamma$ is the Gamma function. However, one can convince numerically that much more simple and suitable for analytics expression
\begin{align}\label{eq:I_Gleb}
 I(t)=2 \log\left[\frac{\sinh^2\left(\pi T t\right)+\left(\pi T \alpha_0/v_g\right)^2\cosh^2\left(\pi T t\right)}{\left(\pi T \alpha_0/ v_g \right)^2}\right]\,.
\end{align}
- perfectly (i.e. with great numerical accuracy) approximates the exact result of Eq.(A1) at all values of argument $ t $ in the whole range of parameters. One can obtain the latter expression (A2) by means of complex contour integration neglecting the infinitesimally small contributions from certain parts of complex contour. From Eq.(A2) it is evident that basic correlator $ \left\langle\left[\varphi_{\pm}(t)-\varphi_{\pm}(0)\right]^2\right\rangle $ is properly regularized at $ t=0 $ (i.e. at equal times $ \tau=\tau' $ ) as it should be by the definition. Formulas (A1) and (A2) tell that concrete form of such regularization of $ I(s) $ at $ s=0 $ is a matter of taste to some extent, since the  behavior of an electron in the leads on corresponding short-time scale  $ \vert\tau-\tau'\vert =\vert t \vert \lesssim \Lambda_{g}^{-1} $ is beyond the control in the "long wavelenght" approach of Luttinger liquid model\cite{12}. Thus, for our purposes, it is enough to take only a "large" $ t $ asymptotic of basic correlator (A2) (or (A1)). As the result, one obtains

\begin{align}\label{eq:Corr_1}
\begin{split}
 \left\langle \Delta\varphi_+  \left(t \right)^2 \right\rangle= \left\langle\left[\varphi_{+}(t)-\varphi_{+}(0)\right]^2\right\rangle \\
 =\frac{ I(t)}{g}  \approx \frac{2}{g} \log\left(\frac{\sinh^2\left(\pi T t\right)}{\left(\pi T/\Lambda_{g}\right)^2}\right)\,.
\end{split}
\end{align}

at $ \vert t \vert \gg \Lambda_{g}^{-1} = \frac{\pi \alpha_{0}}{v_{g}}$ ($ \Lambda_{g}^{-1} \ll 1 $). The latter simple expression for the time-correlator for bosonic fields has been used frequently in the framework of Luttinger liquid calculations (see e.g. Giamarchi's book of Ref.[43]). However, writing  similar real-time basic pair correlator $ f_{g}(\tau-\tau')=\langle e^{i\varphi_-(\tau)}e^{-i\varphi_-(\tau')}\rangle =e^{\frac{-\left\langle\left[\varphi_-(\tau)-\varphi_-(\tau')\right]^2\right\rangle}{2}}e^{\frac{\left[\varphi_-(\tau),\varphi_-(\tau')\right]}{2}} =u_{g}(\tau - \tau')e^{\frac{\left[\varphi_-(\tau),\varphi_-(\tau')\right]}{2}}$ but under the sequence of time-integrations in the expansion of r.h.s. of Eq.(22), one should keep a proper regularization for equal times  $ \tau=\tau' $  even on the level of long-time asymptote of $ u_{g}(\tau - \tau') $ ($ \vert \tau - \tau' \vert > \delta_{g}$) since such a regularization provides correct analytical continuation in calculation of basic time integrals (see Appendix C below). Thus, one needs to rewrite a basic Luttinger liquid real time pair correlator $ f_{g}(\tau-\tau') $ in the form

\begin{align}\label{eq:Corr_2}
\begin{split}
f_{g}(\tau-\tau')=\langle e^{i\varphi_-(\tau)}e^{-i\varphi_-(\tau')}\rangle \\ =e^{\frac{-\left\langle\left[\varphi_-(\tau)-\varphi_-(\tau')\right]^2\right\rangle}{2}}e^{\frac{\left[\varphi_-(\tau),\varphi_-(\tau')\right]}{2}} \\
  =\lim_{\delta \rightarrow 0}\left[ \frac{-i\pi T/\Lambda_{g}}{\sinh\left[\pi T (\tau - \tau' -i \delta_{g} \sgn{\left[\tau-\tau'\right])}\right]}\right]^{2/g}\,
\end{split}
\end{align}

where positive $ \delta \simeq \Lambda_{g}^{-1}\geq 0$ - is supposed to be infinitesimally small ($ \vert\tau-\tau'\vert > \delta_{g} $). Formula (A4) for Luttinger liquid real time pair correlator is well-known in the literature (see e.g. the appendix of the Giamarchi's book of Ref.[43]) and is used here for derivation of basic time-dependent integrals in Appendix C. 

\section{The proof of SE-theorem and SE-lemma}\label{sec:Reexp_tunn}

Let us generalize S-Theorem and S-lemma from Ref.[38] on the case of time-dependent quantum potential $ A_{\xi}(t) $ coupled to a counting field $ \xi(\tau) $ which is dependent on the branch of corresponding complex Keldysh contour $ \mathcal{C}_K \in (0 \pm i0;t) $ 

$ \lozenge $ \textit{SE-Theorem}: \textit{"The case of exact re-exponentiation of average from Keldysh contour-ordered T-exponent with counting field-dependent non-linear bosonic quantum potential"}

$\blacktriangle $ \textit{For any exponential bosonic operator of the form:}
\begin{align}
 A_{\xi(\tau)}(\tau)=\cos{\left[\varphi_{-}(\tau)+f_{\xi(\tau)}(\tau)\right]}.
\end{align}
\textit{where}
\begin{align}
\begin{split}
f_{\xi(\tau)}(\tau)=\left\{ 
 \begin{matrix}
f_{+\xi}(\tau), \textit{Im}\lbrace \tau \rbrace > 0 \\
f_{-\xi}(\tau), \textit{Im}\lbrace \tau \rbrace < 0
        \end{matrix} \right.
 \end{split}
\end{align}
\textit{with property}
 \begin{align}
\begin{split} 
\vert f_{i}(\tau_{1})-f_{j}(\tau_{2})\vert=\vert f_{i}(\tau_{2})-f_{j}(\tau_{1})\vert
\end{split}
 \end{align} 
\textit{ $ i,j=\pm\xi $ - fulfilled for $f_{\pm\xi}(\tau)$, which are certain functions of time defined as $ f_{+\xi}(\tau) $ ($ f_{-\xi}(\tau) $) on the upper (lower) branch of Keldysh contour in the complex plane and $ \varphi_{-}(\tau) $ is time-dependent bosonic field with zero mean $ \langle \varphi_{-}(\tau)\rangle = 0 $ , which fulfils commutation relation of the form:}
\begin{align}
\begin{split}
\left[\varphi_{-}(\tau_{n}),\varphi_{-}(\tau_{n'})\right]=-2i\vartheta_{g}\sgn{\left[\tau_{n}-\tau_{n'}\right]}
\end{split}
 \end{align}
 \textit{where $ \vartheta_{g}=const $ (notice, that in our particular case: $ \vartheta_{g}=\pi/g $.)}

$ \blacktriangledown $ \textit{it follows for time-dependent average from Keldysh contour-ordered  T-exponent with Keldysh contour-dependent counting field $ \xi(\tau) $ and non-linear bosonic quantum potential}
\begin{align}
\begin{split}
\tilde{\chi}(t,\xi)=\left\langle \mathcal{T}_K \exp(-i\tilde{\lambda}\int_{\mathcal{C}_K(t)}A_{\xi(\tau)}(\tau){\rm d}\tau)\right\rangle \\
=\exp \left\lbrace - \mathcal{F}(\xi,t) \right\rbrace
\end{split}
\end{align} 
\textit{where for the functional $\mathcal{F}(\xi,t)$ one has}
\begin{align}
\begin{split}
\mathcal{F}(\xi,t)=\left(\tilde{\lambda}\right)^{2}\times \\
\left\lbrace\frac{1}{2} \int\int_{\mathcal{C}_K(t)} d \tau_{1}d \tau_{2}\langle \mathcal{T}_K A_{\xi(\tau)}(\tau_{1}) A_{\xi(\tau)}(\tau_{2})\rangle\right\rbrace 
\end{split}
\end{align}
\textit{or, alternatively} 
\begin{align}
\begin{split}
\mathcal{F}(\xi,t)=\left(\tilde{\lambda}\right)^{2}\frac{1}{2}\int^{t}_{0}d \tau_{1} \int^{t}_{0} d \tau_{2}u(\tau_{1}-\tau_{2})\times \\ \left\lbrace {e^{i \vartheta_{g}}\Delta\kappa_{+}(\tau_{1}-\tau_{2})+e^{-i \vartheta_{g}}\Delta\kappa_{-}(\tau_{1}-\tau_{2})}\right\rbrace
\end{split}
\end{align}
\textit{with}
\begin{align}
 \begin{split} 
\Delta\kappa_{\pm}(\tau_{1}-\tau_{2})=[\tilde{\kappa}_{\pm}(\tau_{1}-\tau_{2})-\kappa_{\pm}(\tau_{1}-\tau_{2})]
\end{split}
 \end{align} 
\begin{align}
 \begin{split} 
\tilde{\kappa}_{\pm}(\tau_{1}-\tau_{2})=\cos{[f_{\pm\xi}(\tau_{1})-f_{\mp\xi}(\tau_{2})]}\\
\kappa_{\pm}(\tau_{1}-\tau_{2})=\cos{[f_{\pm\xi}(\tau_{1})-f_{\pm\xi}(\tau_{2})]}
\end{split}
 \end{align}
\textit{with function $ u(\tau_{1}-\tau_{2}) $ being a pair correlator: }
\begin{align}
 \begin{split} 
u_{g}(\tau_{1}-\tau_{2})=e^{\frac{-\left\langle \left[\varphi_{-}(\tau_{1})-\varphi_{-}(\tau_{2}) \right]^2\right\rangle}{2}}
\end{split}
 \end{align}
 \textit{properly regularized at the point $\tau_{1}=\tau_{2} $.}
 $ \blacksquare $

\medskip

$ \triangledown $ \textit{The proof of SE-Theorem.}

In what follows we will modify the proof of S-theorem for Keldysh propagator $ \tilde{\chi}(t) $ with respect to a new quantum potential $ A_{\pm\xi}(\tau) $ of Eq.(23) and(B1,B2). First, as well as for usual S-theorem and S-lemma of Ref.[38], let us consider only the pair-correlator as the building block for the rest of material. Especially, in our case, according to Eqs.(A4,B1,B2) we will have
\begin{align}
\begin{split}
\langle  A_{\xi(\tau)}(\tau_{n}) A_{\xi(\tau')}(\tau_{n'})\rangle = \langle  A_{\xi(\tau)}(\tau_{n}) A_{\xi(\tau')}(\tau_{n'})\rangle_{R} \\ 
\times \left\{ 
 \begin{matrix}
    e^{-i\vartheta_{g}} &, \tau_{n} > \tau_{n'} \\
    e^{i\vartheta_{g}}  &,  \tau_{n} < \tau_{n'} \\
    1 &, \tau_{n} = \tau_{n'}
    \end{matrix} \right.
\end{split}    
\end{align}
where the "real" (but not necessary symmetrized) pair-correlator: $ \langle  A_{\xi(\tau)}(\tau_{n}) A_{\xi(\tau')}(\tau_{n'})\rangle_{R}$ reads
\begin{align}
\begin{split}
\langle A_{\pm\xi}(\tau_{l}) A_{\pm\xi}(\tau_{l+1})\rangle_{R}=u(\tau_{l}-\tau_{l+1})\kappa_{\pm}(\tau_{l}-\tau_{l+1}) 
\end{split}
 \end{align}
with
\begin{align}
\begin{split} 
\kappa_{\pm}(\tau_{l}-\tau_{l+1})=\cos{[f_{\pm\xi}(\tau_{l})-f_{\pm\xi}(\tau_{l+1})]}.  
\end{split}
 \end{align}
Also, here one should consider a "mixed" average
\begin{align}
\begin{split}
\langle A_{\pm\xi}(\tau_{l}) A_{\mp\xi}(\tau_{l+1})\rangle_{R}=u(\tau_{l}-\tau_{l+1})\tilde{\kappa}_{\pm}(\tau_{l}-\tau_{l+1}) 
\end{split}
 \end{align}
where
\begin{align}
\begin{split} 
\tilde{\kappa}_{\pm}(\tau_{l}-\tau_{l+1})=\cos{[f_{\pm\xi}(\tau_{l})-f_{\mp\xi}(\tau_{l+1})]}
\end{split}
 \end{align}    
and $ u(\tau_{l}-\tau_{l+1}) $ is the correlator of Eq.(B10) being properly regularized in its denominator at $ \tau_{l}=\tau_{l+1} $ according to Eq.(A4). Functions $ \kappa_{\pm}(\tau_{l}-\tau_{l+1}) $ and $ \tilde{\kappa}_{\pm}(\tau_{l}-\tau_{l+1})  $ -correspond to different "cos" terms generated by $ f(\tau) $ dependence of quantum potential (B1). These factors take care about two possible arrangements of a pair of time arguments $ \tau_{l} $ and $ \tau_{l+1} $ on a complex Keldysh contour: $ \mathcal{C}_K(t) \in (0\pm i0;t)$. Here $ \kappa_{\pm}(\tau_{l}-\tau_{l+1}) $ -correspond to the case where $ \tau_{l} $ and $ \tau_{l+1} $ are on the same (upper or lower) branch of the Keldysh contour, while $ \tilde{\kappa}_{\pm}(\tau_{l}-\tau_{l+1}) $ -to the case when those  time arguments are placed on different branches of the contour.

Using Eqs.(A4,B4), analogously to the case of Ref.[38], general infinte power series of Keldysh partition function of Eq.(26) is straightforward 
\begin{align}\label{eq:comb}
\begin{split}
\tilde{\chi}(t,\xi)=1+\sum_{n=2}^{even}(-i\tilde{\lambda})^{n}  
\sum_{j,k=0}^{n}C^{n}_{k} 
 e^{i \vartheta_{g} k}e^{-i \vartheta_{g}(n/2-k)}\\
 \times \int^{t}_{0} d \tau_{1}\int^{\tau_{1}}_{0} d \tau_{2}\ldots \int^{\tau_{k-2}}_{0} d \tau_{k-1}\int^{\tau_{k-1}}_{0} d \tau_{k} \\
 \times \int^{t}_{0} d \tau'_{1}\int^{\tau'_{1}}_{0} d \tau'_{2}\ldots \int^{\tau'_{(n-k)-2}}_{0} d \tau'_{n-k-1} \int^{\tau'_{(n-k)-1}}_{0} d \tau'_{n-k}\\
\times \langle A_{+\xi}(\tau_{k})\ldots A_{+\xi}(\tau_{1})A_{-\xi}(\tau'_{n-k})\ldots A_{-\xi}(\tau'_{1})\rangle_{S} . 
\end{split}
\nonumber
\end{align}
\begin{equation}
\end{equation}
Here $C^{n}_{k}=\frac{n!}{k!(n-k)!}$ are standard binomial coefficients and $\left\langle A_{+\xi}(\tau_{m})\ldots A_{-\xi}(\tau'_{n})\right\rangle_{S}$ one should understand as the "symmetric" part of the correlator $\left\langle A_{+\xi}(\tau_{m})\ldots A_{-\xi}(\tau'_{n})\right\rangle$ with respect to exchange $\tau_{m} \leftrightarrow \tau'_{n} $ (or $\tau_{m} \leftrightarrow \tau_{n} $, or $\tau'_{m} \leftrightarrow \tau_{n} $) for any pair of time variables $\tau_{m}$ and $\tau'_{n}$. In Eq.(B16) a following modification of quantum potential $ A_{0}(\tau) $ from Ref.[38] is performed
\begin{align}
\begin{split}
A_{\pm\xi}(\tau)=\cos{\left[\varphi_{-}(\tau)+f_{\pm\xi}(\tau)\right]}.
\end{split}
 \end{align}  

Importantly, to apply the FCS-modification of S-Theorem from Ref.[38] to present case, where functions $f_{\pm\xi}(\tau) $ are defined by Eq.(B2) one needs to modify these functions somehow to make them fulfill the constraint (B3). 

Such a modification is straightforward: first, applying the same \textit{"plugging procedure"} as in the proof of S-lemma in Ref.[38] for all \textit{"non-time-ordered"} averages in the r.h.s. of Eq.(B16) one can reduce all $ n $ different time integrations to one sequence of $ n $ \textit{time-ordered} time integrations in each $ n $-th order in $\tilde{\lambda}_{1(2)}$ (see this also below in this section). One might convince from the proof of S-Lemma in the Appendix B of Ref.[38] (one can see this also from calculations below in this section) that such rearrangement can be made \textit{before} the proof of "crossing" diagrams cancellation without any loss of generality of S-lemma. Second, since all time variables of integration $ \tau_{l},\tau_{l+1}$ ($ l=1..n $) become time-ordered after such the procedure , the "signum" function $\sgn{[\tau_{l}-\tau_{l'}]}  $  for any $ \tau_{l} $ and $ \tau_{l'} $ will be always equal to $ 1 $ if $ \tau_{l}>\tau_{l'} $ ($ l>l' $). Therefore, nothing changes in time-integrations of Eq.(B16) if one replace the difference: $ [eV(\tau_{l}-\tau_{l+1})+2\xi] $ in Eq.(B17) by $ [eV(\tau_{l}-\tau_{l+1})+2\xi\sgn{[\tau_{l}-\tau_{l+1}] } $ in the arguments of the "cos" function (B17) in the r.h.s. of Eq.(B16).  With respect to this replacement one can rewrite Eqs.(B13,B15) as follows:
\begin{align}
\begin{split} 
\kappa_{\pm}(\tau_{l}-\tau_{l+1})=\cos{[eV(\tau_{l}-\tau_{l+1})]}.  
\end{split}
 \end{align}
 and
\begin{align}
\begin{split} 
\tilde{\kappa}_{\pm}(\tau_{l}-\tau_{l+1})=\cos{\left[ eV(\tau_{l}-\tau_{l+1})\pm 2\xi\sgn{(\tau_{l}-\tau_{l+1})}\right] }.
\end{split}
 \end{align}
Obviously, functions (B18,B19) fulfill the condition (B3) of the validity of FCS-versions of S-Theorem and S-lemma from Ref.[38]. After such redefinition of quantum potentials in expansion (B16), one can straightforwardly apply the proof of more simple S-Theorem and S-lemma of Ref.[38] to more general FCS (full counting statistics) case of interest. 
 
Now, evidently, for each sequence of $ k $ (and $ n-k $) connected integrations in each term of $ k $-th (and $ (n-k) $-th) order in the infinite sum of r.h.s. of Eq.(B16) one can repeat the same line of logic steps as in the proof of S-Lemma from the Appendix B of Ref.[38] just replacing everywhere $ f(\tau) $ by $ f_{+\xi}(\tau) $ in the sequence of $ k $- and by $ f_{-\xi}(\tau) $ in the sequence of $n-k $ integrations, correspondingly. Below it will be proven for any even natural $ l $ that

\begin{align}\label{eq:series}
\begin{split}
\int^{t}_{0} d \tau_{1}\int^{\tau_{1}}_{0} d \tau_{2}\ldots \int^{\tau_{l-2}}_{0} d \tau_{l-1}\int^{\tau_{l-1}}_{0} d \tau_{l} \\
\times \langle A_{\pm\xi}(\tau_{1})A_{\pm\xi}(\tau_{2})\ldots A_{\pm\xi}(\tau_{l-1})A_{\pm\xi}(\tau_{l})\rangle_{S} \\
=\int^{t}_{0} d \tau_{1}\int^{\tau_{1}}_{0} d \tau_{2}\ldots \int^{\tau_{l-2}}_{0} d \tau_{l-1}\int^{\tau_{l-1}}_{0} d \tau_{l} \\
\times\langle A_{\pm\xi}(\tau_{1}) A_{\pm\xi}(\tau_{2})\rangle_{S}\ldots \langle A_{\pm\xi}(\tau_{j-1}) A_{\mp\xi}(\tau_{j})\rangle_{S}\\
\ldots \langle A_{\pm\xi}(\tau_{l-1}) A_{\pm\xi}(\tau_{l})\rangle_{S} .
 \end{split}
\end{align}
(here $ j<l $ and all binary products of the type $ A_{\pm\xi}(\tau)A_{\pm\xi}(\tau') $ stand for any among four possible distinct combinations $ A_{+\xi}(\tau)A_{+\xi}(\tau') $, $ A_{+\xi}(\tau)A_{-\xi}(\tau') $, $ A_{-\xi}(\tau)A_{+\xi}(\tau') $ and $ A_{-\xi}(\tau)A_{-\xi}(\tau') $ ) where any binary average is a following pair correlator
\begin{align}
\begin{split}
\langle A_{\pm\xi}(\tau_{l}) A_{\pm\xi}(\tau_{l+1})\rangle_{S}=u(\tau_{l}-\tau_{l+1})\kappa_{\pm}(\tau_{l}-\tau_{l+1}) 
\end{split}
 \end{align}
with
\begin{align}
\begin{split} 
\kappa_{\pm}(\tau_{l}-\tau_{l+1})=\cos{[f_{\pm\xi}(\tau_{l})-f_{\pm\xi}(\tau_{l+1})]}.  
\end{split}
 \end{align}
Also, in the framework of Eqs.(B16,B20) one should consider a "mixed" average
\begin{align}
\begin{split}
\langle A_{\pm\xi}(\tau_{l}) A_{\mp\xi}(\tau_{l+1})\rangle_{S}=u(\tau_{l}-\tau_{l+1})\tilde{\kappa}_{\pm}(\tau_{l}-\tau_{l+1}) 
\end{split}
 \end{align}
where
\begin{align}
\begin{split} 
\tilde{\kappa}_{\pm}(\tau_{l}-\tau_{l+1})=\cos{[f_{\pm\xi}(\tau_{l})-f_{\mp\xi}(\tau_{l+1})]}.
\end{split}
 \end{align}    
As it was already mentioned in the above, in Eq.(B22,B24) functions $ \kappa_{\pm}(\tau_{l}-\tau_{l+1}) $ and $ \tilde{\kappa}_{\pm}(\tau_{l}-\tau_{l+1})  $ take care about two possible types of arrangement for a pair of time arguments $ \tau_{l} $ and $ \tau_{l+1} $ on a complex Keldysh contour  $ \mathcal{C}_K\in (0\pm i0;t)$ ($ \kappa_{\pm}(\tau_{l}-\tau_{l+1}) $ -correspond to the case where $ \tau_{l} $ and $ \tau_{l+1} $ are on the same (upper or lower) branch of the Keldysh contour and $ \tilde{\kappa}_{\pm}(\tau_{l}-\tau_{l+1}) $ -to the case where those time arguments are placed on different branches of the contour). Thus, equality (B20) represents a generalized Wick theorem for quantum potential with branch-dependent counting field $ \xi(\tau) $. Eq.(B20) also means that \textit{all} "crossing" diagrams from the expansion (B16) \textit{do not affect} the result of corresponding time integration and the "linked cluster approximation" (in its modified form of Eq.(B20))remains \textit{ exact} also for quantum potential $ A_{\pm\xi}(\tau) $ with contour branch-dependent counting field $ \xi(\tau) $ in the framework of full counting statistics. 

Since the structure of the upcoming proof of Eq.(B20) (which is also the proof of S-lemma) will touch only the products of the type $ A_{\pm\xi}(\tau_{l}) A_{\pm\xi}(\tau_{l+1}) $ (whereas the contribution of "mixed" products $ A_{\pm\xi}(\tau_{l}) A_{\mp\xi}(\tau_{l+1}) $ will be clarified after the proof of Eq.(B20)) nothing will depend on indices $ \pm \xi $. Hence, in order to shorten notations we can safely re-define
\begin{align}
\begin{split} 
A_{\pm\xi}(\tau_{l}) \Leftrightarrow A_{0}(\tau_{l})
\end{split}
 \end{align}
and
\begin{align}
\begin{split} 
f_{\pm\xi}(\tau_{l}) \Leftrightarrow f(\tau_{l})
\end{split}
 \end{align}
in order to return to original notations after the proof of Eq.(B20) (or S-lemma).

$ \triangledown $ \textit{The proof of S-Lemma.} Obviously, one can write the most general result of the averaging $\left\langle A_{0}(\tau_{k})\ldots A_{0}(\tau_{1})\right\rangle_{S}$ using only two properties: i)the symmetry of the correlator: $\left\langle A_{0}(\tau_{k})\ldots A_{0}(\tau_{1})\right\rangle_{S}$ with respect to exchange $\tau_{m} \leftrightarrow \tau_{n} $ for any $ n $ and $ m $; ii) the fact that only "neutral" combinations  $ (\varphi_{-}(\tau_{k})- \varphi_{-}(\tau_{k-1})+\ldots+\varphi_{-}(\tau_{k})- \varphi_{-}(\tau_{1}))$  of bosonic field $\varphi_{-}(\tau_{k})$ with the same field at another moments of time $\varphi_{-}(\tau_{k-1})\ldots\varphi_{-}(\tau_{1})$  do survive in the exponent\cite{17,38} when one calculates the average $\left\langle A_{0}(\tau_{k})\ldots A_{0}(\tau_{1})\right\rangle_{S}$. Thus, applying the Baker-Hausdorff formula (equality between first and second lines of Eq.(A4))to each symmetrized average of the type $\left\langle A_{0}(\tau_{k})\ldots A_{0}(\tau_{1})\right\rangle_{S}$ one can obtain following structure (notice, that $ k $ is even natural number everywhere )
\begin{align}\label{eq:series}
\begin{split}
\langle A_{0}(\tau_{k})A_{0}(\tau_{k-1})\ldots A_{0}(\tau_{2}) A_{0}(\tau_{1})\rangle_{S}= \\
e^{-\frac{1}{2}\left\langle \left[\pm\varphi_{-}(\tau_{k})\mp \varphi_{-}(\tau_{k-1})\pm\ldots \pm \varphi_{-}(\tau_{2})\mp\varphi_{-}(\tau_{1}) \right]^2\right\rangle}\\
\times \prod_{l,m=1}^{k} \cos{[f(\tau_{l})-f(\tau_{l+1})]}\mid_{(l\neq m)} 
 \end{split}
\end{align}
or, performing explicitly the squaring of the expression $[\pm\varphi_{-}(\tau_{k})\mp \varphi_{-}(\tau_{k-1})\pm\ldots \pm \varphi_{-}(\tau_{2})\mp\varphi_{-}(\tau_{1})]$ in the average $ \left\langle \left[\pm\varphi_{-}(\tau_{k})\mp \varphi_{-}(\tau_{k-1})\pm\ldots \pm \varphi_{-}(\tau_{2})\mp\varphi_{-}(\tau_{1}) \right]^2 \right\rangle $ in the exponent of Eq.(B27) one can obtain following structure
\begin{align}\label{eq:series}
\begin{split}
\langle A_{0}(\tau_{k})A_{0}(\tau_{k-1})\ldots A_{0}(\tau_{2}) A_{0}(\tau_{1})\rangle_{S}= \\
u(\tau_{k},\tau_{k-1},\ldots,\tau_{2},\tau_{1})C(\vert\tau_{k}-\tau_{k-1}\vert,\ldots,\vert\tau_{2}-\tau_{1}\vert). 
 \end{split}
\end{align}
Obviously, the expansion (B28) corresponds to the procedure where propagator $ \langle A_{0}(\tau_{k})\ldots A_{0}(\tau_{1})\rangle_{S} $  can be represented as the product of its "vertex" part ($ u(\tau_{k},\tau_{k-1},\ldots,\tau_{2},\tau_{1}) $ term) of $ k $-th order which describes all "crossing" diagrams (or propagators) while the rest is a product of $ k $ "free" propagators ($ C(\vert\tau_{k}-\tau_{k-1}\vert,\ldots,\vert\tau_{2}-\tau_{1}\vert)$ term) corresponding to a "linked cluster" expansion of $ k $ "non-crossing" diagrams. Further, one can check that mentioned "vertex" part $ u(\tau_{k},\tau_{k-1},\ldots,\tau_{2},\tau_{1}) $ from Eqs.(B27,B28) can always be arranged with respect to $ k $ time arguments in the following way
\begin{align}\label{eq:series}
\begin{split}
u(\tau_{k},\tau_{k-1},\ldots,\tau_{2},\tau_{1})=\prod_{l'=1}^{k-2} v(\tau_{l'},\tau_{l'+1};k)\\
=\prod_{l'=1}^{k-2}\left\lbrace \prod_{m'=1}^{l'}\frac{u(\tau_{l'}-\tau_{m'})}{u(\tau_{l'+1}-\tau_{m'+1})}\right\rbrace . 
 \end{split}
\end{align}
Here we define following function $ v(\tau_{l'},\tau_{i};k) $ 
\begin{align}\label{eq:series}
\begin{split}
v(\tau_{l'},\tau_{l'+1};k)=\left\lbrace \prod_{m'=1}^{l'}\frac{u(\tau_{l'}-\tau_{m'})}{u(\tau_{l'+1}-\tau_{m'+1})}\right\rbrace 
 \end{split}
\end{align}
with an evident property
\begin{align}\label{eq:series}
\begin{split}
v(\tau_{l'},\tau_{l'+1};k)=\frac{1}{v(\tau_{l'+1},\tau_{l'};k)}. 
 \end{split}
\end{align}
 In turn, the "linked cluster" (or "non-crossing diagrams") factor $C(\vert\tau_{k}-\tau_{k-1}\vert,\ldots,\vert\tau_{2}-\tau_{1}\vert)$ reads
\begin{align}\label{eq:series}
\begin{split}
C(\vert\tau_{k}-\tau_{k-1}\vert,\ldots,\vert\tau_{2}-\tau_{1}\vert)\\
=\prod_{l,m=1}^{k}u(\tau_{l}-\tau_{m})\cos{[f(\tau_{l})-f(\tau_{l+1})]}\mid_{(l\neq m)}.
 \end{split}
\end{align}
In Eqs.(B29,B30) we have
\begin{align}\label{eq:series}
\begin{split}
u(\tau)=u(-\tau)=e^{-I(\tau)/g}
 \end{split}
\end{align}    
where $ I(\tau)=I(-\tau) $ is the symmetric "free" pair-correlator from Eq.(A1). 
Substitution of Eq.(B29) into Eq.(B28) gives us
\begin{align}\label{eq:series}
\begin{split}
\langle A_{0}(\tau_{k})A_{0}(\tau_{k-1})\ldots A_{0}(\tau_{2}) A_{0}(\tau_{1})\rangle_{S}= \\
= \left\lbrace \prod_{l'=1}^{k-2}v(\tau_{l'},\tau_{l'+1};k)\right\rbrace \times C(\vert\tau_{k}-\tau_{k-1}\vert,\ldots,\vert\tau_{2}-\tau_{1}\vert)
 \end{split}
\end{align}
Expression (B34) is, in fact, the result of Wick theorem application in order to extract absolute value from the average of the product  $ \langle A_{0}(\tau_{n})\ldots A_{0}(\tau_{1})\rangle_{S} $ of any $n $ operators $ A_{0}(\tau_{j}) $ being exponential in bosonic fields $ \varphi_{-}(\tau_{j}) $.
This way, using Eqs.(B28,B29,B34), one can write following identity
\begin{align}\label{eq:series}
\begin{split}
\int^{t}_{0} d \tau_{1}\int^{\tau_{1}}_{0} d \tau_{2}\ldots \int^{\tau_{n-2}}_{0} d \tau_{n-1}\int^{\tau_{n-1}}_{0} d \tau_{n} \\
\times \langle A_{0}(\tau_{1})A_{0}(\tau_{2})\ldots A_{0}(\tau_{n-1})A_{0}(\tau_{n})\rangle_{S} \\
=\int^{t}_{0} d \tau_{1}\int^{\tau_{1}}_{0} d \tau_{2}\ldots \int^{\tau_{n-2}}_{0} d \tau_{n-1}\int^{\tau_{n-1}}_{0} d \tau_{n} \\
\times u(\tau_{1},\tau_{2},\ldots,\tau_{n-1},\tau_{n})C(\vert\tau_{1}-\tau_{2}\vert,\ldots,\vert\tau_{n-1}-\tau_{n}\vert).
 \end{split}
\end{align} 
Now, on one hand, if in the l.h.s. of Eq.(B35) one exchanges two indices simultaneously in all pairs of "neighbouring" time arguments $ \tau_{r-1},\tau_{r} $ (notice only one possible choice of such pairs in the product of given structure) as well as in the upper limits of corresponding time integrations, nothing will change since both sides of Eq.(B35) do not depend on any time arguments except $ t $ and since the average $ \langle A_{0}(\tau_{1})\ldots A_{0}(\tau_{n})\rangle_{S} $ is invariant under such exchange procedure by its definition (see above). This allows us to write
\begin{align}\label{eq:series}
\begin{split}
\int^{t}_{0} d \tau_{1}\int^{\tau_{1}}_{0} d \tau_{2}\ldots \int^{\tau_{n-2}}_{0} d \tau_{n-1}\int^{\tau_{n-1}}_{0} d \tau_{n} \\
\times \langle A_{0}(\tau_{1})A_{0}(\tau_{2})\ldots A_{0}(\tau_{n-1})A_{0}(\tau_{n})\rangle_{S} \\
=\int^{t}_{0} d \tau_{2}\int^{\tau_{2}}_{0} d \tau_{1}\ldots \int^{\tau_{n-3}}_{0} d \tau_{n}\int^{\tau_{n}}_{0} d \tau_{n-1} \\
\times \langle A_{0}(\tau_{2})A_{0}(\tau_{1})\ldots A_{0}(\tau_{n})A_{0}(\tau_{n-1})\rangle_{S} \\
=\int^{t}_{0} d \tau_{2}\int^{\tau_{2}}_{0} d \tau_{1}\ldots \int^{\tau_{n-3}}_{0} d \tau_{n}\int^{\tau_{n}}_{0} d \tau_{n-1} \\
\times \langle A_{0}(\tau_{1})A_{0}(\tau_{2})\ldots A_{0}(\tau_{n-1})A_{0}(\tau_{n})\rangle_{S}  .
 \end{split}
\end{align}
or 
\begin{align}\label{eq:series}
\begin{split}
\int^{t}_{0} d \tau_{2}\int^{\tau_{2}}_{0} d \tau_{1}\ldots \int^{\tau_{n-3}}_{0} d \tau_{n}\int^{\tau_{n}}_{0} d \tau_{n-1} \times \\
\langle A_{0}(\tau_{2})A_{0}(\tau_{1})\ldots A_{0}(\tau_{n})A_{0}(\tau_{n-1})\rangle_{S} \\
=\int^{t}_{0} d \tau_{2}\int^{\tau_{2}}_{0} d \tau_{1}\ldots \int^{\tau_{n-3}}_{0} d \tau_{n}\int^{\tau_{n}}_{0} d \tau_{n-1} \times\\
 u(\tau_{1},\tau_{2},\ldots,\tau_{n-1},\tau_{n})C(\vert\tau_{1}-\tau_{2}\vert,\ldots,\vert\tau_{n-1}-\tau_{n}\vert) .
 \end{split}
\end{align}  
 On the other hand, one can rewrite the l.h.s. of Eq.(B37) using Eq.(B34) in the form
\begin{align}\label{eq:series}
\begin{split}
\int^{t}_{0} d \tau_{2}\int^{\tau_{2}}_{0} d \tau_{1}\ldots \int^{\tau_{n-3}}_{0} d \tau_{n}\int^{\tau_{n}}_{0} d \tau_{n-1}\\ 
\times\langle A_{0}(\tau_{2})A_{0}(\tau_{1})\ldots A_{0}(\tau_{n})A_{0}(\tau_{n-1})\rangle_{S}= \\
\int^{t}_{0} d \tau_{2}\int^{\tau_{2}}_{0} d \tau_{1}v(\tau_{2},\tau_{1};n)\ldots \\ 
\ldots \int^{\tau_{n-3}}_{0} d \tau_{n}\int^{\tau_{n}}_{0} d \tau_{n-1}v(\tau_{n},\tau_{n-1};n) \\
\times C(\vert\tau_{1}-\tau_{2}\vert,\ldots,\vert\tau_{n-1}-\tau_{n}\vert) .
 \end{split}
\end{align}  
which in turn can be rewritten using property (B31) as
\begin{align}\label{eq:series}
\begin{split}
\int^{t}_{0} d \tau_{2}\int^{\tau_{2}}_{0} d \tau_{1}v(\tau_{2},\tau_{1};n)\ldots \\ 
\ldots \int^{\tau_{n-3}}_{0} d \tau_{n}\int^{\tau_{n}}_{0} d \tau_{n-1}v(\tau_{n},\tau_{n-1};n) \\
\times C(\vert\tau_{1}-\tau_{2}\vert,\ldots,\vert\tau_{n-1}-\tau_{n}\vert) \\
=\int^{t}_{0} d \tau_{2}\int^{\tau_{2}}_{0} d \tau_{1}\frac{1}{v(\tau_{1},\tau_{2};n)}\ldots \\ 
\ldots \int^{\tau_{n-3}}_{0} d \tau_{n}\int^{\tau_{n}}_{0} d \tau_{n-1}\frac{1}{v(\tau_{n-1},\tau_{n};n)}\\
\times C(\vert\tau_{1}-\tau_{2}\vert,\ldots,\vert\tau_{n-1}-\tau_{n}\vert) .
 \end{split}
\end{align} 
Now, substituting Eq.(B39) into the r.h.s. of Eq.(B37) and using definition of Eq.(B29) one can perform Eq.(B37) in the form
\begin{align}\label{eq:series}
\begin{split}
\int^{t}_{0} d \tau_{2}\int^{\tau_{2}}_{0} d \tau_{1}\ldots \int^{\tau_{n-3}}_{0} d \tau_{n}\int^{\tau_{n}}_{0} d \tau_{n-1} \\
\times \langle A_{0}(\tau_{2})A_{0}(\tau_{1})\ldots A_{0}(\tau_{n})A_{0}(\tau_{n-1})\rangle_{S} \\
=\int^{t}_{0} d \tau_{2}\int^{\tau_{2}}_{0} d \tau_{1}\ldots \int^{\tau_{n-3}}_{0} d \tau_{n}\int^{\tau_{n}}_{0} d \tau_{n-1} \\
\times \frac{1}{u(\tau_{1},\tau_{2},\ldots,\tau_{n-1},\tau_{n})}C(\vert\tau_{1}-\tau_{2}\vert,\ldots,\vert\tau_{n-1}-\tau_{n}\vert).
 \end{split}
\end{align}   
 Finally, changing back indices in all pairs of "neighbouring" time arguments $ \tau_{r-1},\tau_{r}\rightarrow \tau_{r},\tau_{r-1}$ in Eq.(40) with respect to symmetry property of Eq.(B36)and comparing the left- and right-hand sides of Eqs.(B35,B37,B40) one can conclude that
\begin{align}\label{eq:series}
\begin{split}
\int^{t}_{0} d \tau_{1}\int^{\tau_{1}}_{0} d \tau_{2}\ldots \int^{\tau_{n-2}}_{0} d \tau_{n-1}\int^{\tau_{n-1}}_{0} d \tau_{n} \\
\times \left\lbrace \textbf{K}_{\lbrace n\rbrace}-{\textbf{K}_{\lbrace n\rbrace}}^{-1}\right\rbrace  C(\vert\tau_{1}-\tau_{2}\vert,\ldots,\vert\tau_{n-1}-\tau_{n}\vert) =0.
 \end{split}
\end{align} 
In Eq.(B41) the "kernel" function
\begin{align}\label{eq:series}
\begin{split}
\textbf{K}_{\lbrace n\rbrace}=u(\tau_{1},\tau_{2},\ldots,\tau_{n-1},\tau_{n})
 \end{split}
\end{align}
is a kind of generalized function which can act on any function $ f(\tau_{1},\ldots,\tau_{n}) $ only under $ n $-fold time-integration over $\tau_{1},.., \tau_{n} $ with the following evident property
\begin{align}\label{eq:series}
\begin{split}
\textbf{K}_{\lbrace n\rbrace}{\textbf{K}_{\lbrace n\rbrace}}^{-1}={\textbf{K}_{\lbrace n\rbrace}}^{-1}\textbf{K}_{\lbrace n\rbrace}=1.
 \end{split}
\end{align} 
Obviously, from Eq.(B41) it follows that
\begin{align}\label{eq:series}
\begin{split}
\int^{t}_{0} d \tau_{1}\int^{\tau_{1}}_{0} d \tau_{2}\ldots \int^{\tau_{n-2}}_{0} d \tau_{n-1}\int^{\tau_{n-1}}_{0} d \tau_{n} \\
\textbf{K}_{\lbrace n\rbrace} \times C(\vert\tau_{1}-\tau_{2}\vert,\ldots,\vert\tau_{n-1}-\tau_{n}\vert) \\
= \int^{t}_{0} d \tau_{1}\int^{\tau_{1}}_{0} d \tau_{2}\ldots \int^{\tau_{n-2}}_{0} d \tau_{n-1}\int^{\tau_{n-1}}_{0} d \tau_{n} \\
{\textbf{K}_{\lbrace n\rbrace}}^{-1}\times C(\vert\tau_{1}-\tau_{2}\vert,\ldots,\vert\tau_{n-1}-\tau_{n}\vert).
 \end{split}
\end{align} 
Then since by definition $ C(\vert\tau_{1}-\tau_{2}\vert,\ldots,\vert\tau_{n-1}-\tau_{n}\vert)> 0 $ (see Eqs.(B32,B33)), it follows that the only possibility for kernel generalized function of Eq.(B42) to fulfil both Eq.(B43) and Eq.(B44) simultaneously is
\begin{align}\label{eq:series}
\begin{split}
\textbf{K}_{\lbrace n\rbrace}={\textbf{K}_{\lbrace n\rbrace}}^{-1}=\textbf{1}_{\lbrace n\rbrace}=\tilde{1}(\tau_{1},\tau_{2},\ldots,\tau_{n-1},\tau_{n}).
 \end{split}
\end{align} 
Here by means of Eq.(B45) I defined the $ n $ -dimensional "generalized unit function": $ \tilde{1}(\tau_{1},\ldots,\tau_{n}) $ which is a sort of generalized function (or operator) being "unit" in the sense that the result of expression $ \textbf{1}_{\lbrace n\rbrace}C(\vert\tau_{1}-\tau_{2}\vert,\ldots,\vert\tau_{n-1}-\tau_{n}\vert) $ after all the integrations over all $ \tau_{1},\ldots,\tau_{n} $ time arguments in Eq.(B44) will be the same as if one would integrate only the function $ C(\vert\tau_{1}-\tau_{2}\vert,\ldots,\vert\tau_{n-1}-\tau_{n}\vert)$  over those time arguments   
\begin{align}\label{eq:series}
\begin{split}
\int^{t}_{0} d \tau_{1}\int^{\tau_{1}}_{0} d \tau_{2}\ldots \int^{\tau_{n-2}}_{0} d \tau_{n-1}\int^{\tau_{n-1}}_{0} d \tau_{n} \\
\times\textbf{K}_{\lbrace n\rbrace}C(\vert\tau_{1}-\tau_{2}\vert,\ldots,\vert\tau_{n-1}-\tau_{n}\vert) \\
=\int^{t}_{0} d \tau_{1}\int^{\tau_{1}}_{0} d \tau_{2}\ldots \int^{\tau_{n-2}}_{0} d \tau_{n-1}\int^{\tau_{n-1}}_{0} d \tau_{n} \\
\times C(\vert\tau_{1}-\tau_{2}\vert,\ldots,\vert\tau_{n-1}-\tau_{n}\vert).
 \end{split}
\end{align}
In turn, the latter equality Eq.(B46) means that, without the loss of generality, one can perform Eq.(B35) simply as
\begin{align}\label{eq:series}
\begin{split}
\int^{t}_{0} d \tau_{1}\int^{\tau_{1}}_{0} d \tau_{2}\ldots \int^{\tau_{n-2}}_{0} d \tau_{n-1}\int^{\tau_{n-1}}_{0} d \tau_{n} \\
\times \langle A_{0}(\tau_{1})A_{0}(\tau_{2})\ldots A_{0}(\tau_{n-1})A_{0}(\tau_{n})\rangle_{S} \\
=\int^{t}_{0} d \tau_{1}\int^{\tau_{1}}_{0} d \tau_{2}\ldots \int^{\tau_{n-2}}_{0} d \tau_{n-1}\int^{\tau_{n-1}}_{0} d \tau_{n} \\
\times C(\vert\tau_{1}-\tau_{2}\vert,\ldots,\vert\tau_{n-1}-\tau_{n}\vert).
 \end{split}
\end{align}
Finally, evident properties (see Eqs.(B32,B33))
\begin{align}\label{eq:series}
\begin{split}
 \langle A_{0}(\tau_{1})A_{0}(\tau_{2})\rangle_{S}= C(\vert\tau_{1}-\tau_{2}\vert)
 \end{split}
\end{align}
and
\begin{align}\label{eq:series}
\begin{split}
C(\vert\tau_{1}-\tau_{2}\vert,\ldots,\vert\tau_{n-1}-\tau_{n}\vert) \\
=C(\vert\tau_{1}-\tau_{2}\vert)\ldots C(\vert\tau_{n-1}-\tau_{n}\vert)
 \end{split}
\end{align}
allow for the \textit{exact} factorization of the average under the integrals in the l.h.s. of Eq.(B47) on a product of $ n/2 $ pair-correlators (recall that $ n $ is an arbitrary even number everywhere)
\begin{align}\label{eq:series}
\begin{split}
\int^{t}_{0} d \tau_{1}\int^{\tau_{1}}_{0} d \tau_{2}\ldots \int^{\tau_{n-2}}_{0} d \tau_{n-1}\int^{\tau_{n-1}}_{0} d \tau_{n} \\
\times \langle A_{0}(\tau_{1})A_{0}(\tau_{2})\ldots A_{0}(\tau_{n-1})A_{0}(\tau_{n})\rangle_{S} \\
=\int^{t}_{0} d \tau_{1}\int^{\tau_{1}}_{0} d \tau_{2}\ldots \int^{\tau_{n-2}}_{0} d \tau_{n-1}\int^{\tau_{n-1}}_{0} d \tau_{n} \\
\times\langle A_{0}(\tau_{1}) A_{0}(\tau_{2})\rangle_{S}\ldots \langle A_{0}(\tau_{n-1}) A_{0}(\tau_{n})\rangle_{S} .
 \end{split}
\end{align}

Now, it is time to consider all "non-time ordered" contributions to the Keldysh contour-ordered time-integrals from Eq.(B16,B20). In particular, taking a closer look on the averages of the kind $\langle \mathcal{T}_K A_{0}(\tau_{k})\ldots A_{0}(\tau_{1})A_{0}(\tau'_{n-k})\ldots A_{0}(\tau'_{1})\rangle_{S}$ under time integrals over $ \tau_{j} $ ($ j'=1,..,k $) and $ \tau'_{j'} $ ($ j'=1,..,(n-k) $)- where two sets $ \tau  $ and $ \tau'  $ of time arguments are disconnected from each other - in the T-exponent expansion in r.h.s. of Eq.(B16), one can notice that application of Keldysh-contour ordering procedure to such the averages with respect to Eq.(B20) derived above - leads to the appearance of all possible "mixed" pair-correlators of the kind $ \int_{0}^{t} \int_{0}^{\tau_{j+1}} d \tau_{j}d \tau'_{j'} \langle A_{0}(\tau_{j})A_{0}(\tau'_{j'})\rangle_{S} $ in corresponding factorization formulas of Eqs.(B20,B50). The latter involve operators $ A_{0}(\tau_{j}) $ and $ A_{0}(\tau'_{j'}) $ from different branches of Keldysh contour (or, alternatively, from both time- and anti-time-ordered sequences of such operators in the r.h.s. of Eq.(B16)). At first glance, the correlators of such type should "break" the sequence of time-ordered integrations in the l.h.s. of Eqs.(B35) because, for example, for the "non-time-oredered" average of the kind $\int_{0}^{t} \int_{0}^{\tau_{j+1}} d \tau_{j}d \tau'_{j'}\langle A_{0}(\tau_{j})A_{0}(\tau'_{j'})\rangle_{S} $ two corresponded integrations (over $ \tau_{j} $ and over $ \tau'_{j'} $) to appear in the r.h.s. of Eq.(B16) are disconnected. However, this obstacle can be circumvented by decomposing each contribution of the kind $ \int_{0}^{t} \int_{0}^{\tau_{j+1}} d \tau_{j}d \tau'_{j'} $ (in the expansion of the r.h.s. of Eq.(B16)) on its "time-" and "anti-time" -ordered parts (with respect to the cases $ \tau_{j}>\tau'_{j'}$ and $ \tau_{j}<\tau'_{j'}$, correspondingly) and, then, by "assigning" each (anti-)time-ordered "part" of this double integral to the (anti-)time-ordered sequence of integrations of Eq.(B35) in order to "restore" the "broken" sequence  of time-ordered integrations to the $ n $-fold integral over $ n $ time variables $ \tau_{1},\ldots \tau_{n} $. 
Obviously, as the result, all sequences of $ n $ time-ordered integrations being obtained in such a way will be the same as one in the r.h.s. of Eq.(B50). Hence, afterwards, one will need only to count all these sequences properly, extracting a correct combinatoric pre-factor in front of the sequence of $ n $ time-integrations similar to one from the r.h.s. of Eq.(B50). 
Applying this procedure, one can easily convince that
\begin{align}\label{eq:comb}
\begin{split}
\int^{t}_{0} d \tau_{1}\int^{\tau_{1}}_{0} d \tau_{2}\ldots \int^{\tau_{k-2}}_{0} d \tau_{k-1}\int^{\tau_{k-1}}_{0} d \tau_{k}\times  \\
 \int^{t}_{0} d \tau'_{1}\int^{\tau'_{1}}_{0} d \tau'_{2}\ldots \int^{\tau'_{(n-k)-2}}_{0} d \tau'_{n-k-1}\int^{\tau'_{(n-k)-1}}_{0} d \tau'_{n-k}\\
\times \langle \mathcal{T}_K A_{0}(\tau_{k})\ldots A_{0}(\tau_{1}) A_{0}(\tau'_{n-k})\ldots A_{0}(\tau'_{1})\rangle_{S} \\
=\textbf{D}_{(n)}\times \int^{t}_{0} d \tau_{1}\int^{\tau_{1}}_{0} d \tau_{2}\langle A_{0}(\tau_{1})A_{0}(\tau_{2})\rangle_{S}\ldots \\
\ldots \int^{\tau_{n-2}}_{0} d \tau_{n-1} \int^{\tau_{n-1}}_{0} d \tau_{n}\langle A_{0}(\tau_{n-1})A_{0}(\tau_{n})\rangle_{S}. 
\end{split}
\end{align}
where
\begin{align}\label{eq:rel}
\begin{split}
\textbf{D}_{(n)}=\sum_{j=0}^{k}C_{j}^{k}\sum_{j'=0}^{n-k}C_{j'}^{n-k}=2^{k}\cdot 2^{n-k}=2^{n}
\end{split}
\end{align} 
is the desired combinatoric pre-factor. This factor gives us the number of ways in which one could "compose" all "mixed" correlators from two disconnected sequences of time arguments $ \tau_{1}\ldots \tau_{k}$ and $ \tau'_{1}\ldots \tau'_{n-k}$. In Eq.(B52) I  used a usual binomial formula $\sum_{j=0}^{k}C_{j}^{k}=2^{k} $ for binomial coefficients $ C_{j}^{k}=\frac{k!}{j!(k-j)!} $ and $ C_{j'}^{k}=\frac{(n-k)!}{j'!(n-k-j')!} $ which counts the numbers of different ways one could put $ j $ and $ j' $ "plugs" into the sequences of $ k $ "time-ordered" and $ n-k $ "anti-time ordered" integrations, correspondingly, in the r.h.s. of Eq.(B50). 
 At last, taking into account that according to Eqs.(B21-B24,B28) $ \langle A_{0}(\tau_{l+1}) A_{0}(\tau_{l})\rangle_{S}=\langle A_{0}(\tau_{l}) A_{0}(\tau_{l+1})\rangle_{S} $  and also recalling the fact that everywhere in the above formulas $ n=2m $ (i.e. $ n $ is even natural number), with the help of the obvious property for $ m=n/2 $-fold double integral \cite{55} 
\begin{align}\label{eq:rel}
\begin{split}
\int^{t}_{0} d \tau_{1}\int^{\tau_{1}}_{0} d \tau_{2}\langle A_{0}(\tau_{1})A_{0}(\tau_{2})\rangle_{S}\ldots \\
\ldots \int^{\tau_{n-2}}_{0} d \tau_{n-1} \int^{\tau_{n-1}}_{0} d \tau_{n}\langle A_{0}(\tau_{n-1})A_{0}(\tau_{n})\rangle_{S}\\ 
= \frac{1}{(n/2)!} \prod_{l=1}^{n/2} \left\lbrace  \int^{t}_{0} d \tau_{l+1}\int^{\tau_{l+1}}_{0} d \tau_{l}\langle A_{0}(\tau_{l+1}) A_{0}(\tau_{l})\rangle_{S}\right\rbrace \\
= \frac{1}{(n/2)!} \prod_{l=1}^{n/2} \left\lbrace \frac{1}{2} \int^{t}_{0} d \tau_{l+1}\int^{t}_{0} d \tau_{l}\langle A_{0}(\tau_{l+1}) A_{0}(\tau_{l})\rangle_{S}\right\rbrace 
\end{split}
\end{align}
and combining Eqs.(B51-B53) one can easily obtain following \textit{exact} equality \cite{55}
\begin{align}\label{eq:comb}
\begin{split}
\int^{t}_{0} d \tau_{1}\int^{\tau_{1}}_{0} d \tau_{2}\ldots \int^{\tau_{k-2}}_{0} d \tau_{k-1}\int^{\tau_{k-1}}_{0} d \tau_{k}\times  \\
 \int^{t}_{0} d \tau'_{1}\int^{\tau'_{1}}_{0} d \tau'_{2}\ldots \int^{\tau'_{(n-k)-2}}_{0} d \tau'_{n-k-1}\int^{\tau'_{(n-k)-1}}_{0} d \tau'_{n-k}\\
\times \langle \mathcal{T}_K A_{0}(\tau_{k})\ldots A_{0}(\tau_{1}) A_{0}(\tau'_{n-k})\ldots A_{0}(\tau'_{1})\rangle_{S} \\
=\frac{2^{n}}{(n/2)!} \prod_{l=1}^{n/2} \left\lbrace \frac{1}{2} \int^{t}_{0} d \tau_{l+1}\int^{t}_{0} d \tau_{l}\langle A_{0}(\tau_{l+1}) A_{0}(\tau_{l})\rangle_{S} \right\rbrace \\
= \frac{1}{(n/2)!} \prod_{l=1}^{n/2} \left\lbrace 2 \int^{t}_{0} d \tau_{l+1}\int^{t}_{0} d \tau_{l}\langle A_{0}(\tau_{l+1}) A_{0}(\tau_{l})\rangle_{S} \right\rbrace . 
\end{split}
\end{align}
which states S-Lemma. Obviously, substituting definitions (B25-B26) into this equation (B54) we prove the validity of Eq.(B20) of interest.

Now, in order to bring the statement (B20) to the form more similar to one of conventional S-lemma of Ref.[38], first let us note that, if 
\begin{align}
\begin{split}
\vert f_{i}(\tau_{l})-f_{j}(\tau_{l+1})\vert=\vert f_{i}(\tau_{l+1})-f_{j}(\tau_{l})\vert 
\end{split}
 \end{align} 
with $ i,j=\pm\xi $- is fulfilled for the functions $f_{\pm\xi}(\tau)$  then from Eqs.(B22,B24) it follows that  
 \begin{align}
 \begin{split} 
\kappa_{\pm}(\tau_{l},\tau_{l+1})=\kappa_{\pm}(\tau_{l+1},\tau_{l}) \\
\tilde{\kappa}_{\pm}(\tau_{l},\tau_{l+1})=\tilde{\kappa}_{\pm}(\tau_{l+1},\tau_{l})
\end{split}
 \end{align}
hence, from Eqs.(B21) 
 \begin{align}
 \begin{split} 
\langle A_{\pm\xi}(\tau_{l}) A_{\pm\xi}(\tau_{l+1})\rangle_{S}=\langle A_{\pm\xi}(\tau_{l+1}) A_{\pm\xi}(\tau_{l})\rangle_{S}
\end{split}
 \end{align}
and as well from (B23)
\begin{align}
 \begin{split} 
\langle A_{\pm\xi}(\tau_{l}) A_{\mp\xi}(\tau_{l+1})\rangle_{S}=\langle A_{\mp\xi}(\tau_{l+1}) A_{\pm\xi}(\tau_{l})\rangle_{S}.
\end{split}
 \end{align} 

Obviously, in the latter case, $\langle A_{\pm\xi}(\tau_{l}) A_{\pm\xi}(\tau_{l+1})\rangle_{S} $ and $\langle A_{\pm\xi}(\tau_{l}) A_{\mp\xi}(\tau_{l+1})\rangle_{S} $ play the role of "symmetrized" pair-correlators $ \langle A_{0}(\tau_{l+1}) A_{0}(\tau_{l})\rangle_{S} $ from the proof of S-lemma in the above. Thus, having in hands Eqs.(B57,B58) one can apply to all disconnected time integrations in the r.h.s. of Eq.(B16) the same "plugging" procedure, which has been described in the proof of S-lemma in the above and in Ref.[38]. Hovewer, in the present case, such the procedure will give a slightly different result because of the presence of new ingredients: $ \kappa_{\pm}(\tau) $ and $ \tilde{\kappa}_{\pm}(\tau) $ which mark four different arrangements of two time arguments  $ \tau_{l} $ and $ \tau_{l+1} $ on two (time- and anti-time -ordered) branches of complex Keldysh contour. In particular, applying to time integrals in the r.h.s. of Eq.(B16) described procedure and making use of formulas (B20,B55-B58) in $ 2m $-fold integral from the symmetrized function, one can obtain (below everywhere $ n=2m $ is an even natural number)
\begin{align}\label{eq:comb}
\begin{split}
\int^{t}_{0} d \tau_{1}\int^{\tau_{1}}_{0} d \tau_{2}\ldots \int^{\tau_{k-2}}_{0} d \tau_{k-1}\int^{\tau_{k-1}}_{0} d \tau_{k}\times  \\
 \int^{t}_{0} d \tau'_{1}\int^{\tau'_{1}}_{0} d \tau'_{2}\ldots \int^{\tau'_{(n-k)-2}}_{0} d \tau'_{n-k-1}\int^{\tau'_{(n-k)-1}}_{0} d \tau'_{n-k}\\
\times\langle A_{+\xi}(\tau_{k})\ldots A_{+\xi}(\tau_{1}) A_{-\xi}(\tau'_{n-k})\ldots A_{-\xi}(\tau'_{1})\rangle_{S} \\
=\frac{1}{(n/2)!} \sum_{j=0}^{k} C_{j}^{k} \sum_{j'=0}^{(n/2-k)} C_{j'}^{(n/2-k)}\times \\
\left\lbrace -\frac{1}{2}\int^{t}_{0} d \tau_{1}\int^{t}_{0} d \tau_{2}u(\tau_{1}-\tau_{2})\tilde{\kappa}_{+}(\tau_{1}-\tau_{2})\right\rbrace ^{j}\times\\
 \left\lbrace \frac{1}{2}\int^{t}_{0} d \tau_{1}\int^{t}_{0} d \tau_{2}u(\tau_{1}-\tau_{2})\kappa_{+}(\tau_{1}-\tau_{2})\right\rbrace ^{(k-j)}\times\\
\left\lbrace -\frac{1}{2}\int^{t}_{0} d \tau_{1}\int^{t}_{0} d \tau_{2}u(\tau_{1}-\tau_{2})\tilde{\kappa}_{-}(\tau_{1}-\tau_{2})\right\rbrace ^{j'}\times \\
\left\lbrace \frac{1}{2}\int^{t}_{0} d \tau_{1}\int^{t}_{0} d \tau_{2}u(\tau_{1}-\tau_{2})\kappa_{-}(\tau_{1}-\tau_{2})\right\rbrace ^{(n/2-k-j')}\\
\end{split}
\end{align}
Here the "minus" sign in the first (and third) brackets in r.h.s. of the latter equation is due to inversion of the limits in one among two time integrations when one "plugs" the "empty place" in the time- (anti-time)-ordered sequence of integrations by one integral from anti-time- (and time-) ordered sequence of integrals originated from one "mixed" term of Eq.(B58). In Eq. (B59) the binomial coefficients $ C_{j}^{k}=\frac{k!}{j!(k-j)!} $ and $ C_{j'}^{(n/2-k)}=\frac{(n/2-k)!}{j'!(n/2-k-j')!} $ - count the number of ways one  can compose $ j $ and $ j' $ "mixed" pair correlators of the form (B58) from the product of $ k $ and $ (n/2-k) $ available terms, correspondingly. Now using twice a binomial formula: $ (x+y)^{m}=\sum_{k=1}^{m} C^{m}_{k}x^{k}y^{m-k} $ one can obtain from Eq.(B59) following expression
\begin{align}\label{eq:comb}
\begin{split}
\int^{t}_{0} d \tau_{1}\int^{\tau_{1}}_{0} d \tau_{2}\ldots \int^{\tau_{k-2}}_{0} d \tau_{k-1}\int^{\tau_{k-1}}_{0} d \tau_{k}\\
 \int^{t}_{0} d \tau'_{1}\int^{\tau'_{1}}_{0} d \tau'_{2}\ldots \int^{\tau'_{(n-k)-2}}_{0} d \tau'_{n-k-1}\int^{\tau'_{(n-k)-1}}_{0} d \tau'_{n-k}\\
\times\langle A_{+\xi}(\tau_{k})\ldots A_{+\xi}(\tau_{1})A_{-\xi}(\tau'_{n-k})\ldots A_{-\xi}(\tau'_{1})\rangle_{S}\\
=\frac{(-1/2)^{n/2}}{(n/2)!}\left\lbrace\int^{t}_{0} d \tau_{1} \int^{t}_{0} d \tau_{2}u(\tau_{1}-\tau_{2})\Delta\kappa_{+}(\tau_{1}-\tau_{2})\right\rbrace^{k}\\
\times \left\lbrace\int^{t}_{0} d \tau_{1} \int^{t}_{0}d \tau_{2}u(\tau_{1}-\tau_{2})\Delta\kappa_{-}(\tau_{1}-\tau_{2})\right\rbrace^{(n/2-k)}
\end{split}
\end{align}
where
\begin{align}
 \begin{split} 
\Delta\kappa_{\pm}(\tau_{1}-\tau_{2})=[\tilde{\kappa}_{\pm}(\tau_{1}-\tau_{2})-\kappa_{\pm}(\tau_{1}-\tau_{2})]
\end{split}
 \end{align} 
Evidently, equations(B60,B61) have been derived above state following 
 
 $ \lozenge $ \textit{SE-Lemma}: \textit{"The case of the exact factorization (Wick theorem) for average of a sequence of connected time-integrals from Keldysh contour-ordered product of non-linear bosonic quantum potentials coupled to Keldysh contour-dependent counting field."}

$\blacktriangle $ \textit{For any exponential bosonic operator of the form:}
\begin{align}
 A_{\xi(\tau)}(\tau)=\cos{\left[\varphi_{-}(\tau)+f_{\xi(\tau)}(\tau)\right]}.
 \nonumber
\end{align}
\textit{where}
\begin{align}
\begin{split}
f_{\xi(\tau)}(\tau)=\left\{ 
 \begin{matrix}
f_{+\xi}(\tau), \textit{Im}\lbrace \tau \rbrace > 0 \\
f_{-\xi}(\tau), \textit{Im}\lbrace \tau \rbrace < 0
        \end{matrix} \right.
 \end{split}
\nonumber 
\end{align}
\textit{where the property}
 \begin{align}
\begin{split} 
\vert f_{i}(\tau_{1})-f_{j}(\tau_{2})\vert=\vert f_{i}(\tau_{2})-f_{j}(\tau_{1})\vert
\end{split}
 \end{align} 
\textit{with $ i,j=\pm\xi $- is fulfilled for  $f_{\pm\xi}(\tau)$ being certain functions of time and defined as $ f_{+\xi}(\tau) $ ($ f_{-\xi}(\tau) $) on the upper (lower) branch of Keldysh contour in the complex plane and $ \varphi_{-}(\tau) $ is time-dependent bosonic field with zero mean $ \langle \varphi_{-}(\tau)\rangle = 0 $ , which fulfils commutation relation of the form:}
\begin{align}
\begin{split}
\left[\varphi_{-}(\tau),\varphi_{-}(\tau')\right]=-2i\vartheta_{g}\sgn{\left[\tau-\tau'\right]}
 \nonumber
\end{split}
 \end{align}
 \textit{where $ \vartheta_{g}=const $ (notice, that in our particular case: $ \vartheta_{g}=\pi/g $)}
 
$ \blacktriangledown $ \textit{it follows for the sequence of connected time integrals from the averaged Keldysh contour-ordered product of non-linear bosonic quantum potentials coupled to Keldysh contour-dependent counting field $ \xi(\tau) $}
\begin{align}
\begin{split}
\int^{t}_{0} d \tau_{1}\int^{\tau_{1}}_{0} d \tau_{2}\ldots \int^{\tau_{k-2}}_{0} d \tau_{k-1}\int^{\tau_{k-1}}_{0} d \tau_{k}\\
 \int^{t}_{0} d \tau'_{1}\int^{\tau'_{1}}_{0} d \tau'_{2}\ldots \int^{\tau'_{(n-k)-2}}_{0} d \tau'_{n-k-1}\int^{\tau'_{(n-k)-1}}_{0} d \tau'_{n-k}\\
\times\langle A_{+\xi}(\tau_{k})\ldots A_{+\xi}(\tau_{1})A_{-\xi}(\tau'_{n-k})\ldots A_{-\xi}(\tau'_{1})\rangle_{S}\\
=\frac{(-1/2)^{n/2}}{(n/2)!}\left\lbrace\int^{t}_{0} d \tau_{1} \int^{t}_{0} d \tau_{2}u(\tau_{1}-\tau_{2})\Delta\kappa_{+}(\tau_{1}-\tau_{2})\right\rbrace^{k}\\
\times \left\lbrace\int^{t}_{0} d \tau_{1} \int^{t}_{0}d \tau_{2}u(\tau_{1}-\tau_{2})\Delta\kappa_{-}(\tau_{1}-\tau_{2})\right\rbrace^{(n/2-k)}
\nonumber
\end{split}
\end{align}
\textit{where}
\begin{align}
 \begin{split} 
\Delta\kappa_{\pm}(\tau_{1}-\tau_{2})=[\tilde{\kappa}_{\pm}(\tau_{1}-\tau_{2})-\kappa_{\pm}(\tau_{1}-\tau_{2})]
\nonumber
\end{split}
 \end{align} 
\begin{align}
 \begin{split} 
\tilde{\kappa}_{\pm}(\tau_{1}-\tau_{2})=\cos{[f_{\pm\xi}(\tau_{1})-f_{\mp\xi}(\tau_{2})]}\\
\kappa_{\pm}(\tau_{1}-\tau_{2})=\cos{[f_{\pm\xi}(\tau_{1})-f_{\pm\xi}(\tau_{2})]}
\nonumber
\end{split}
 \end{align}
\textit{with function $ u(\tau_{1}-\tau_{2}) $ being a pair correlator:}
\begin{align}
 \begin{split} 
u_{g}(\tau_{1}-\tau_{2})=e^{\frac{-\left\langle \left[\varphi_{-}(\tau_{1})-\varphi_{-}(\tau_{2}) \right]^2\right\rangle}{2}}
\nonumber
\end{split}
 \end{align}
 \textit{properly regularized at the point $\tau_{1}=\tau_{2} $.}

\textit{Therefore, by means of Eqs.(B60,B61) the SE-Lemma is proven.} $ \blacksquare $

Now using the SE- Lemma and substituting the r.h.s. of Eq.(B60) into the r.h.s. of Eq.(B16) one  immediately obtains
\begin{align}
\begin{split}
\tilde{\chi}(t,\xi)=1+\sum_{n=2}^{even}(-i\tilde{\lambda})^{n}  
\sum_{j,k=0}^{n}C^{n}_{k}e^{i \vartheta_{g} k}e^{-i \vartheta_{g}(n/2-k)}\\
\times \frac{(-1/2)^{n/2}}{(n/2)!}\left\lbrace\int^{t}_{0} d \tau_{1} \int^{t}_{0} d \tau_{2}u(\tau_{1}-\tau_{2})\Delta\kappa_{+}(\tau_{1}-\tau_{2})\right\rbrace^{k}\\
\times \left\lbrace\int^{t}_{0} d \tau_{1} \int^{t}_{0}d \tau_{2}u(\tau_{1}-\tau_{2})\Delta\kappa_{-}(\tau_{1}-\tau_{2})\right\rbrace^{(n/2-k)}. 
\end{split}
\end{align}
Expansion (B63) automatically takes care about all possible combinations of pair correlators (B57) and (B58) which can appear in formula (B16) and it remains valid in \textit{all} orders of $ n $ giving rise to nonperturbative calculation of $ \tilde{\chi}(t,\xi) $. Indeed, recalling that in Eq.(B63) $ n=2m  $, ($ m=1,2,3.. $) and using a standard binomial formula: $ (x+y)^{m}=\sum_{k=1}^{m} C^{m}_{k}x^{k}y^{m-k} $ one obtains from Eq.(B63) following infinite power series for time-dependent (i.e. nonequilibrium) Keldysh partition function (KPF)
\begin{align}
\begin{split}
\tilde{\chi}(t,\xi)=1+\sum_{m=1}^{\infty}\frac{(-\tilde{\lambda}^{2})^{m}}{m!}\times \\
\left\lbrace \frac{1}{2}\int^{t}_{0}d \tau_{1} \int^{t}_{0} d \tau_{2}u(\tau_{1}-\tau_{2})F_{C}((\tau_{1}-\tau_{2}),\xi)
\right\rbrace^{m}.
\end{split}
\end{align}
where
\begin{align}
\begin{split}
F_{C}((\tau_{1}-\tau_{2}),\xi)= \\
\left\lbrace {e^{i \vartheta_{g}}\Delta\kappa_{+}(\tau_{1}-\tau_{2})+e^{-i \vartheta_{g}}\Delta\kappa_{-}(\tau_{1}-\tau_{2})}\right\rbrace 
\end{split}
\end{align}
is the "kernel" function being responsible for the four different layouts of two time arguments $ \tau_{1}$ and $ \tau_{2}$ on the complex Keldysh contour: $ \mathcal{C}_K \in (0 \pm i0;t) $. 
Obviously, infinite series (B64,B65) re-exponentiates exactly into the following compact form
\begin{eqnarray}\label{eq:reex}
\tilde{\chi}(t,\xi)=\exp \left\lbrace - \mathcal{F}(\xi, t ) \right\rbrace 
\end{eqnarray}
where for the generating functional $\mathcal{F}(\xi,t)$ one has following \textit{exact} formula
\begin{align}
\begin{split}
\mathcal{F}(\xi,t)=\left(\tilde{\lambda}\right)^{2}\times \\
\left\lbrace \frac{1}{2}\int^{t}_{0}d \tau_{1} \int^{t}_{0} d \tau_{2}u(\tau_{1}-\tau_{2})F_{C}((\tau_{1}-\tau_{2}),\xi)\right\rbrace \\
=\left(\tilde{\lambda}\right)^{2}\frac{1}{2}\int^{t}_{0}d \tau_{1} \int^{t}_{0} d \tau_{2}u(\tau_{1}-\tau_{2})\times \\ \left\lbrace {e^{i \vartheta_{g}}\Delta\kappa_{+}(\tau_{1}-\tau_{2})+e^{-i \vartheta_{g}}\Delta\kappa_{-}(\tau_{1}-\tau_{2})}\right\rbrace. 
\end{split}
\end{align}
Remarkably, Eq.(B67) can be also rewritten in the form of a double integral from our "initial" (non-symmetrized) average $\langle A_{\xi(\tau)}(\tau_{1}) A_{\xi(\tau)}(\tau_{2})\rangle$ over complex Keldysh contour $ \mathcal{C}_K(t) \in (0 \pm i0;t) $  
\begin{align}
\begin{split}
\mathcal{F}(\xi,t)=\left(\tilde{\lambda}\right)^{2}\times \\
\left\lbrace \frac{1}{2}\int\int_{\mathcal{C}_K(t)} d \tau_{1}d \tau_{2}\langle \mathcal{T}_K A_{\xi(\tau)}(\tau_{1}) A_{\xi(\tau)}(\tau_{2})\rangle \right\rbrace.
\end{split}
\end{align}
Evidently, equations (B66-B68) complete the proof of the SE-Theorem. \textit{Therefore, the SE-Theorem is proven.} $\blacksquare$

\section{Exact calculation of a basic time integral in the long-time limit}\label{sec:time_int}

To calculate real-time integrals $ J_{C} $ and $ J_{S} $ from Eqs.(45,46) in their "long-time" limit: $ t \rightarrow \infty $ let us perform it as following sum
\begin{align}
 J_{C(S)}= \frac{(J_{d}^{+} \pm J_{d}^{-})}{4}
\end{align}
where 
\begin{eqnarray}
 \nonumber
 J_{d}^{\pm}= \left[\frac{\pi T}{\Lambda_{g}}\right]^{2/g} \lim_{\delta \rightarrow 0} \int^{+\infty}_{-\infty} ds \frac{e^{\pm i(eVs)}e^{-i(\pi/g) \sgn(s)}}{\sinh^{2/g}\left[\pi T(s-i \delta_{g} \sgn(s))\right]} .\\
  \label{eq:intd_pm}
\end{eqnarray} 
In the latter formula one can replace infinitesimal $ \delta_{g} \rightarrow 0$ introducing instead the principal value of corresponding integral "without" the infinitesimal vicinity of the point $ s=0 $. The result is
\begin{align}
\begin{split}
 J_{d}^{\pm}= \left[\frac{2\pi T}{\Lambda_{g}}\right]^{2/g}\times \\
  \mathcal{P} \int^{+\infty}_{-\infty} ds \frac{e^{\pm i(eVs)}}{\left[e^{\pi Ts+i(\pi/2)\sgn(s)}+e^{-\pi Ts-i(\pi/2)\sgn(s)}\right]^{2/g}}
 \end{split} 
\end{align}

Now changing variable of integration to $ z=e^{\pi Ts+i\pi/2} $ and $ ds=z^{-1}dz/\pi T $ one obtains
\begin{eqnarray}
\nonumber
 J_{d}^{\pm}=\frac{e^{\pm eV/2T}}{\pi T} \left[\frac{2\pi T}{\Lambda_{g}}\right]^{2/g} \mathcal{P} \int^{+i\infty}_{+i0}  \frac{z^{(2/g-1) \pm i eV/\pi T}}{\left[z^{2}+1\right]^{2/g}}dz .\\
  \label{eq:intd_pm}
\end{eqnarray}
And, finally, changing $ z $ to $ \tilde{z}=z^{2} $ we arrive the expression 
\begin{eqnarray}
 J_{d}^{\pm}=J_{C}^{\pm}\cdot \frac{e^{\pm eV/2T}}{\Lambda_{g}} \left[\frac{2\pi T}{\Lambda_{g}}\right]^{(2/g-1)}   
  \label{eq:intd_pm}
\end{eqnarray}
where we define following complex integral
\begin{eqnarray}
\nonumber
 J_{C}^{\pm}=-\mathcal{P} \int^{0}_{-\infty}u_{\pm}(\tilde{z})d \tilde{z}
  =-\mathcal{P} \int^{0}_{-\infty}  \frac{\tilde{z}^{[1/g-1 \pm i eV/2\pi T]}}{\left[\tilde{z}+1\right]^{2/g}}d \tilde{z}. \\
  \label{eq:intd_pm}
\end{eqnarray}
Notice, that symbol $ \mathcal{P} $ in r.h.s. of Eq.(C6) means that we have excluded the point $\tilde{z}=-1$ with its infinitesimal  vicinity from  the integration along the real axis.
 
Let us consider the integral in the r.h.s. of Eq.(C6) on the complex plane. Function $u_{+}(\tilde{z}) $ ($u_{-}(\tilde{z}) $)under the integral is analytical everywhere on the upper (lower) half of a complex plane, except the point: $\tilde{z}=-1 $, where the function $u_{+}(\tilde{z_{k}}) $ (or $u_{-}(\tilde{z_{k}}) $) diverges. Further, let us introduce for the integrals $ J_{C}^{\pm} $ two closed complex contours $ \mathcal{C}_{\pm} $ consisting of free parts each. Let first part of both contours $  \mathcal{C}_{\pm} $ goes from $ \tilde{z}=-\infty $ to $ \tilde{z}=0 $ along the real axis, being infinitesimally bent around the point $ \tilde{z}=-1 $  in the upper- (for the contour $  \mathcal{C}_{+} $ ) and in the lower (for the contour $  \mathcal{C}_{-} $ ) halves of complex plane. Then, let the second part of contours $  \mathcal{C}_{\pm} $ goes from  $ \tilde{z}=0 $ to $ \tilde{z}=+\infty $ along the real axis. And, finally, let us "close"  our contours $  \mathcal{C}_{+} $ ( $  \mathcal{C}_{-} $) by its third part i.e. by the semi-circle of infinite radius ($ R_{\pm}\rightarrow \infty $) in the upper (lower) half of a complex plane. 
 
 Both introduced closed contours $ \mathcal{C}_{\pm} $ contain no special points for functions $u_{\pm}(\tilde{z}) $ correspondingly. It means that, by Residue theorem, the integral from the function $u_{+}(\tilde{z}) $ (or $u_{-}(\tilde{z}) $) along closed complex contour $ \mathcal{C}_{+} $ (or $ \mathcal{C}_{-} $, correspondingly) - is equal to zero. Thus, one can write
\begin{eqnarray}
\nonumber
 \oint_{\mathcal{C}_{\pm}}u_{\pm}(\tilde{z})d \tilde{z}=\int_{I(\pm)}u_{\pm}(\tilde{z})d \tilde{z}+\int_{II(\pm)}u_{\pm}(\tilde{z})d \tilde{z}+ \\
 \nonumber
 \int_{III(\pm)}u_{\pm}(\tilde{z})d \tilde{z}=0. \\
  \label{eq:int_cc}
\end{eqnarray} 
 Now, from the form of the third parts of our contours as well as from the analytic properties of complex functions $u_{\pm}(\tilde{z}) $ one can conclude that
\begin{eqnarray}
\int_{III(\pm)}u_{\pm}(\tilde{z})d \tilde{z}=0.
  \label{eq:int_a}
\end{eqnarray}
Evidently, from Eqs.(C7,C8) it follows
\begin{eqnarray}
\int_{I(\pm)}u_{\pm}(\tilde{z})d \tilde{z}+\int_{II(\pm)}u_{\pm}(\tilde{z})d \tilde{z}=0
  \label{eq:int_b}
\end{eqnarray} 
But, on the other hand, by the definition of chosen contours $ \mathcal{C}_{\pm} $ we have
\begin{eqnarray}
\nonumber
\int_{I(\pm)}u_{\pm}(\tilde{z})d \tilde{z}=-J_{C}^{\pm}. \\
  \label{eq:int_c}
  \end{eqnarray} 
Equations (C9,C10) allow us to re-define our  "initial" complex integral (C6) as
\begin{eqnarray}
J_{C}^{\pm}=\int_{II(\pm)}u_{\pm}(\tilde{z})d \tilde{z}=\int^{+\infty}_{0}u_{\pm}(\tilde{z})d \tilde{z}
  \label{eq:int_d}
\end{eqnarray} 
It means that
\begin{eqnarray}
J_{C}^{\pm}=\int^{+\infty}_{0}\frac{\tilde{z}^{[1/g-1 \pm i eV/2\pi T]}}{\left[\tilde{z}+1\right]^{2/g}}d \tilde{z}.
  \label{eq:int_e}
\end{eqnarray}
The latter integral, in turn, is well-tabulated and is nothing more than the integral representation of beta-function $ B(p;q) $ of two complex arguments $ p,q $
\begin{eqnarray}
B(p;q)=\int^{+\infty}_{0}\frac{\tilde{z}^{[p-1]}}{\left[\tilde{z}+1\right]^{p+q}}d \tilde{z}.
  \label{eq:int_eb}
\end{eqnarray}
Comparing Eqs.(C12,C13) one can conclude that in our case $ p=q^{\ast}=1/g+i(eV/2\pi T) $ and, hence,
\begin{eqnarray}
J_{C}^{+}=J_{C}^{-}=B\left( \left[ \frac{1}{g}+i\frac{eV}{2\pi T}\right]  ; \left[ \frac{1}{g}-i\frac{eV}{2\pi T}\right]  \right).
  \label{eq:int_ej}
\end{eqnarray}
Finally, using well-known identity
\begin{eqnarray}
B(p;q)=B(q;p)=\frac{\Gamma(p)\Gamma(q)}{\Gamma(p+q)}
  \label{eq:int_eb}
\end{eqnarray}
with Euler's gamma-function $ \Gamma(z) $ of complex argument, one can obtain following explicit formula for the integral $ J_{C}^{\pm} $, which is just a real number
\begin{eqnarray}
J_{C}^{+}=J_{C}^{-}=\frac{\vert\Gamma\left(1/g+i\left[eV/2\pi T\right]\right)\vert^{2}}{\Gamma\left(2/g\right)}.
  \label{eq:int_m}
\end{eqnarray}
Substituting formula (C16) into Eq.(C5) we will have
\begin{eqnarray}
\nonumber
 J_{d}^{\pm}=\frac{e^{\pm eV/2T}}{\Lambda_{g}}\left[\frac{2\pi T}{\Lambda_{g}}\right]^{(2/g-1)}\frac{\vert\Gamma\left(1/g+i\left[eV/2\pi T\right]\right)\vert^{2}}{\Gamma\left(2/g\right)}.\\
  \label{eq:int_fur}
\end{eqnarray}
In general, formulas being similar to Eq.(C17) are known in the literature on Luttinger liquid effects(see Ref.[43]).

At last, substituting formulas (C17) into Eq.(C1) we obtain the desired expressions for the "long-time" asymptotics ($ t\rightarrow \infty $) of our two basic real time integrals from Eqs.(45,46)
\begin{eqnarray}
\nonumber
 J_{C}=\frac{\cosh(eV/2T)}{2\Lambda_{g}}\left[\frac{2\pi T}{\Lambda_{g}}\right]^{(2/g-1)}\frac{\vert\Gamma\left(1/g+i\left[eV/2\pi T\right]\right)\vert^{2}}{\Gamma\left(2/g\right)}.\\
  \label{eq:int_basic}
  \nonumber
\end{eqnarray}
\begin{equation}
\end{equation}
and
\begin{eqnarray}
\nonumber
 J_{S}=\frac{\sinh(eV/2T)}{2\Lambda_{g}}\left[\frac{2\pi T}{\Lambda_{g}}\right]^{(2/g-1)}\frac{\vert\Gamma\left(1/g+i\left[eV/2\pi T\right]\right)\vert^{2}}{\Gamma\left(2/g\right)}.\\
  \label{eq:int_basic}
  \nonumber
\end{eqnarray}
\begin{equation}
\end{equation}
Obviously, formulas (C17-C19) are exact, they give the expressions (37,38) from the main text and represent main "building blocks", carrying  voltage- and temperature-dependent "steady-flow" (i.e. steady state-equilibrated) Luttinger liquid effects in our model.


\begin{references}
\bibitem{53} Such Gibbs factors are usually generalized by the integration over imaginary part of Keldysh contour for arbitrary Hamiltonians. However, within taken model of two QPC electrodes as two independent gases of free bosons (or, equally, plasmons), such imaginary axis integration becomes completely equivalent to the common thermal averaging over ground state of Luttinger liquid electrodes being performed here. This is because respective non-interacting Luttinger liquid Hamiltonian $ H_{LL} $ of QPC leads is quadratic in bosonic variables (see Eq.(18)) or, in other words, because taken model of QPC electrodes can be perfectly described in terms of well-defined quasi-particles ( i.e. in terms of two sorts ($ \pm $) of non-interacting non-local plasmons, see Eqs.(17-19)) within the bosonization picture. However, the latter fact does not mean one should put any additional restrictions on the type of thermal fluctuations in the system, instead the "gaussianity" of relevant thermal fluctuations in the system follows automatically from the fact, that non-interacting part $ H_{LL} $ of the system's Hamiltonian is quadratic, i.e. it allows for the existence of quasi-particles (chiral plasmons) with well-defined energy spectrum of Eq.(18).
\bibitem{55} In order to obtain the r.h.s. of Eqs.(B53,B54) and (42) from the l.h.s. of these equations one should notice a following FCS scheme-induced property for the basic involved symmetric "free" pair correlator: $ u(\tau_{l+1} - \tau_{l} + t) = \langle A_{0}(\tau_{l+1} + t ) A_{0}(\tau_{l})\rangle_{S} = \langle A_{0}(\tau_{l+1}) A_{0}(\tau_{l})\rangle_{S} = u(\tau_{l+1} - \tau_{l}) $  where $ t $ is a fixed time interval of transport measurements performed on given QPC (notice, that in the l.h.s. of Eqs.(B53,B54) this time interval $ t $ is also an upper limit for the $ m $-th time integration). This property, in turn, follows from quite obvious fact that, in practice, one needs to "destroy" the junction by "cutting" it at the moment $ t $ of interest, in order to be able to measure the distribution function $ \mathcal{P}( N, t ) $ and all its moments with well-defined classical variable $ N $ - the number of electrons being transmitted through a given tunnel junction up to a given moment of time $ t $ - by means of "detector" Hamiltonian $ H_{\sigma}(t) $ (see also Section 3 of the main text). Therefore, two  moments of time $ \tau'=0 + \tau  $ and $ \tau'' = t + \tau $ should be equivalent with respect to the accepted FCS scheme. Hence, for  pair correlator $ u(\tau) $ the equality $ u(\tau')=u(\tau'') $ should also be fulfilled at these two moments of time. Notice, though, that the latter property does not imply any "real" periodicty on the basic free pair correlator $ u(\tau) $ defined by e.g. Eq.(34). Instead, the above property one can think of as just a proper continuation of a function $ u( \tau )  $ to the region $ \tau > t $, which, however, remains beyond any experimentalist's interest concerning the FCS scheme of tunnel electrons measurement we are discussing here. Nevertheless, this inherent property of the QPC model in the FCS regime of transferred charge measurements allows one to "factorize"  the  $ m $-folded time-integrals of Eqs.(B53,B54,42).
\end{references}
\end{document}